\documentclass[a4paper,11pt]{article}
\pdfoutput=1

\usepackage[geometry=arxiv,physics=compatible,caption=lily,subfigure=subfig,tikz=lily,quiver=off,biblatex=lily,hyperref=lily,style=off]{lily}

\begin{document}
\title{The Shadow Formalism of Galilean CFT\secmath{_2}}
\vspace{14mm}
\author{
Bin Chen$^{1,2,3}$, Reiko Liu$^1$\footnote{bchen01@pku.edu.cn, reiko\_liu@pku.edu.cn}
}
\date{}

\maketitle

\begin{center}
	{\it
		$^{1}$School of Physics and State Key Laboratory of Nuclear Physics and Technology,\\Peking University, No.5 Yiheyuan Rd, Beijing 100871, P.~R.~China\\
		\vspace{2mm}
		$^{2}$Collaborative Innovation Center of Quantum Matter, No.5 Yiheyuan Rd, Beijing 100871, P.~R.~China\\
		$^{3}$Center for High Energy Physics, Peking University, No.5 Yiheyuan Rd, Beijing 100871, P.~R.~China
	}
	\vspace{10mm}
\end{center}

\begin{abstract}
In this work, we develop the shadow formalism for two-dimensional Galilean conformal field theory (GCFT$_2$). We define the principal series representation of Galilean conformal symmetry group and find its relation with the Wigner classification, then we determine the shadow transform of local operators. Using this formalism we derive the OPE blocks, Clebsch-Gordan kernels, conformal blocks and conformal partial waves. A new feature is that the conformal block admits additional branch points, which would  destroy the convergence of OPE for certain parameters. We establish another inversion formula different from the previous one, but get the same result when decomposing the four-point functions in the mean field theory (MFT). We also construct a continuous series of bilocal actions of MFT, and an exceptional series of local actions, one of which is the BMS free scalar model. We notice that there is an outer automorphism of the Galilean conformal symmetry, and the GCFT$_2$ can be regarded as null defect in higher dimensional CFTs.

\end{abstract}

\baselineskip 18pt

\newpage

\tableofcontents{}

\newpage

\section{Introduction}

The conformal bootstrap program \cite{Ferrara:1973yt,Polyakov:1974gs} provides a nonperturbative framework to study the conformal field theories, without using the Lagrangian. By solving the crossing equation, it allows us to extract the CFT data, the operator spectrum and the three-point coefficients, with the help of unitarity and symmetry \cite{Poland:2018epd}. The revival of conformal bootstrap in the past decade since the seminal work in \cite{Rattazzi:2008pe} has brought up huge developments in both numerical and analytic studies \cite{ElShowk:2012ht,ElShowk:2014dwa,Komargodski:2012ek,Fitzpatrick:2012yx,Kaviraj:2015cxa,Kaviraj:2015xsa,Alday:2016njk,Caron-Huot:2017vep,Simmons-Duffin:2017nub,Mazac:2016qev,Mazac:2018mdx,Mazac:2018ycv,Mazac:2019shk,Paulos:2019gtx,Bissi:2022mrs}, and has shed light on the AdS/CFT correspondence \cite{Heemskerk:2009pn,Penedones:2010ue,Fitzpatrick:2011ia,Alday:2016htq} and S-matrix bootstrap \cite{Paulos:2016fap,Paulos:2016but,Paulos:2017fhb,Homrich:2019cbt}.

One interesting question is whether the conformal bootstrap can be extended to the theories with other types of conformal-like symmetries. Such conformal-like symmetries usually come from non-Lorentzian geometries invariant under dilatation\footnote{The superconformal field theories on superspace, conformal defects on stratified space and $p-$adic CFT on $\Q_{p}$ can also be regarded as examples.}, including e.g. Schrodinger conformal symmetry \cite{Henkel:1993sg,Nishida:2007pj,Goldberger:2014hca,Golkar:2014mwa,Kravec:2018qnu,Karananas:2021bqw,Shimada:2021xsv}, Carrollian conformal symmetry and Galilean conformal symmetry in higher-dimensions ($d\geq 3$) \cite{Bagchi:2017yvj,Bagchi:2019xfx,Bagchi:2019clu,Banerjee:2020qjj,Chen:2021xkw}. For the two-dimensional ($2d$) non-Lorentzian geometries, the conformal-like symmetries are even richer, including e.g. warped conformal symmetry \cite{Hofman:2011zj,Detournay:2012pc,Chen:2020juc}, anisotropic Galilean conformal symmetries \cite{Chen:2019hbj} and $2d$ Galilean conformal symmetry.

In particular, the $2d$ Galilean conformal algebra is isomorphic to the $2d$ Carrollian conformal algebra and the BMS$_3$ algebra, the latter of which plays an important role in $3d$ flat holography \cite{Bagchi:2009my,Bagchi:2012cy,Barnich:2012xq,Bagchi:2012xr,Bagchi:2014iea,Jiang:2017ecm,Hijano:2018nhq,Hijano:2019qmi,Apolo:2020bld,Apolo:2020qjm}. The $2d$ Galilean conformal algebra consists of two sets of generators, satisfying the following commutation relations
\begin{align}
[L_n,L_m]&=(n-m)L_{n+m}+c_L\delta_{n+m,0}(n^3-n),\\
[L_n,M_m]&=(n-m)M_{n+m}+c_M\delta_{n+m,0}(n^3-n),\nn\\
[M_n,M_m]&=0.\nn
\end{align}
The global subalgebra is generated by $\set{L_{\pm 1}, L_0, M_{\pm 1}, M_0}$ and is isomorphic to $\isolie(2,1)$. It can be obtained in several different ways: by taking the non-relativistic and ultra-relativistic contractions of the relativistic conformal algebra; by considering the conformal structure of the $2d$ Newton-Cartan spacetime and the $2d$ Carrollian spacetime; as the $3d$ Poincare algebra acting on the null infinity of the flat spacetime; as part of the residual symmetry of null defect in $3d$ Lorentzian CFT. 

In \cite{Chen:2020vvn}, we initiated the study on bootstrapping the $2d$ Galilean conformal field theory (GCFT$_2$) based on the global Galilean conformal algebra, and mainly focused on $\xi\neq 0$ sector\footnote{For the $\xi=0$ sector, the study will appear soon in \cite{Chen:2022jhx}.}. We have studied the decomposition of the Hilbert space into the quasi-primary states, have computed the conformal partial waves and checked the consistency of the program by studying the mean field theory in different ways. It turns out that $2d$ Galilean conformal bootstrap is viable, even though the theory is non-unitary. Our study has revealed a few novel features in Galilean conformal bootstrap. 
Firstly, there exist multiplet representations in the Hilbert space, which share similar features as the logarithmic multiplets in the logarithmic CFT \cite{Hogervorst:2016itc}. To distinguish them, we call these multiplets as boost multiplets. Secondly the boost multiplets satisfy a modified version of the Casimir equations, and appear in the inversion function as the multiple poles rather than the simple poles. Finally, harmonic analysis for the GCFT$_2$ is quite subtle, since the global Galilean conformal algebra is non-semisimple.

In this work, we would like to continue our study on $2d$ Galilean conformal bootstrap. We revisit the harmonic analysis and establish the shadow formalism of GCFT$_2$. In our former study \cite{Chen:2020vvn}, the technical treatment followed closely the one in CFT$_1$ \cite{Maldacena:2016hyu,Polchinski:2016xgd,Murugan:2017eto}, i.e. spectral decomposition of the Casimir operators. To define the Hilbert space properly, we determined the inner product and boundary conditions, and the conformal partial waves supported on the whole cross-ratio plane $(x,y)\in \R^{2}$. Moreover we showed that the conformal partial waves could not be reached by taking non-relativistic limit on the ones of $2d$ conformal group, as the normalizable condition and boundary conditions should be analyzed in a way independent of the non-relativistic limit. On the other hand, when using the inversion function, it is only necessary to work in the region $(x,y)\in (0,1)\times \R$. This inspires us to develop the shadow formalism for a better understanding of analytic Galilean bootstrap.

The shadow formalism relies heavily on the representation theory of the conformal group. It was firstly developed in the early 1970s
\cite{Ferrara:1972ay,Ferrara:1972uq,Ferrara:1972xe,Ferrara:1973vz,Mack:1975je,Mack:1976pa,Dobrev:1976vr,Dobrev:1977qv}, 
and was revisited in the modern bootstrap
\cite{Osborn_2012,SimmonsDuffin:2012uy,Karateev:2017jgd,Karateev:2018oml,Kravchuk:2018htv,Liu:2018jhs}.
% The references here are kept same as those at the beginning of section 3.
To develop the shadow formalism for GCFT$_2$, we have to understand the unitary irreducible representations of $2d$ Galilean conformal group. It is isomorphic to the $3d$ Poincare group, whose unitary irreducible representations are classified by the Wigner-Mackey method. We will construct the unitary principal series representation of the $2d$ Galilean conformal group, and then define the shadow transform in GCFT$_2$. With the shadow transform, we compute the operator product expansion (OPE) blocks, the conformal blocks and conformal partial waves in GCFT$_2$ in the framework of shadow formalism. Furthermore, we study several applications of the shadow formalism, including the decomposition of four-point functions in mean field theory, the construction of bilocal actions with Galilean conformal symmetry.

The remaining parts of the paper are organized as follows. In section 2, we give a brief review of $2d$ Galilean conformal field theory. In section 3, we discuss the representations of $2d$ Galilean conformal group and define the shadow transform. In section 4, we derive the OPE blocks and Clebsch-Gordan kernels. In section 5, we discuss the conformal blocks and conformal partial waves in the shadow formalism. In section 6, we discuss several applications of the shadow formalism. In section 7, we end with conclusions and discussions. There are a few appendices. In appendix A, we summarize the conventions and notations in this work. In appendix B, we provide a review on the kinematics and shadow formalism of CFT$_1$. In appendix C, we show how to get a Carrollian CFT$_2$ on a null conformal defect in Lorentzian CFT$_3$. In appendix D, we present the details of some calculations.

\section{Review of Galilean/Carrollian CFT\secmath{_2}}

In this section, we briefly review the kinematical aspects of the two dimensional Galilean conformal field theory (GCFT$_2$), including the symmetry algebra, local operators and correlation functions. For more complete discussions, see \cite{Chen:2020vvn,Chen:2022jhx}. In this work, we are going to consider the global Galilean conformal algebra $\isolie(2,1)$, and in the following will refer to the quasi-primary operators with respect to the local Galilean conformal algebra $\lie{gca}_2\simeq \lie{bms}_3$ as the primary operators for short. 

The global Galilean conformal algebra $\isolie(2,1)$ singularly acts on the plane $\R^2$ with coordinates $(\xy)$, as summarized in table \ref{table:global_gca}. It is generated by $\{L_{0,\pm 1},M_{0,\pm 1}\}$, with the commutation relations
\begin{align}
[L_{n},L_{m}]& =(n-m)L_{n+m},\\
[L_{n},M_{m}]& =(n-m)M_{n+m},\nn\\
[M_{n},M_{m}]& =0,\hs{3ex}n,m=-1,0,1.\nn
\end{align}
In two dimensions the Galilean conformal symmetries and the Carrollian conformal symmetries are isomorphic due to the coincidence of $2d$ Carrollian structures and $2d$ Newton-Cartan structures\footnotemark{} \cite{Duval:2014lpa,Duval:2014uva}. From physical point of view, the $x$ coordinate serves as the temporal direction in the Galilean geometry. In contrast, the $y$ coordinate serves as the temporal one in the Carrollian case. Hence we may use the terms Galilean and Carrollian interchangeably.

\footnotetext{This Carrollian structure was revisited in the framework of $G$-structures recently \cite{Figueroa-OFarrill:2019sex,Figueroa-OFarrill:2020gpr,Figueroa-OFarrill:2021sxz}.}

\begin{table}[htbp]
    \centering
    \tablevspace{2.4}
    \begin{tabular}{l L| L|L}
    name                & \text{charge} &\text{vector field}     & \text{finite transformation}   \\
    \hline
    $x$-translation     & L_{-1}    & \p_x              & \makecell[l]{x'=x+a\\y'=y}\\
    dilation            & L_{0}     & x\p_x+y\p_y       & \makecell[l]{x'=\lm\,x\\y'=\lm\,y}\\
    $x$-SCT             & L_{1}     & x^2\p_x+2x y\p_y          & \makecell[l]{x'=x/(1-\mu x)\\y'=y/(1-\mu x)^{2}}\\
    $y$-translation     & M_{-1}    & -\p_y                & \makecell[l]{x'=x\\y'=y+b}\\
    boost               & M_{0}     & -x\p_y             & \makecell[l]{x'=x\\y'=y+v\,x}\\
    $y$-SCT             & M_{1}     & -x^2\p_y                & \makecell[l]{x'=x\\y'=y+\nu x^{2}}\\
    inversion           & I         &                          & \makecell[l]{x'=-1/x\\y'=y/x^2}\\
    \end{tabular}
    \caption{The generators of global Galilean conformal group in 2D. The last line is the inversion which is useful to check conformal covariance. Notice that because of the reversed order of successive actions $[Q_{u},[Q_{v},\op(x)]]=D_{v}D_{u} \op(x)$, we have $[Q_{v},Q_{u}]=Q_{-[v,u]}$.}
    \label{table:global_gca}
\end{table}

\subsection{Local operators: singlet and boost multiplet}\label{sec:localoperators}

\textbf{Singlet and boost multiplet.} In GCFT$_2$, the primary operators in a boost multiplet inserted at the origin $\op^{a}=\op^{a}(0,0)$ can be characterized by the eigenvalues $(\D,\xi)$ of $(L_0,M_0)$ and the rank $r$ of the boost multiplet, where the superscript runs from $a=1,2,\dots, r$. When $r=1$ it reduces to the singlet case and the trivial index will be dropped.
The definition of a boost multiplet is as follows: the action of dilatation is diagonalized, the two translations act as derivative operators and the two special conformal transformations (SCTs) annihilate the primaries, namely
\begingroup
\allowdisplaybreaks
\begin{alignat}{2}
& [L_0,\op^{a}]=\D \op^{a},          &\hs{3ex}  & \forall a=1,\cdots r, \\
& [L_{-1},\op^{a}]=\p_{x}\op^{a},    &\hs{3ex}  & [M_{-1},\op^{a}]=-\p_{y}\op^{a},\nn\\
& [L_1,\op^{a}]=0,                   &\hs{3ex}  & [M_1,\op^{a}]=0.\nn
\end{alignat}
\endgroup
Simply speaking, the primary operators in a boost multiplet share the same scaling dimension.
The action of boost $M_{0}$ gives a rank-$r$ upper Jordan block\footnote{Another convention of the $\xi$-matrix is lower Jordan block: the subscript of $\op_{a}$ runs from $a=0,1,\dots ,r-1$. The two conventions are related by $\op^{a}=\op_{r-a}$.}, and equivalently we have
\begin{equation}
[M_{0},\op^{a}]=\xi^{a}_{b}\op^{b}=\xi \op^{a} + \op^{a+1}
\label{eq:actionofM0}
\end{equation}
with the conditions $\op^{a}=0$ if $a\leq 0$ or $a>r$. The relation \eqref{eq:actionofM0} is formally solved by
\begin{equation}
\op^{a}=\frac{1}{(r-a)!}\p_{\xi}^{r-a}\op_{\D,\xi},
\end{equation}
implying that the primary operators in a multiplet can be treated as the $\xi$-derivatives of a singlet operator. The descendant operators are $(-1)^{m}\p^{n}_{x}\p^{m}_{y}\op^{a}$, in which the minus sign is due to $[M_{-1},\op^{a}]=-\p_{y}\op^{a}$. The primary operators in a boost multiplet together with their descendants form a (generalized) highest weight representation with weight $(\D,\xi)$. This defines a rank-$r$ boost multiplet and we denote it as $\rep{\D,\xi,r}$ in the following.

The infinitesimal transformations of the primary operators are
\begingroup
\allowdisplaybreaks
\begin{align}
    [L_n,\op^{a}(\xy)]&=\ppair{( x^{n+1}\p_x + (n+1)\D x^n + (n+1) x^n y \p_y)\delta^a_b - n(n+1)\xi_{b}^{a} x^{n-1} y)}\op^{b}(\xy),\nn\\
    [M_n,\op^{a}(\xy)]&=\ppair{-x^{n+1}\delta^a_b\p_y  + (n+1)\xi_{b}^{a} x^n}\op^{b}(\xy), \hs{3ex}n=\pm1,0,
    \label{eq:localoperatortrans}
\end{align}
\endgroup
and the finite transformations are
\begin{equation}
    % U(f,g)\cdot \op^{a}(\xy)=|f'|^{\D}\sum^{r-a}_{n=0}\frac{1}{n!}\p_{\xi}^{n}\exp\ppair{-\xi\frac{g'+y f''}{f'}}\,\op^{a+n}(x',y'),
    U(f,g)\op^{a}(\xy)U^{-1}(f,g)=|f'|^{\D}\sum^{r-a}_{n=0}\frac{1}{n!}\p_{\xi}^{n}\exp\ppair{-\xi\frac{g'+y f''}{f'}}\,\op^{a+n}(x',y'),
\label{eq:globaltransformationsmultiplet}
\end{equation}
where $x'=f(x),\,y'=f'(x)y+g(x)$ are the global Galilean conformal transformations as shown in table \ref{table:global_gca}. The convention of the translation operator is\footnote{Another convention is $U(\xy)=e^{xL_{-1}+yM_{-1}}$, and the replacement rule $y\to -y$ swaps them.} $U(\xy)=e^{xL_{-1}-yM_{-1}}$. We see that $M_{0},M_{1},L_{1}$ mix $\op^{a}(\xy)$ with $\op^{a+1}(\xy)$.

\textbf{State-operator correspondence.} Assuming the conformal invariance of vacuum state $\vac$,  the state-operator correspondence (SOC) for a single operator\footnote{The single-operator SOC is kinematical, and the multiple-operator SOC relies on the convergence and operator content of the OPE.} is given by
\begin{equation}
\ket{\op^{a}}=\lim_{x\to 0}\lim_{y\to 0}\op^{a}(\xy)\vac=\lim_{\substack{k \to 0 \\ x \to 0}}\op^{a}(x,k x)\vac,\label{operatorstate}
\end{equation}
then one can switch between the states and the operators interchangeably. Notice that the order of taking $x$-limit and $y$-limit cannot be changed in some circumstances. In the last equality of \eqref{operatorstate}, the slope coordinates $(x,k)$ with
\begin{equation}
k=y/x,\label{slope}
\end{equation}
were adopted to resolve the singularity at $(\xy)=(0,0)$.

There are two types of complete bases of a boost multiplet. The first one consists of the primaries and their descendant operators inserted at the origin $\set{\op,\p_{x}\op,-\p_{y}\op, \dots}$. By the state-operator correspondence they are mapped to the states $\set{\ket{\op},L_{-1}\ket{\op},M_{-1}\ket{\op},\dots}$. The second one consists of the primary operators at different points $\set{\op(\xy)\vac:(\xy)\in\R^{2}}$, which are related by the mode expansion acting on the vacuum state
\begin{equation}
\op^{a}(\xy)\vac=\sum_{n=0}^{\inf}\frac{x^{n}y^{m}}{n!m!}\p^{n}_{x}\p^{m}_{y} \op^{a}(0,0)\vac.
\end{equation}

For a rank-$r$ boost multiplet $\rep{\D,\xi,r}$, the descendant states are
\begin{equation}
\ket{a,n,m}_r=L_{-1}^n M_{-1}^m\ket{\op^{a}_{\D,\xi,r}}=(-1)^{m}\p_{x}^{n}\p_{y}^{m}\op^{a}(0,0)\vac,\hs{3ex}n,m\in \Zpositive,
\end{equation}
and $l=n+m$ is called the level since $L_{0}\ket{a,n,m}_r=(\D+l)\ket{a,n,m}_r$. The actions of the generators of $\isolie(2,1)$ on the descendant states are
\begingroup
\allowdisplaybreaks
\begin{align}
\numberthis
\label{eq:actionsondescendants}
L_{k}\ket{a,n,m}_r  & = \frac{n!}{(n-k)!}\big[ (n+m+\D+k(-1+m+\D))\ket{a,n-k,m}_r\\
                    & \peq +\xi \frac{k(1+k)m}{(n-k+1)}\ket{a,n-k+1,m-1}_r\nn\\
                    & \peq +\frac{k(1+k)m}{(n-k+1)}\ket{a+1,n-k+1,m-1}_r \big],\nn\\
M_{k}\ket{a,n,m}_r  & = \frac{n!}{(n-k)!}\big[\xi(k+1)\ket{a,n-k,m}_r
                    +(n-k)\ket{a,n-k-1,m+1}_r\nn\\
                    & \peq +(k+1)\ket{a+1,n-k,m}_r \big],\nn
\end{align}
\endgroup
where $k=-1,0,1$ and $\ket{a,n,m}_r=0$ if $a>r$. This is equivalent to the commutation relation of primary operators \eqref{eq:localoperatortrans}.  The mixing between the descendants of $\ket{a}_r$ and $\ket{a+1}_r$ happens for $M_{0},M_{1},L_{1}$.

\textbf{Out-state and inner product.} The physical conjugation relation is the BPZ-like conjugation
\begin{equation}
L^{\dagger}_{n}=L_{-n},\hspace{3ex} M^{\dagger}_{n}=M_{-n}.
\end{equation}
The out-state can be defined as
\begin{equation}
\bra{\op^{a}}= \lim_{\substack{k\to 0\\ x\to \inf}} |x|^{2\D} \exp(-2\xi k) \bra{0} \sum_{n=0}^{r-a}\frac{(-2k)^{n}}{n!} \op^{a+n}(x,k x).
\end{equation}
And the inner product of primary states
\begin{equation}
\braket*{\op^{a}}{\op^{b}}=\d_{a+b,r+1}
\end{equation}
is anti-diagonal and contains $\lfloor\frac{r}{2}\rfloor$ negative norms, which is also a common feature in Logarithmic CFTs. Such an indefinite inner product on the highest weight representation is called the Shapovalov form \cite{humphreys2021representations}, and has been used to analyse the null states in relativistic CFTs, see e.g. \cite{Penedones2016,Yamazaki:2016vqi,Pasterski:2021fjn}. In GCFT$_2$, the Gramian matrix of this inner product and the null states for boost multiplets are obtained in \cite{Chen:2022jhx}, and it turns out that the $\xi=0$ boost multiplets behave drastically different from the ones with $\xi\neq 0$. In most places of this paper we assume the boost charges of exchanged operators are nonvanishing.

\textbf{Bosonic vs. fermionic.} Similar to CFT$_1$, the operators can be commutative or anti-commutative, and the infinitesimal transformations \eqref{eq:localoperatortrans} cannot distinguish them. The finite transformations of fermionic primaries are modified by the multiplier $c(f,g)=\sign\left(1- \mu x\right)$ if the $x$-SCT is involved,
\begin{equation}
    U(f,g)\op^{a}(\xy)U^{-1}(f,g)=c(f,g)|f'|^{\D}\sum^{r-a}_{n=0}\frac{1}{n!}\p_{\xi}^{n}\exp\ppair{-\xi\frac{g'+y f''}{f'}}\,\op^{a+n}(x',y').
\end{equation}
Accordingly the power factors $|x_{ij}|^{\a}$ in correlation functions should be replaced by $\sign(x_{ij})|x_{ij}|^{\a}$. In most of the following sections we only consider the bosonic operators, and there is no essential difference when discussing fermionic operators.

\subsection{Ward identities of two-point functions}\label{sec:wardidentities}
In this subsection we list the Ward identities of two-point functions, since they will reappear in several circumstances later. The primary operators are denoted as $\op_i^{a}\in \rep{\D_i,\xi_i,r}$ with $\xi$-matrices $(\xi_{i})^{a}_{b}$, and the default position of $\op_i^{a}$ is $(\xy{i})$ unless otherwise specified. The two-point functions are denoted as $K^{ab}:=K^{ab}(\xy{12})=\vev{\op_1^{a}(\xy{1}) \op_2^{b}(\xy{2})}$, where $x_{12}= x_1-x_2,\, y_{12}= y_1-y_2$ and $k_{12}=\frac{y_{12}}{x_{12}}$. In the following Ward identities, the corresponding generators are $M_{0},\, L_{0},\, M_{1},\, L_{1}$ respectively. 
\begingroup
\allowdisplaybreaks
\begin{align}
\intertext{Singlet:}
&\ppair{x_{12}\p_{y_1}-\xi_{1}-\xi_{2}}K(\xy{12})=0,\label{eq:wardsinglet}\\
&\ppair{x_{12}\p_{x_1}+y_{12}\p_{y_1}+\D_1+\D_2}K(\xy{12})=0,\nn\\
&\bpair{(x_{1}^{2}-x_{2}^{2})\p_{y_1} - 2(\xi_1 x_1 +\xi_2 x_2)}K(\xy{12})=0,\nn\\
&\bpair{(x_{1}^{2}-x_{2}^{2})\p_{x_1} + 2(x_{1}y_{1}-x_{2}y_{2})\p_{y_1} +2(\D_1 x_1 +\D_2 x_2 - \xi_1 y_1 -\xi_2 y_2)}K(\xy{12})=0.\nn\\
\intertext{Boost Multiplet:}
&x_{12}\p_{y_1}K^{ab}-(\xi_{1})^{a}_{c}K^{cb}-(\xi_{2})^{b}_{c}K^{ac}=0,\label{eq:wardmultiplet}\\
&\ppair{x_{12}\p_{x_1}+y_{12}\p_{y_1}+\D_1+\D_2}K^{ab}=0,\nn\\
&(x_{1}^{2}-x_{2}^{2})\p_{y_1}K^{ab} - 2 x_1 (\xi_{1})^{a}_{c}K^{cb}-2x_2(\xi_{2})^{b}_{c}K^{ac} =0,\nn\\
&\bpair{(x_{1}^{2}-x_{2}^{2})\p_{x_1}+ 2(x_{1}y_{1}-x_{2}y_{2})\p_{y_1}+2(\D_1 x_1 +\D_2 x_2 )}K^{ab}- 2 y_1(\xi_{1})^{a}_{c}K^{cb} -2 y_2(\xi_{2})^{b}_{c}K^{ac}=0.\nn
\end{align}
\endgroup

There are three types of solutions of these equations, which will be referred to as continuous, exceptional and discrete types for later convenience. The continuous type of solutions is a linear combination of $|x_{12}|^{-2\D_1}$ $k_{12}^{i}e^{2\xi_1 k_{12}}$ and will be reviewed in the next section \ref{sec:type1}. Similar to relativistic CFTs, the two conformal families are forced by the equations of $L_{1},\, M_{1}$ to be identical, $\D_{12}=0,\, \xi_{12}=0$. We mainly focus on the continuous type in this work.

The exceptional type of solutions exists only when $\xi_1+\xi_2=0$ and is a linear combination of $\d^{(i)}(x_{12})|y_{12}|^{-\D_1-\D_2+1+i}$. The two conformal families are not necessarily identical. They appear in e.g. the bilocal actions in section \ref{sec:bilocalsinglet}. This type of solutions also appears in higher dimensional Carrollian and Galilean CFTs \cite{Chen:2021xkw}, and is relevant to the proposed relations between Carrollian CFT and celestial CFT \cite{Donnay:2022aba,Bagchi:2022emh}. 

The discrete type of solutions exists when $\xi_1+\xi_2=0,\, \D_1+\D_2\in\Z$ and is a linear combination of $\d^{(i)}(x_{12})\d^{(\D_1+\D_2-2-i)}(y_{12})$. The further restrictions on the weights from the Ward identities are different from the ones in the exceptional type. They appear in e.g. the inner product of the principal series representations in section \ref{sec:derivationshadow}. More complicatedly, the three types of solutions can mix with each other when the conditions on the weights and the charges in different types are satisfied simultaneously.

\subsection{Correlation functions}\label{sec:type1}

\textbf{Singlets.} The two-point functions of singlets are diagonalized as
\begin{equation}
\vev{\op_1 \op_2}=\d_{12}|x_{12}|^{-2\D}\exp(2\xi k_{12}), \qquad \D=\D_1=\D_2,\,\xi=\xi_1=\xi_2
,
\end{equation}
where $k_{12}=\frac{y_{12}}{x_{12}}$. The three-point functions are of the form
\begin{equation}
\vev{\op_1 \op_2 \op_3}
=c_{123}|x_{12}|^{-\D_{12,3}}|x_{23}|^{-\D_{23,1}}|x_{31}|^{-\D_{31,2}}\,
\exp\ppair{\xi_{12,3}k_{12}+\xi_{23,1}k_{23}+\xi_{31,2}k_{31}}
,
\end{equation}
where $c_{123}$ are the three-point coefficients and
\begin{equation}
\D_{ij,k}\equiv\D_i+\D_j-\D_k,\hs{3ex}\xi_{ij,k}\equiv\xi_i+\xi_j-\xi_k.
\end{equation}
The four-point function can be written as a product of the stripped four-point function $\fourpt^{\schannel}(\set{\op_i},\xy)$ containing the dynamical information  and a kinematical factor $K^{\schannel}(x_i,y_i)$ compensating the conformal covariance of the four-point function
\begin{equation}
\vev{\op_1 \op_2 \op_3 \op_4}= 
K^{\schannel}(x_i,y_i)
\fourpt^{\schannel}(\xy)
.
\label{eq:schannel1gcft}
\end{equation}
Here we are considering the \schannel $\op_1\times \op_2\to \op_3\times \op_4$, and we find the following kinematical factor 
\begin{align}
K^{\schannel}(x_i,y_i)
& =
|x_{12}|^{-(\D_1+\D_2)}|x_{34}|^{-(\D_3+\D_4)} 
\abs{\frac{x_{24}}{x_{14}}}^{\D_{12}}
\abs{\frac{x_{14}}{x_{13}}}^{\D_{34}}
\nn
\\
& \peq \cdot\exp\bpair{(\xi_1+\xi_2)k_{12}+(\xi_3+\xi_4)k_{34}-\xi_{12}(k_{24}-k_{14})-\xi_{34}(k_{14}-k_{13})}
\label{eq:kinematical1gcft}
\end{align}
is convenient for \schannel OPE. As a result the stripped conformal blocks \eqref{eq:strippedblock1} depend only on $\D_{ij}\equiv \D_i-\D_j,\xi_{ij}\equiv\xi_i-\xi_j$. For clarity the \tchannel $\op_2\times \op_3\to \op_1\times \op_4$ stripped four-point function is obtained by the permutation $(13)$,
\begin{equation}
\vev{\op_1 \op_2 \op_3 \op_4}= 
K^{\tchannel}(x_i,y_i)
\fourpt^{\tchannel}(1-x,-y),
\label{eq:tchannel1gcft}
\end{equation}
where the kinematical factor is,
\begin{align}
K^{\tchannel}(x_i,y_i)
& =
|x_{23}|^{-(\D_2+\D_3)}|x_{14}|^{-(\D_1+\D_4)} 
\abs{\frac{x_{24}}{x_{34}}}^{\D_{32}}
\abs{\frac{x_{34}}{x_{13}}}^{\D_{14}}
\nn
\\
& \peq \cdot \exp\bpair{
(\xi_2+\xi_3)k_{23}+(\xi_1+\xi_4)k_{14}-\xi_{23}(k_{24}-k_{34})-\xi_{14}(k_{34}-k_{13})
}.
\end{align}
The $s-t$ crossing equation from \eqref{eq:schannel1gcft} and \eqref{eq:tchannel1gcft} leads to 
\begin{equation}
x^{-(\D_1+\D_2)}\exp[(\xi_1+\xi_2)\frac{y}{x}]\fourpt^{\schannel}(\xy)=(1-x)^{-(\D_2+\D_3)}\exp[(\xi_2+\xi_3)\frac{-y}{1-x}]\fourpt^{\tchannel}(1-x,-y), \label{eq:stcrossinggcft} 
\end{equation}
where the crossing region is $(\xy)\in (0,1)\times \R$.

The standard conformal frame of four points can be chosen as $\set{(0,0),(\xy),(1,0),(\inf,0)}$. Then the inner product interpretation of the four-point function is
\begin{equation}
\bra{\op_{4}}\op_{3}(1,0)\op_{2}(\xy)\ket{\op_{1}}
=\lim_{\substack{x_4\to \inf\\ k_4\to 0}}\vev{\op_{1}(0) \op_{2}(\xy) \op_{3}(1,0) \op_{4}(x_{4},k_{4}x_{4})}|x_4|^{2\D_4}e^{-2\xi_4 k_4},
\label{eq:innerproductfourpoint}
\end{equation}
and its relation to the \schannel stripped four-point function is
\begin{equation}
\bra{\op_{4}}\op_{3}(1,0)\op_{2}(\xy)\ket{\op_{1}}=
x^{-(\D_1+\D_2)}\exp[(\xi_1+\xi_2)\frac{y}{x}]\fourpt^{\schannel}(\xy).
\end{equation}

\textbf{Boost multiplet.} The two-point functions of different boost multiplets vanish. For the same multiplet $\rep{\D,\xi,r}$ its two-point functions form an left-upper triangular matrix
\begin{equation}
\label{eq:twopointrankr}
\vev{\op^{a} \op^{b}}=\vev{\op\op}_{\singletstructure} 
\case{
\frac{1}{n!} \ppair{2 k_{12}}^{n} & \qif n=r+1-a-b\geq 0
\\
0& \qelse
}
\end{equation}
where $\vev{\op\op}_{\singletstructure}$ is the two-point structure\footnote{For $n=2,3$ by $n$-point structure we mean the conformal-covariant functions appearing in correlation functions without $\d_{12}$ or $c_{123}$.} of a singlet.

The three-point functions of $\op_i^{a_{i}}\in \rep{\D_i,\xi_i,r_i}$ are
\begin{equation}
\vev{\op_1^{a_{1}} \op_2^{a_{2}} \op_3^{a_{3}}}
=\vev{\op_1 \op_2 \op_3}_{\singletstructure}K^{a_{1}a_{2}a_{3}}(k_{23,1},k_{31,2},k_{12,3}),
\label{eq:threepoint1}
\end{equation}
where $\vev{\op_1 \op_2 \op_3}_{\singletstructure}$ is the three-point structure of singlets, $k_{ij,l}=k_{li}+k_{jl}-k_{ij}$, and
\begin{equation}
K^{a_{1}a_{2}a_{3}}(k_{23,1},k_{31,2},k_{12,3})=
\sum_{n_{1}=0}^{r_{1}-a_{1}}\sum_{n_{2}=0}^{r_{2}-a_{2}}\sum_{n_{3}=0}^{r_{3}-a_{3}}
\frac{(k_{23,1})^{n_{1}}}{n_{1}!}
\frac{(k_{31,2})^{n_{2}}}{n_{2}!}
\frac{(k_{12,3})^{n_{3}}}{n_{3}!}
c_{123}^{(a_{1}+n_{1}),(a_{2}+n_{2}),(a_{3}+n_{3})}.
\end{equation}
There are $r_{1}r_{2}r_{3}$ independent three-point coefficients $c_{123}^{abc}$ if no further constraints are imposed. As an example, the three-points functions of two singlets and a rank-$r$ multiplet are
\begin{align}
\label{eq:threepoint11r}
\vev{\op_1 \op_2 \op_3^{1}}
& =\vev{\op_1 \op_2 \op_3}_{\singletstructure}
\ppair{c^r\frac{k_{12,3}^{r-1}}{(r-1)!}+\dots +c^3 \frac{k_{12,3}^{2}}{2!}+c^2 k_{12,3}+c^1}
\\
\vev{\op_1 \op_2 \op_3^{2}}
& =\vev{\op_1 \op_2 \op_3}_{\singletstructure}
\ppair{c^{r}\frac{k_{12,3}^{r-2}}{(r-2)!}+\dots +c^3 k_{12,3}+c^2}
\nn
\\
\vdots\nn
\\
\vev{\op_1 \op_2 \op_3^{r}}
& =\vev{\op_1 \op_2 \op_3}_{\singletstructure}
\hspace{1.2ex} c^{r}
\nn
\end{align}
in which there are $r$ three-point coefficients $c^{a}:=c_{123}^{11a},\,a=1,\dots,r$.

\section{Shadow Transforms}
In this section we briefly review the ideas of the shadow formalism in relativistic CFT, and then discuss its analog in GCFT$_2$. The shadow formalism was developed in \cite{Ferrara:1972ay,Ferrara:1972uq,Ferrara:1972xe,Ferrara:1973vz,Mack:1975je,Mack:1976pa,Dobrev:1976vr,Dobrev:1977qv} and was applied to the modern bootstrap in e.g. \cite{Osborn_2012,SimmonsDuffin:2012uy,Karateev:2017jgd,Karateev:2018oml,Kravchuk:2018htv,Liu:2018jhs}. It's based on the representation theory and harmonic analysis of the conformal algebras and groups.
% The references here are kept same as those in the introduction section.

\textbf{Symmetry.} 
The Euclidean conformal algebra $\glie_E=\solie(d+1,1)$ and the Lorentzian one $\glie_L=\solie(d,2)$, are different real slices of the complex Lie algebra $\solie(d+2,\mathbb{C})$, hence one's complex\footnotemark{} representation is naturally the other's representation. 

\footnotetext{In the representation theory, we are interested in the complex representations due to the complex nature of Hilbert space in physics and technical simplifications in mathematics.}

For a classical symmetry group $G$, the physical projective representation on the Hilbert space corresponds to the linear representation of the universal covering group $\wave{G}$. The two groups are related by modding out the fundamental group, $G=\wave{G}/\pi_{1}(G)$. 
The classical Euclidean conformal group is $G_E=SO(d+1,1)$ with $\pi_{1}(G_E)=\Z_2,\, d\geq 2$, and if spinors are involved we need to consider the double covering group $\spin(d+1,1)$. 

The fundamental group of the classical Lorentzian conformal group $G_L=SO(d,2)$ is a little bigger: $\pi_{1}(G_L)\simeq \pi_{1}(SO(d))\times \pi_{1}(SO(2))=\Z_2\times \Z,\, d\geq 3$, and unlike the spin group the universal covering group $\wave{G}_L$ is not a linear Lie group, i.e. it cannot be embedded as a linear subgroup of $GL(n,\C)$ for any finite $n$.

\textbf{Representation.} 
There are various types of representations appearing in relativistic CFT. The first type describes physical operators. The operators located in Lorentzian region $\op_L(x),\,x\in \mathbb{R}^{d-1,1}$ and the one in Euclidean region $\op_E(x),\,x\in \mathbb{R}^{d}$ are mapped to the same state $\ket{\op}$ by the state-operator correspondence, then the conformal family containing complex linear combinations of $\ket{\op}$ and its descendants is simultaneously the representation of the Lorentzian and Euclidean conformal algebras. 

In a Lorentzian CFT, the physical requirement of unitarity is that the conformal family is unitary with respect to $\glie_L$. By the Wick rotation the physical unitarity is transformed into the reflection positivity in the Euclidean theory. This leads to restrictions on the conformal dimension and the spin of the primary operator, named as the unitary bound. At the Lie group level, the conformal family satisfying the unitary bound is a unitary representation of the Lorentzian conformal group $\wave{G}_L$, named as the discrete series representation, and is a non-unitary representation of the Euclidean conformal group $G_E$.

Another type of representations is the not-necessarily-unitary principal series representation of the Euclidean conformal group $G_E$, taking arbitrary complex conformal dimension $\D\in \mathbb{C}$. And they are irreducible for generic values of $\D$. This type of representation is neither the highest nor the lowest weight module, and does not correspond to a physical operator. Imposing the unitarity condition with respect to $G_E$, i.e. the existence of $G_E$-invariant positive-definite inner product, the unitary principal series $\D=\frac{d}{2}+is$ are picked out. 

\textbf{Harmonic analysis.} 
Back to the field theory, inserting a complete basis of the physical Hilbert space into the correlation function, we get a summation of inner products labelled by the exchanged states. The exchanged states are organized into conformal families of $\wave{G}_L$, hence in the summation we can separate the contributions from different exchanged conformal families and obtain the conformal block expansion. 

On the other hand, the correlation function is also covariant function on some homogeneous space of $G_E$. Since the unitary principal series representations of $G_E$ provide a complete basis for decomposing normalizable functions on $G_E$\footnotemark{} and its homogeneous spaces, using the Euclidean inversion formula the correlation function can be decomposed into conformal partial waves corresponding to the unitary principal series.

\footnotetext{In odd dimensions there are also discrete series appearing in the reduced unitary dual.}

The reason that the two different aspects, conformal blocks corresponding to the unitary representations of $\wave{G}_L$, and conformal partial waves corresponding to the unitary representations of $G_E$, are simply related by analytic continuation of $\D$ and linear combination, can be traced back to the fact that the cyclic vectors (corresponding to the primary operators) in these two types of the representations share the same transformation rules under the conformal transformations
\begin{equation}\op^a(x)\to \det(\pdv{x'}{x})^{\D/d} M^a_b\op^b(x').\end{equation}
In the following for convenience we will call the ``operators" with analytic continued weight as virtual operators, since they are not in the physical Hilbert space, only providing a complete basis in the decomposition of correlation functions.

\textbf{GCFT$_2$.} Different from the relativistic conformal algebras, the ``Wick rotation" of the Galilean conformal algebra $\isolie(2,1)$ is isomorphic to itself. This is similar to the case of CFT$_1$, as reviewed in section \ref{sec:cft1}.

The first type of representations in GCFT$_2$ includes the singlet and multiplet representations with real weight $(\D,\xi) \in \R$. Despite of being non-unitary generically they describe physical operators, like the conformal families in Euclidean CFTs not satisfying the unitary bound. This is acceptable since non-unitary theories are common in Euclidean CFTs, e.g., all the logarithmic CFTs and most of the $2d$ minimal models. 

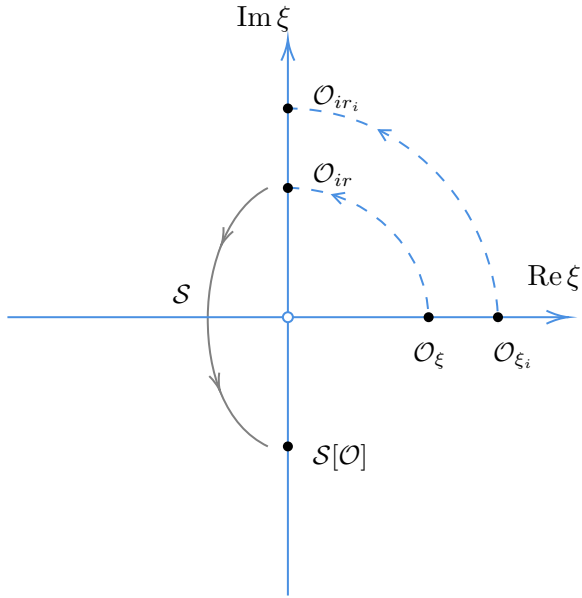
\begin{figure}[htbp]
\centering

\tikzset{every picture/.style={line width=0.75pt}} %set default line width to 0.75pt        

\begin{tikzpicture}[x=0.75pt,y=0.75pt,yscale=-1,xscale=1]
%uncomment if require: \path (0,539); %set diagram left start at 0, and has height of 539

%Straight Lines [id:da8246282135716239] 
% \draw [color={rgb, 255:red, 255; green, 255; blue, 255 }  ,draw opacity=1 ]   (203.5,245) -- (198.5,245) ;
%Shape: Arc [id:dp9764296629802391] 
\draw  [draw opacity=0][dash pattern={on 4.5pt off 4.5pt}] (199.7,115) .. controls (238.15,114.71) and (269.69,144.24) .. (270.28,181.23) .. controls (270.28,181.29) and (270.29,181.36) .. (270.29,181.42) -- (200.15,182.35) -- cycle ; \draw  [color={rgb, 255:red, 74; green, 144; blue, 226 }  ,draw opacity=1 ][dash pattern={on 4.5pt off 4.5pt}] (199.7,115) .. controls (238.15,114.71) and (269.69,144.24) .. (270.28,181.23) .. controls (270.28,181.29) and (270.29,181.36) .. (270.29,181.42) ;
\draw  [color={rgb, 255:red, 74; green, 144; blue, 226 }  ,draw opacity=1 ] (226.55,123.02) -- (222.31,118.47) -- (228.59,117.44) ;

%Shape: Arc [id:dp8335762260103781] 
\draw  [draw opacity=0][dash pattern={on 4.5pt off 4.5pt}] (198.83,74.99) .. controls (256.33,74.42) and (303.75,122.5) .. (304.96,182.88) .. controls (304.96,182.94) and (304.96,183) .. (304.96,183.06) -- (200,184.98) -- cycle ; \draw  [color={rgb, 255:red, 74; green, 144; blue, 226 }  ,draw opacity=1 ][dash pattern={on 4.5pt off 4.5pt}] (198.83,74.99) .. controls (256.33,74.42) and (303.75,122.5) .. (304.96,182.88) .. controls (304.96,182.94) and (304.96,183) .. (304.96,183.06) ;
\draw  [color={rgb, 255:red, 74; green, 144; blue, 226 }  ,draw opacity=1 ] (248.8,91.34) -- (245.45,85.94) -- (251.8,85.98) ;

%Curve Lines [id:da28164540560248796] 
\draw [color={rgb, 255:red, 128; green, 128; blue, 128 }  ,draw opacity=1 ]   (190,115) .. controls (171.31,124.58) and (160.2,152) .. (160,180) .. controls (159.8,208) and (168.68,234.6) .. (190,245) ;
\draw [shift={(166.74,143.77)}, rotate = 293.71] [color={rgb, 255:red, 128; green, 128; blue, 128 }  ,draw opacity=1 ][line width=0.75]    (8.74,-2.63) .. controls (5.56,-1.12) and (2.65,-0.24) .. (0,0) .. controls (2.65,0.24) and (5.56,1.12) .. (8.74,2.63)   ;
\draw [shift={(166.08,217.02)}, rotate = 250.97] [color={rgb, 255:red, 128; green, 128; blue, 128 }  ,draw opacity=1 ][line width=0.75]    (8.74,-2.63) .. controls (5.56,-1.12) and (2.65,-0.24) .. (0,0) .. controls (2.65,0.24) and (5.56,1.12) .. (8.74,2.63)   ;
%Straight Lines [id:da3889976272790474] 
\draw [color={rgb, 255:red, 74; green, 144; blue, 226 }  ,draw opacity=1 ]   (200,42) -- (200,320) ;
\draw [shift={(200,40)}, rotate = 90] [color={rgb, 255:red, 74; green, 144; blue, 226 }  ,draw opacity=1 ][line width=0.75]    (10.93,-3.29) .. controls (6.95,-1.4) and (3.31,-0.3) .. (0,0) .. controls (3.31,0.3) and (6.95,1.4) .. (10.93,3.29)   ;
%Straight Lines [id:da4164818279010085] 
\draw [color={rgb, 255:red, 74; green, 144; blue, 226 }  ,draw opacity=1 ]   (338,180) -- (60,180) ;
\draw [shift={(340,180)}, rotate = 180] [color={rgb, 255:red, 74; green, 144; blue, 226 }  ,draw opacity=1 ][line width=0.75]    (10.93,-3.29) .. controls (6.95,-1.4) and (3.31,-0.3) .. (0,0) .. controls (3.31,0.3) and (6.95,1.4) .. (10.93,3.29)   ;
%Shape: Circle [id:dp7580682987313128] 
\draw  [color={rgb, 255:red, 74; green, 144; blue, 226 }  ,draw opacity=1 ][fill={rgb, 255:red, 255; green, 255; blue, 255 }  ,fill opacity=1 ] (197.5,180) .. controls (197.5,178.62) and (198.62,177.5) .. (200,177.5) .. controls (201.38,177.5) and (202.5,178.62) .. (202.5,180) .. controls (202.5,181.38) and (201.38,182.5) .. (200,182.5) .. controls (198.62,182.5) and (197.5,181.38) .. (197.5,180) -- cycle ;

% Text Node
\draw (188,30) node    {$\Im \xi $};
% Text Node
\draw (261,191.4) node [anchor=north west][inner sep=0.75pt]  [font=\small]  {$\mathcal{O}_{\xi}$};
% Text Node
\draw (301,191.4) node [anchor=north west][inner sep=0.75pt]  [font=\small]  {$\mathcal{O}_{\xi_{i}}$};
% Text Node
\draw (211,101.4) node [anchor=north west][inner sep=0.75pt]  [font=\small]  {$\mathcal{O}_{i r}$};
% Text Node
\draw (211,242.4) node [anchor=north west][inner sep=0.75pt]  [font=\small]  {$\shadow[\op]$};
% Text Node
\draw (211,62.4) node [anchor=north west][inner sep=0.75pt]  [font=\small]  {$\mathcal{O}_{i r_{i}}$};
% Text Node
\draw (141,162.4) node [anchor=north west][inner sep=0.75pt]  [font=\small]  {$\shadow$};
% Text Node
\draw (333,160) node    {$\Re \xi $};

\draw [fill={rgb, 255:red, 0; green, 0; blue, 0 }  ,fill opacity=1 ]  (200, 115) circle [x radius= 2, y radius= 2]   ;
\draw [fill={rgb, 255:red, 0; green, 0; blue, 0 }  ,fill opacity=1 ]  (270.25, 180) circle [x radius= 2, y radius= 2]   ;
\draw [fill={rgb, 255:red, 0; green, 0; blue, 0 }  ,fill opacity=1 ]  (200, 74.99) circle [x radius= 2, y radius= 2]   ;
\draw [fill={rgb, 255:red, 0; green, 0; blue, 0 }  ,fill opacity=1 ]  (304.87, 180) circle [x radius= 2, y radius= 2]   ;
\draw [fill={rgb, 255:red, 0; green, 0; blue, 0 }  ,fill opacity=1 ]  (200, 245) circle [x radius= 2, y radius= 2]   ;
\end{tikzpicture}

\caption{Virtual operators lie on the principal series $\xi=i\Rnot$. The external and exchanged operators should be analytic continued simultaneously keeping the ratios $R_i$ real. The case that the exchanged operator is degenerate $\xi=0$ should be handled separately.}
\label{fig:analyticcontinuation1}
\end{figure}

The second type is the unitary principal series representation of the Galilean conformal group with complex weight $\D=1+is,\, \xi=ir$. The cyclic vectors in the two types of representations follow the same transformation rule, suggesting the viability of the shadow formalism in GCFT$_2$. The procedure of analytic continuation of weight $(\D,\xi)$ is shown in Figure \ref{fig:analyticcontinuation1}. In the rest of this section we will discuss the principal series representations and the shadow transforms as the starting point of the shadow formalism.

\subsection{Unitary principal series representations}\label{sec:ups}

Since the $2d$ Galilean conformal group is isomorphic to the $3d$ Poincare group, the ``unitary principal series" representations should be identified as unitary irreducible representations of the Poincare group, which has been classified by using the Wigner-Mackey method \cite{Wigner:1939cj,bargmann1997group,mackey1958unitary}, see also e.g. \cite{Bekaert:2006py,Campoleoni:2016vsh}. To make the shadow transform rigorous, we firstly construct the unitary principal series, then in the next subsection identify them with the tachyonic unitary representation of the Poincare group.

\textbf{Definition.} We define the unitary principal series representation $\urep{\D,\xi}$ of $ISO(2,1)$ as follows: the representation space is $L^2(\R^2)\ni f(\xy)$, with the inner product
\begin{equation}
    (f_1,f_2)=\int_{\R^2}dxdy\,f_1^*(\xy)f_2(\xy),
    \label{eq:innerproductshadow}
\end{equation}
and the group action is the same as the one on the singlet primary operators \eqref{eq:globaltransformationsmultiplet} but with complex weight $(\D=1+is,\xi=ir),\,r \in \Rneq,\,s\in\R$,
\begin{equation}
    U(a,b)\cdot f(\xy)=|a'|^{\D}\,\exp\ppair{-\xi\frac{b'+ya''}{a'}}\,f(x',y'),
    \label{eq:globaltransformations}
\end{equation}
where the global Galilean conformal transformations $(f,g)$ are relabeled as $(a,b)$. The infinitesimal transformations are the same as those of primary operators \eqref{eq:localoperatortrans}. The inner product is invariant under the action due to the selected weight, hence defining a unitary representation. 

We emphasize that this unitarity is not the physical unitarity, and the conjugation relation on generators is not the BPZ conjugation $L^{\dagger}_{n}=L_{-n},\, M^{\dagger}_{n}=M_{-n}$. Instead, the ``Euclidean" conjugation relation is the default anti-Hermitian one,
\begin{equation}
L^{\dagger}_{n}=- L_{n},\hspace{3ex} M^{\dagger}_{n}=-M_{n},
\label{eq:euclideanconjugation}
\end{equation}
which can be checked directly. In section \ref{sec:derivationshadow} we will verify this point again.

\textbf{Irreducibility.} For infinite dimensional representations on Hilbert spaces $\H_{1},\,\H_{2}$, the intertwining map from $\H_1$ to $\H_2$ can be written as a distributional kernel by the Schwartz kernel theorem
\begin{equation}
K:  f(x)\in \H_1 \mapsto \int dx f(x) K(x,x') \in \H_{2},
\end{equation}
and the commutativity with the group action is
\begin{equation}
\int dx_1dx_2\, K(x_2,x_3) U_1(g,x_1,x_2) f(x_1) = \int dx_1dx_2\, U_2(g,x_2,x_3) K(x_1,x_2) f(x_1).
\end{equation}
A representation is irreducible if any bounded self-intertwining map is proportional to the identity map, and in the following the concept of irreducibility is in this sense, see e.g. \cite{folland2016course,Dobrev:1977qv}. There are other definitions of irreducibility, and there can be further subtleties from finite transformations down to the infinitesimal ones. We omit these technical issues for simplicity.

To check $\urep{\D,\xi}$ is irreducible or not, we determine the self-intertwining map $K:\urep{\D,\xi}\to \urep{\D,\xi}$,
\begin{equation}
(K\cdot f)(\xy{1})=\inttt{dx_{2} dy_{2}}{\R^2} K(\xy{12}) f(\xy{2})
\label{eq:selfintertwiner}
\end{equation}
by requiring $K$ commute with infinitesimal transformations $X\in \isolie(2,1)$ 
\begin{equation}
\intt{dx_2 dy_2} K(\xy{12}) X_{2}f(\xy{2})=X_{1}\intt{dx_2 dy_2} K(\xy{12}) f(\xy{2})
\end{equation}
in which $K(\xy)$ is the distributional kernel. In the above relations, we have used the translation $L_{-1},\,M_{-1}$ to restrict the kernel depending on $(\xy{12})$. After doing integration by parts, the generators $M_{0},\, L_{0},\, M_{1},\, L_{1}$ lead to four equations of $K(\xy)$, and they are related to the two-point Ward identities \eqref{eq:wardsinglet} by the replacement 
\begin{equation}
(\D_1,\xi_1,\D_2,\xi_2)\to(\D,\xi,2-\D,-\xi),
\end{equation}
due to the conformal covariance property of \eqref{eq:selfintertwiner}. The distributional solutions of the equations with respect to $L_{0},\, M_{0}$ are
\begin{equation}
K(\xy)=\d(x)\d(y)+a_{1}\d(x)y^{-1}+a_{2}x^{-2}+a_{3}\d'(x).
\end{equation}
In the case $\xi\neq0$, the equation of $L_{1}$ restricts $a_{1}=0$ and the one of $M_{1}$ restricts $a_{2}=a_{3}=0$ such that
\begin{equation}
K(\xy)=\d(x)\d(y).
\end{equation}
Hence we conclude that $\urep{1+is,ir}$, $s \in \R,\,r\in\Rnot$ is a unitary irreducible representation of the Galilean conformal group.

\subsection{Relation to tachyonic representations}
In this subsection we identify the unitary principal series representation constructed in section \ref{sec:ups} with the tachyonic representation. By the Wigner classification, the unitary irreducible representation of $ISO(2,1)$ is induced from the representation of the stabilizer group of the mass-shell, and can be distinguished by the eigenvalues of Casimir elements.
For the principal series representation $\urep{\D=1+is,\xi=ir}$, we find that the eigenvalues of the Casimirs take $m=i r$, hence it corresponds to the tachyonic representation.
Contrary to the massive and the massless representations, the mass-shell of the tachyonic representation is an one-sheeted hyperboloid and the two masses $m=\pm ir$ label the same mass-shell, indicating the existence of self-intertwining map from $m=-ir$ to $m=ir$.

Following the convention in the appendix \ref{sec:app1}, the signature of $\R^{2,1}\ni (x^{0},x^{1},x^{2}) $ is $(-1,1,1)$, and the commutation relations of the Poincare algebra are
\begin{align}
& [M_{ab},M_{cd}]=g_{ad}M_{bc}+g_{bc}M_{ad}-g_{ac}M_{bd}-g_{bd}M_{ac},\\
& [M_{ab},P_{c}]=-g_{ac}P_{b}+g_{bc}P_{a}.
\end{align}
The conjugation relation is $Q^{\dagger}=-Q$, and $M_{ab}$'s acting on the projective nullcone serve as the generators of $1d$ conformal algebra,
\begin{equation}
L_{0}=-M_{01}, \hs L_{-1}=-M_{20}-M_{12}, \hs L_{1}=M_{20}-M_{12}.
\end{equation}
Then extending the above relations to the whole Galilean conformal algebra, we get
\begin{equation}
    M_{0}=-P_{2}, \hs M_{-1}=P_{0}-P_{1}, \hs  M_{1}=P_{0}+P_{1},
\end{equation}
and the conjugation relation \eqref{eq:euclideanconjugation} is preserved. This identification is exactly the same as \eqref{eq:inf}. Then the Casimirs are\footnote{Notice that the momentum $P_{a}$ is anti-Hermitian in our convention.}
\begin{align}
C_1& =m^2=P_{a}P^{a}=M^2_0-M_{-1}M_1,\\
C_2& =\e^{abc}M_{ab}P_{c}=2(L_0-1) M_0-L_{-1}M_1-M_{-1}L_1.
\label{eq:casimirPoincare}
\end{align}

The Casimirs act on the principal series representation $\urep{\D=1+is,\xi=ir}$ as scalars,
\begin{equation}
C_1=-r^2<0, \hs C_2=-2 sr,
\end{equation}
hence from the first Casimir we find $\urep{\D=1+is,\xi=ir}$ is tachyonic.

\subsection{Shadow transforms}

In this subsection we try to establish the shadow transform of $ISO(2,1)$, the global conformal group in Galilean CFT$_2$. The construction is similar to that of the Euclidean conformal group $SO(d+1,1)$ \cite{Dobrev:1977qv,knapp1971interwining,SimmonsDuffin:2012uy,Karateev:2018oml}. When discussing the shadow formalism, the notations $\vev{\op_1\op_2}$ and $\vev{\op_1 \op_2 \op_3}$ mean the two-point and three-point structures of the singlets, and the default position of the operator $\op_i$ is $(\xy{i})$ unless otherwise specified.

For the unitary principal series $\urep{\D=1+is,\xi=ir},\,s\in \R,r \in \Rneq$, we define the associated shadow representation as $\urep{\wave{\D}=2-\D,\wave{\xi}=-\xi}$, and denote the virtual operator transforming in $\urep{\wave{\D},\wave{\xi}}$ as $\opshadow$.
For a virtual operator $\op$ lying on the unitary principal series $\urep{1+is,ir}$, we construct the shadow transform $\shadow$ as
\begin{align}
    \shadow{\op}(\xy)
    & =\int_{\R^2}dx_0 dy_0\,\vev{\opshadow(\xy) \opshadow(\xy{0})}\op(\xy{0})     \label{eq:shadowgcft2}\\
    & =\int_{\R^2}dx_0 dy_0\,|x-x_0|^{2\D-4} e^{-2\xi \frac{y-y_0}{x-x_0}}\op(\xy{0}),
\end{align}
which is an intertwining map between the two representations
\begin{equation}
\shadow: \urep{\D,\xi} \to \urep{\wave{\D},\wave{\xi}}.
\end{equation}
If the representations $\urep{\D,\xi}$ and $\urep{\wave{\D},\wave{\xi}}$ are unitary and irreducible, by Schur lemma $\shadow$ is an isomorphism, otherwise the kernel subspace $\ker \shadow\in \urep{\D,\xi}$ and the image subspace $\im \shadow\in \urep{\wave{\D},\wave{\xi}}$ can be subrepresentations, or even worse, the integration kernel of $\shadow$ is ill-defined as a tempered distribution. Applying the shadow transform twice $\shadow^2: \urep{\D,\xi} \to \rep{\D,\xi}$,
\begin{eqnarray}
\lefteqn{\int dx_1dy_1\, K(\shadow^2,\xy{1},\xy{2}) \op(\xy{1})}\nn\\
&&=\int dx_1dx_1\, dx_0dy_0\,\vev{\opshadow(\xy{1})\opshadow(\xy{0})}\vev{\op(\xy{0})\op(\xy{2})}\op(\xy{1})
\end{eqnarray}
and in the case that $\shadow$ is an isomorphism, the intertwining kernel $K(\shadow^2)$ should be proportional to the $\d$-distribution,
\begin{equation}
K(\shadow^2)=\cN(\D,\xi) \d(x_{12},y_{12}).
\label{eq:doubleshadow}
\end{equation}
The prefactor $\cN(\D,\xi)$ can be calculated as follows
\begingroup
\allowdisplaybreaks
\begin{align}
    K(\shadow^2,\xy{1},\xy{2})
    & =\int dx_0dy_0\,\vev{\opshadow(\xy{1}) \opshadow(\xy{0})}\vev{\op(\xy{0})\op(\xy{2})}\nn\\
    & =\int_{\R^2}dx_0dy_0\,|x_{01}|^{2\D-4}|x_{02}|^{-2\D} 
    e^{-2ir\frac{y_0 x_{12}}{x_{01}x_{02}}}
    e^{2ir\frac{x_0y_{12}+x_1y_2-x_2y_1}{x_{01}x_{02}}} \nn\\
    & =\frac{\pi}{|\xi|}\d(x_1-x_2)\int_{\R}dx_0\,|x_{01}|^{-2}e^{2ir\frac{y_{12}}{x_{01}}} \nn\\
    & =\frac{\pi^2}{|\xi|^2}\d(x_1-x_2)\d(y_1-y_2),
\end{align}
\endgroup
where in the second line the integration of $y_0$ contributes to $\d(x_1-x_2)$, in the third line the simplification is due to $x_1=x_2$ and in the last line we change the variable $\frac{1}{x_0-x_1}=t$. This determines
\begin{equation}
\cN(\D,\xi)=\frac{\pi^2}{|\xi|^2}.
\label{eq:inversePlancherel}
\end{equation}
In CFT the factor $\cN^{-1}(\D,\xi)$ is proportional to the Plancherel measure \cite{Karateev:2018oml}. In GCFT$_2$ we find that the factor $\cN^{-1}(\D,\xi)$ is in match with the Plancherel measure of the tachyonic representations \cite{ASENS_1972_4_5_3_459_0,ASENS_1973_4_6_1_103_0}.

\subsection{Derivation of shadow transforms}\label{sec:derivationshadow}
In this subsection we give an intrinsic derivation of the shadow transform of GCFT$_2$, then discuss the analytic continuation and the inner product.

% \begin{align}
% \label{eq:shadoweq1}
% & \ppair{x_{12}\p_{y_1}-\xi_{12}}K(\xy{12})=0,\\
% & \ppair{x_{12}\p_{x_1}+y_{12}\p_{y_1}+2+\D_{12}}K(\xy{12})=0,\nn\\
% & \ppair{(x_{1}^{2}-x_{2}^{2})\p_{y_1} - 2\xi_1 x_1 + 2\xi_2 x_2}K(\xy{12})=0,\nn\\
% & \ppair{(x_{1}^{2}-x_{2}^{2})\p_{x_1} + 2(x_{1}y_{1}-x_{2}y_{2})\p_{y_1} +2\D_1 x_1 -2\D_2 x_2 - 2\xi_1 y_1 +2\xi_2 y_2 +4x_1}K(\xy{12})=0.\nn
% \end{align}

\textbf{Shadow transform.} The shadow transform is derived in the following way. For a singlet primary operator $\op(\xy)\in\rep{\D_2,\xi_2}$, we may ask the question whether the smeared operator
\begin{equation}
\op(K)=\inttt{dx_2 dy_2}{I}K(\xy{2})\op(\xy{2})
\label{eq:naivetransform}
\end{equation}
can transform as another local primary operator $\op'\in \rep{\D_1,\xi_1}$ or not. Firstly the integral region $I$ should be invariant under the Galilean conformal transformations in table \ref{table:global_gca}, i.e. $I=\R^2$, then the locality requires that the smeared operator depends on a single point $\op(K)=\op'(\xy{1})$, hence the smearing kernel is of the form $K(\xy{1},\xy{2})$. Next the conformal covariance \eqref{eq:globaltransformations} or its infinitesimal version \eqref{eq:localoperatortrans} of $\op'(\xy{1})$ leads to a set of equations of the kernel $K$. The translations $L_{-1},\,M_{-1}$ lead to $K=K(x_{12},y_{12})$. 
After doing integration by parts, the generators $M_{0},\, L_{0},\, M_{1},\, L_{1}$ give rise to equations  related to the two-point Ward identities \eqref{eq:wardsinglet} by 
\begin{equation}
(\D_1,\xi_1,\D_2,\xi_2)\to(\D_1,\xi_1,2-\D_2,-\xi_2),
\end{equation}
and if $\xi_{12}=\D_{12}=0$ we come back to the discussion of self-intertwining map in section \ref{sec:ups}. When $\xi_{12}\neq 0$, the solution of the first two equations is
\begin{equation}
K(x_{12},y_{12})=e^{-\xi_{12}\frac{y_{12}}{x_{12}}}\bpair{c_1|x_{12}|^{\D_{12}-2}+c_2 \sign(x_{12})|x_{12}|^{\D_{12}-2}}.
\label{eq:intertwiningmap}
\end{equation}
Then the third equation restricts $\xi_1=-\xi_2$ and the last one restricts $\D_1=2-\D_2$. The part proportional to $\sign(x_{12})$ cannot be eliminated by infinitesimal transformations. For the bosonic operators $\op$ and $\op'$, the finite transformation restricts $c_2=0$. For the fermionic operators, we do need the part proportional to $\sign(x_{12})$ and find $c_1=0$. In any case, we get the shadow transform \eqref{eq:shadowgcft2}.

\textbf{Inner product.} However, we need to check whether the integral transform \eqref{eq:naivetransform} is well-defined or not, and this requires us to select the correct weight $(\D,\xi)$. As discussed in section \ref{sec:localoperators}, the conformal family $\rep{\D,\xi}$ is generated by the smeared states $\ket{f}=\intt{dx dy} f(\xy)\op(\xy)\vac$ labeled by the wave-function $f$. The normalizable state, after implementing the integral transform \eqref{eq:naivetransform},
\begin{equation}
f'(\xy{1})=\intt{dx_1 dy_1} K^{-1}(\xy{12}) f(\xy{2})
\end{equation}
should be normalizable as well, where $K^{-1}$ is the kernel of the inverse integral transform.

% \begin{align}
% &\ppair{x_{12}\p_{y_1}-\xi^{\dagger}-\xi}K_{\text{ip}}(\xy{12})=0,\label{eq:ipeq}\\
% &\ppair{x_{12}\p_{x_1}+y_{12}\p_{y_1}+\D^{\dagger}+\D}K_{\text{ip}}(\xy{12})=0,\nn\\
% &\bpair{(x_{1}^{2}-x_{2}^{2})\p_{y_1} - 2(\xi^{\dagger} x_1 +\xi x_2)}K_{\text{ip}}(\xy{12})=0,\nn\\
% &\bpair{(x_{1}^{2}-x_{2}^{2})\p_{x_1} + 2(x_{1}y_{1}-x_{2}y_{2})\p_{y_1} +2(\D^{\dagger} x_1 +\D x_2 - \xi^{\dagger} y_1 -\xi y_2)}K_{\text{ip}}(\xy{12})=0.\nn
% \end{align}

To answer this question we need to specify the inner product, and it turn out that there are two choices. The ansatz of the inner product of the wave-functions is
\begin{equation}
(f_1,f_2)=\intt{\dxy{1}\dxy{2}}f_1^*(\xy{1})K_{\text{ip}}(\xy{12})f_2(\xy{2}).
\end{equation}
Following the same trick above, we get the equations of the inner product kernel $K_{\text{ip}}$. They are related to the two-point Ward identities \eqref{eq:wardsinglet} by
\begin{equation}
(\D_1,\xi_1,\D_2,\xi_2)\to(\D^{\dagger},\xi^{\dagger},\D,\xi),
\end{equation}
and have two types of solutions. Combining the equations of $M_{0},\,M_{1}$ we have 
\begin{equation}
(\xi^{\dagger}-\xi)x_{12}K_{\text{ip}}(\xy{12})=0,
\label{eq:delta1}
\end{equation}
and similarly the equations of $M_{0},\,  L_{0},\, L_{1}$ imply that 
\begin{equation}
\ppair{(\xi^{\dagger}-\xi)y_{12}-(\D^{\dagger}-\D)x_{12}}K_{\text{ip}}(\xy{12})=0.
\label{eq:delta2}
\end{equation}

\textbf{Physical inner product.} For physical operators $\D\in\R,\, \xi\in\Rnot$, the equations \eqref{eq:delta1} and \eqref{eq:delta2} are trivial, and the solution is simply the two-point function $\vev{\op\op}$. This inner product
\begin{equation}
(f_1,f_2):=\bra{f_1}\ket{f_2}=\int{dx_1 dy_1 dx_2 dy_2}f_1^*(\xy{1})f_2(\xy{2})|x_{12}|^{-2\D}e^{2\xi \frac{y_{12}}{x_{12}}}
\label{eq:innerproduct1}
\end{equation}
is badly-behaved because of the exponential growth. Recall that in relativistic CFTs \cite{streater2016pct,DSDLorentzian}, the physical inner product is 
\begin{equation}
\bra{f_1}\ket{f_2}=\int{dx_1 dx_2}f_1^*(x_1)f_2(x_2)\vev{\op(x_1)\op(x_2)}_W,
\end{equation}
where $x_i\in \R^{d-1,1}$ and the inner product kernel $\vev{\op(x_1)\op(x_2)}_W$ is the two-point Wightman distribution Wick-rotated from the Euclidean correlator. Inspired by this we can Wick-rotate either the Carrollian time or the Galilean time to the imaginary axis: $y=i\g$ or $x=i\t$, then for physical weight $\xi\in \Rnot$, the exponential factor $e^{2i\xi \frac{\g_{12}}{x_{12}}}=e^{-2i\xi \frac{y_{12}}{\t_{12}}}$ is tamed to a oscillating phase. The two wick-rotations are distinguished by the power factors: $|x_{12}|^{-2\D}$ and $(-\t_{12}^2+i\t_{12}\e)^{-\D}$. One of the Wick-rotated integral transform \eqref{eq:intertwiningmap} should be the analog of the Lorentzian shadow transform, see e.g. \cite{Kravchuk:2018htv,Gillioz:2018mto}, and for this one the inner product \eqref{eq:innerproduct1} cannot be positive-definite since the corresponding highest weight representation contains negative-norm states. This may cause technical difficulties and we leave it for further study.

\textbf{Inner product of unitary principal series.} The equations \eqref{eq:delta1} and \eqref{eq:delta2} admit an distributional solution
\begin{equation}
K_{\text{ip}}(\xy{12})=\d(x_{12})\d(y_{12}),
\end{equation}
and the weight is restricted by the original equations of $K_{\text{ip}}(\xy{12})$ onto the unitary principal series $(\D=1+is,\, \xi=ir)$. This inner product gives an analog of the Euclidean shadow tranform: analytic continuing the weight to the unitary principal series and replacing the physical inner product by the positive-definite one
\begin{equation}
(f_1,f_2)=\int{dx dy}f_1^*(\xy)f_2(\xy).
\end{equation}
This can also be understood as choosing the rewriting of the double shadow transform \eqref{eq:doubleshadow},
\begin{equation}
\vev{\mathcal{S}^2[\op](\xy{1})\op(\xy{2})}=\cN(\D,\xi)\d(x_{12})\d(y_{12}),
\end{equation}
as the inner product kernel. Due to the modification of inner product and the selected weight, the representation is $\urep{\D,\xi}$ instead of $\rep{\D,\xi}$, and the integral transform \eqref{eq:intertwiningmap} preserves the norm, hence is well-defined.

\section{OPE Blocks and Shadow Coefficients}
Before introducing the conformal block expansion and the inversion formula, in this section we discuss the quantities associated with three-point structures, including the OPE blocks\footnote{We would like to thank B. Czech for raising the question how to determine the OPE blocks in GCFT$_2$.}, the Clebsch-Gordan kernels and the shadow coefficients. For the four-point functions, the conformal blocks are the two-point functions of OPE blocks \cite{Czech:2016xec,Gadde:2016fbj}, and the conformal partial waves are the integrals of two Clebsch-Gordan kernels.

\subsection{OPE blocks}
In this subsection, we determine the OPE blocks from the shadow formalism. The idea of OPE blocks are illustrated in CFT$_1$ in the appendix \ref{sec:opeblockcft1}. The OPE relation can be written as
\begin{equation}
\op_1(\xy{1})\op_2(\xy{2})=\sum_{k} c_{12}^{k}\,\cD_{12k}(x_{12},y_{12},\p_{x_2},\p_{y_2}) \op_{k}(\xy{2}),
\end{equation}
where the derivatives are understood as acting on $\op_{k}$ only.
The OPE block $\cD$ encodes all the contributions of the derivative operators
\begin{equation}
\label{eq:opeblockgcft1}
\cD_{123}(x_{12},y_{12},\p_{x_2},\p_{y_2})
=
x_{12}^{-\D_{12,3}}e^{\xi_{12,3}\frac{y_{12}}{x_{12}}} 
\sum_{n,m}
\ppair{\sum_{k=0}^{n+m} a_{n,m,k} x_{12}^{k} y_{12}^{n+m-k}}\cdot  
\p_{x_2}^{n} \p_{y_2}^{m},
\end{equation}
in which the prefactor $x_{12}^{-\D_{12,3}}\exp (\xi_{12,3}k_{12})$ is to give the correct two-point function, and the OPE coefficients and three-point coefficients are related by $c_{123}=c_{12}^{k}\d_{3k}$ and $\d_{12}=c_{12}^{\id}$.

In the shadow formalism, the OPE block should be
\begin{equation}
\label{eq:opeblockshadow1}
\cD_{123} \op_{3}(\xy{2})=N_{123} \inttt{dx_0dy_0}{I}\vev{\op_1(\xy{1}) \op_2(\xy{2}) \opshadow_3(\xy{0})} \op_3(\xy{0})
\end{equation}
where the integral region is $I=(x_1,x_2)\times \R$ and the normalization factor $N_{123}$ is to ensure that the primary operator contributes to one. The calculation is a bit lengthy and we leave it into the appendix \ref{sec:calculationOPEblocks}. In the end, the closed form of the OPE block is
\begin{align}
\label{eq:opeblockgcft3}
\cD_{123}(\xy,\p_{x},\p_{y})
&=x^{-\D_{12,3}}e^{\xi_{12,3}\frac{y}{x}} 
\sum_{n,m} \frac{ (2\xi_3)^{-m}}{n!}
\ppair{\frac{1 + R}{2}}^{n} 
P^{(\D_{32,1}-1,\D_{31,2}+n-1)}_{m}(R)\nn
\\
& \peq\cdot
(x\p_{x}+y\p_{y})^{n} (x\p_{y})^{m}, 
\end{align}
where $R=\frac{\xi_1-\xi_2}{\xi_3}$ and $P_{n}^{(a, b)}(z)$ is the Jacobi polynomial,
\begin{equation}
P_{n}^{(a, b)}(z)
=\frac{(a+1)_{n}}{n !}\Fba(-n, 1+a+b+n ; a+1 ; \half(1-z)).
\end{equation}

For two identical operators, the OPE block gets simplified to
\begin{equation}
\label{eq:opeblockgcft5}
\cD_{113}=\vev{\op_{1}\op_{1}}x^{\D_{3}}e^{-\xi_{3}\frac{y}{x}} \sum_{n,m} \frac{2^{-n - m} \xi_3^{-m}}{n!} P^{(\D_{3}-1,\D_{3}+n-1)}_{m}(0) (x\p_{x}+y\p_{y})^{n} (x\p_{y})^{m}. 
\end{equation}
In the appendix of \cite{Bagchi:2017cpu}, the low-level OPE block coefficients of two identical external operators with respect to the BMS algebra was computed by using the recursion relations. Our results of $R=0$, $\D_{12}=0$ should match theirs with $c_{M}\to\inf$, and this is indeed true.

\textbf{Boost multiplets in OPE.} Suppose there is a rank-$r$ boost multiplet $\op^{a}_{3}$ in the singlet-singlet OPE $\op_1\times \op_2$, the leading term from the primaries is
\begin{equation}
\op_1(\xy{1})\op_2(\xy{2})=|x_{12}|^{-\D_{12,3}}e^{\xi_{12,3}\frac{y_{12}}{x_{12}}}\sum_{a=1}^{r}d_{123,a}\op^{a}_{3}(\xy{2})+\dots,
\label{eq:opelimit1}
\end{equation}
then inserting the OPE into three-point functions \eqref{eq:threepoint11r} we get the relation between three-point coefficients $c^{a}$ and the OPE coefficients $d_{a}:=d_{123,a}$,
\begin{equation}
d_{a}=\d_{a+b,r+1}c^{b}=c^{r+1-a}.
\label{eq:opecoefficients1}
\end{equation}

\subsection{Clebsch-Gordan kernels and shadow coefficients}\label{sec:cgkernel}
In this subsection we discuss the Clebsch-Gordan kernel and the shadow coefficient. From the representation theory perspective, the three-point structure is the Clebsch-Gordan kernel \cite{Dobrev:1976vr,Gadde:2017sjg}. Denoting $\urep{i}=\urep{\D_i,\xi_i}$ and $\urep{\wave{i}}=\urep{\wave{\D}_i,\wave{\xi}_i}$, for two principal series representations, the tensor product contains the vectors like $f_1(\xy{1})\otimes f_2(\xy{2})\in \urep{1}\ox \urep{2}$, and the irreducible decomposition is
\begin{equation}
f_3(\xy{3})=\intt{dx_1 dy_1 dx_2 dy_2}f_1(\xy{1}) f_2(\xy{2}) K(\xy{1},\xy{2},\xy{3})\in \urep{3},
\label{eq:cgkernel1}
\end{equation}
in which the kernel $K$ is the infinite-dimensional version of the Clebsch-Gordan coefficient
\begin{equation}
\ket{j,m}=\sum_{m_1,m_2}\braket{j_1m_1j_2m_2}{jm} \ket{j_1,m_1}\otimes \ket{j_2,m_2}.
\end{equation}
Intuitively the coordinates $(\xy)$ serve as the magnetic quantum numbers and the weight $(\D,\xi)$ serve as the angular momentum quantum numbers. By comparing the conformal covariance of both sides, the kernel $K$ is proportional to the three-point structure $\vev{\opshadow_1\opshadow_2\op_3}$ and transforms as in the representation $\urep{\wave{1}}\otimes \urep{\wave{2}}\otimes \urep{3}$. Since the shadow transform is an isomorphic intertwining map, the shadow-transformed three-point structure $\vev{\shadow{\op_1}\shadow{\op_2}\op_3}$ is also in the same representation, hence should be proportional to $\vev{\opshadow_1\opshadow_2\op_3}$. Similarly the three-point structure $\vev{\op_1\op_2\shadow{\op_3}}$ is expected being proportional to $\vev{\op_1\op_2\opshadow_3}$. The relative coefficient is named as the shadow coefficient $\shadow(\op_1\op_2[\op_3])$ \cite{Karateev:2018oml,Liu:2018jhs},
\begin{align}
\vev{\op_1\op_2 \shadow{\op_3}(\xy{4})}
& =
\intt{dx_3 dy_3} \vev{\op_1\op_2\op_3(\xy{3})}\vev{\opshadow_3(\xy{3})\opshadow_3(\xy{4})}
\label{eq:star1}
\\
& =
\shadow(\op_1\op_2[\op_3]) \vev{\op_1\op_2\opshadow_3(\xy{4})}.
\label{eq:star2}
\end{align}
In relativistic CFTs, this is also known as the vertex-graph identity or the star-triangle relation \cite{Ferrara:1972xe,DEramo:1971hnd}. The integral in \eqref{eq:star1} can be evaluated explicitly
\begin{equation}
\vev{\op_1\op_2 \shadow{\op_3}(\xy{4})}
=
\inttt{dx_3 dy_3}{\R^2} F_x\exp[ \xi_3 y_3 J_0(x_3-X) + A_0],
\end{equation}
where 
\begingroup
\allowdisplaybreaks
\begin{align}
F_x& =|x_{12}|^{-\D_{12,3}}|x_{23}|^{-\D_{23,1}}|x_{31}|^{-\D_{31,2}}|x_{34}|^{-2(2-\D_3)},\nn\\
J_0& = \frac{R\, x_{12}+ x_1 + x_2 -2 x_4}{x_{13}x_{23}x_{34}},\nn\\
A_0& = \frac{\xi_{12,3} y_{12}}{x_{12}}+\frac{\xi_{13,2} y_1}{x_{13}}+\frac{\xi_{23,1} y_2}{x_{23}}+\frac{2\xi_{3} y_{4}}{x_{34}},\nn\\
X& = \frac{-\xi_{23,1} x_1 x_4 -\xi_{13,2} x_2 x_4 + 2 \xi _3 x_1 x_2}{\xi_{13,2} x_1 + \xi_{23,1} x_2 -2 \xi_3 x_4},
\label{eq:localizationpoint1}
\end{align}
\endgroup
with 
\begin{equation}
   R=\frac{\xi_1-\xi_2}{\xi_3}. 
\end{equation}
The $y_3$-dependent part in the integrand is a pure phase, hence gives rise to a $\delta$-distribution of $x_3$, and the integral gives
\begin{equation}
\vev{\op_1\op_2 \shadow{\op_3}(\xy{4})}=\frac{2\pi F_x}{|\xi_3 J_0|} e^{A_0}\eval_{x_3=X},
\end{equation}
in which $e^{A_0}$ gives exactly the exponential part in $\vev{\op_1\op_2\opshadow_3}$, and $\frac{F_x}{|J_0|}$ is proportional to the power-law part, hence the shadow coefficient can be determined to be
\begin{equation}
\shadow(\op_1\op_2[\op_3]) = 2^{-2+2\D_3}\frac{\pi}{|\xi_3|}|1+R|^{1-\D_{13,2}}|1-R|^{1-\D_{23,1}}.
\end{equation}
Notice that we have the useful identity $\shadow(\opshadow_1\opshadow_2[\op_3])=\shadow(\op_1\op_2[\op_3])$.

\textbf{Properties of the shadow coefficient.} The shadow coefficient is related to the normalization factor of the OPE block \eqref{eq:opeblockshadow1} and the one of double shadow transform \eqref{eq:doubleshadow}. Consider the three-point structure $\vev{\op_1\op_2\shadow{\opshadow_3}}$, using \eqref{eq:star1} and then inserting the OPE block \eqref{eq:opeblockshadow1} into the three-point function $\vev{\op_1 \op_2 \op_3}$, we get
\begin{align}
\vev{\op_1\op_2\shadow{\opshadow_3}(\xy{4})}
& =\shadow(\op_1\op_2[\opshadow_3])\vev{\op_1\op_2\op_3(\xy{4})}\nn\\
& =\shadow(\op_1\op_2[\opshadow_3])N_{123}\int_I \vev{\op_1\op_2\opshadow_3}\vev{\op_3\op_3(\xy{4})}\label{eq:shadowco1}\\
& =\int_{\R^2} \vev{\op_1\op_2\opshadow_3}\vev{\op_3\op_3(\xy{4})},\label{eq:shadowco2}
\end{align}
where the last line comes from expanding the definition of $\shadow{\opshadow_3}$ in $\vev{\op_1\op_2\shadow{\opshadow_3}}$. Notice that the integral regions in \eqref{eq:shadowco1} and \eqref{eq:shadowco2} are not the same. In the appendix \ref{sec:relationshadowcoefficientandopeblock} we determine that if the weights $\xi_i$ satisfy $R\in(-1,1)$, the two integral expressions \eqref{eq:shadowco1} and \eqref{eq:shadowco2} hold simultaneously and the normalization factor is
\begin{equation}
\label{eq:opeblocknormalization1}
N_{123}^{-1}=\shadow(\op_1\op_2[\opshadow_3])= 2^{2\D_3-2}\frac{\pi}{|\xi_3|}(1-R)^{-1+\D_{23,1}}(1+R)^{-1+\D_{31,2}},
\end{equation}
matching with the result \eqref{eq:opeblocknormalization}.

The relation between the shadow coefficient and the factor \eqref{eq:doubleshadow} is determined as follows. Consider the doubly shadow-transformed three-point structure $\vev{\op_1\op_2\shadow^2[\op_3]}$, by \eqref{eq:doubleshadow} we have
\begin{equation}
\vev{\op_1\op_2\shadow^2[\op_3]}=\cN(\D_3,\xi_3)\vev{\op_1\op_2\op_3},
\end{equation}
and by applying \eqref{eq:star2} twice we get
\begin{equation}
\vev{\op_1\op_2\shadow^2[\op_3]}=\shadow(\op_1\op_2[\shadow{\op_3}])\shadow(\op_1\op_2[\op_3]) \vev{\op_1\op_2\op_3}.
\end{equation}
Notice that $\shadow(\op_1\op_2[\shadow{\op_3}])=\shadow(\op_1\op_2[\opshadow_3])$, thus the relation between the shadow coefficients and the factor $\cN(\D,\xi)$ is
\begin{equation}
\shadow(\op_1\op_2[\op_3])\shadow(\op_1\op_2[\opshadow_3])=\cN(\D_3,\xi_3),
\label{eq:shadowCoefficientsToPlancherelMeasure}
\end{equation}
which is expected.

\subsection{Orthogonality of the Clebsch-Gordan kernels}\label{sec:cgkernelO}
The orthogonality and completeness relations of the Clebsch-Gordan coefficients are respectively
\begin{align}
& \sum_{m_1,m_2}\braket{j'm'}{j_1m_1j_2m_2}\braket{j_1m_1j_2m_2}{jm}=\d_{jj'}\d_{mm'},
\label{eq:cgorthogonality1}
\\
& \sum_{j}\sum_{m\in j}\braket{j_1m'_1j_2m'_2}{jm}\braket{jm}{j_1m_1j_2m_2}=\d_{m_1,m'_1}\d_{m_2,m'_2},
\label{eq:cgorthogonality2}
\end{align}
where $\braket{jm}{j_1m_1j_2m_2}$ is the complex conjugate of $\braket{j_1m_1j_2m_2}{jm}$. The infinite-dimensional version of \eqref{eq:cgorthogonality1} for the Clebsch-Gordan kernel should be
\begin{equation}
\intt{dx_1 dy_1 dx_2 dy_2} \vev{\op_1\op_2 \opshadow_4}\vev{\opshadow_1\opshadow_2\op_3}\sim \d(r_{34})\d(s_{34})\d(x_{34})\d(y_{34})+\text{shadow term},
\label{eq:bubble0}
\end{equation}
in which $\D_i=1+is_i,\,\xi_i=ir_i$. The shadow term is proportional to $\d(s_{3}+s_4)\d(r_3+r_4)$ due to the equivalence of $\urep{4}$ and $\urep{\wave{4}}$, and can be determined by the shadow transform once the first term is known. Following the convention of \cite{Karateev:2018oml} we swap the operators in \eqref{eq:bubble0} and define the bubble integral of two three-point structures as
\begin{align}
B(\op_3,\op_4)
& =
\intt{dx_{1}dy_{1}dx_{2}dy_{2}} \vev{\op_{1}\op_{2}\op_{3}}\vev{\opshadow_{1}\opshadow_{2}\opshadow_{4}}
\label{eq:bubble1}
\\
& =
\intt{dx_{1}dy_{1}dx_{2}dy_{2}}F_x e^{iA_1y_1+iA_2y_2+iA_0},
\end{align}
where
\begin{align}
A_0& =\frac{r_{14,2}}{x_{14}}y_4+\frac{r_{24,1}}{x_{24}}y_4-\frac{r_{13,2}}{x_{13}}y_3-\frac{r_{23,1}}{x_{23}}y_3
,
\nn\\
A_1& =-\frac{r_{34}}{x_{12}}+\frac{r_{13,2}}{x_{13}}-\frac{r_{14,2}}{x_{14}}, 
\hs 
A_2 =\frac{r_{34}}{x_{12}}+\frac{r_{23,1}}{x_{23}}-\frac{r_{24,1}}{x_{24}},
\nn\\
F_x& =|x_{12}|^{\D_{34}-2}|x_{23}|^{-\D_{23,1}}|x_{13}|^{-\D_{13,2}}|x_{24}|^{-2+\D_{24,1}}|x_{14}|^{-2+\D_{14,2}}
.\nn
\end{align}

\textbf{The first term.} To separate the first term from the shadow term, we suppose $r_3r_4>0$. The integration with respect to $y_1,\,y_2$ is of the form like 
\begin{equation}
\intt{dy_{1}dy_{2}}e^{iA_1y_1+iA_2y_2}=(2\pi)^2\d(A_1)\d(A_2).
\end{equation}
The two equations $A_1=A_2=0$ decide an algebraic variety with two irreducible components in the space $\R^{8} \ni(x_i,r_i)$, and the condition $r_3r_4>0$ selects the component $\set{r_{34}=0,x_{34}=0}$, hence
\begin{equation}
\d(A_1)\d(A_2)=\frac{|x_{12}x_{13}x_{23}|}{2|r_3|}\d(r_{34})\d(x_{34}).
\end{equation}
After using $\d(r_{34})\d(x_{34})$ to simplify the rest part, we find that the substitutions
\begin{equation}
X_1=\frac{x_1}{x_{13}}-\frac{x_2}{x_{23}}, \qquad X_2=\frac{x_1}{x_{13}}+\frac{x_2}{x_{23}},
\end{equation}
reduce the exponential factor to $A_0=\frac{r_3 y_{34}}{x_3}(X_2-2) +\frac{y_{34}r_{12}}{x_3}$ and non-exponential part is independent of $X_2$. Hence the integration with respect to $X_2$ gives rise to $\d(y_{34})$, and we get
\begin{equation}
B(\op_3,\op_4)
=\frac{2\pi^{3}}{r_3^{2}}\d(r_{34})\d(x_{34})\d(y_{34})\inttt{dX_1}{\R}|X_1|^{-1+is_{34}}|x_3|^{-is_{34}}
=\frac{8\pi^{4}}{r_3^{2}}\d(r_{34})\d(s_{34})\d(x_{34})\d(y_{34}),
\end{equation}
justifying the first term in \eqref{eq:bubble0}. 

\textbf{The shadow term.} Relaxing the assumption $r_3r_4>0$, the shadow term comes from the integration localized on the second component of $A_1=A_2=0$,
\begin{equation}
r_3+r_4=0,  \qquad  r_{12}x_{12}x_{34}+r_3 \ppair{(x_1+x_2)(x_3+x_4)-2x_1 x_2 -2 x_3 x_4}=0,
\label{eq:secondvariety}
\end{equation}
and can be determined by the following procedure\footnote{As a crosscheck, we provide a direct calculation in the appendix \ref{sec:secondtermbubble}.}. Denoting the bubble integral as
\begin{equation}
B(\op_3,\op_4)=\d(\op_3,\op_4)\d(x_{34},y_{34})+ B_1(\op_3,\op_4,x_{34},y_{34}),
\label{eq:bubble2}
\end{equation}
where $\d(\op_3,\op_4)=8\pi^2\cN(\D_3,\xi_3)\d(r_{34})\d(s_{34})$ and applying the shadow transform of $\opshadow_4$ on \eqref{eq:bubble2}, the two terms should be switched. The right-hand side becomes
\begin{equation}
\rhs=\d(\op_3,\op_4)\vev{\op_4(\xy{3})\op_4(\xy{5})}+ \intt{dx_4dy_4}B_1\vev{\op_4(\xy{4})\op_4(\xy{5})},
\end{equation}
while the left-hand side can be calculated using the shadow coefficient
\begin{align}
\lhs
& =\int{dx_1dy_1dx_2dy_2}\vev{\op_{1}\op_{2}\op_{3}}\vev{\opshadow_{1}\opshadow_{2}\shadow{\opshadow_{4}}(\xy{5})}\nn
\\
& =\int{dx_1dy_1dx_2dy_2}\vev{\op_{1}\op_{2}\op_{3}}\vev{\opshadow_{1}\opshadow_{2}\op_4(\xy{5})}\shadow(\opshadow_1\opshadow_2[\opshadow_4])\nn
\\
& =
\shadow(\opshadow_1\opshadow_2[\opshadow_4])\ppair{\d(\op_3,\opshadow_4)\d(x_{35})\d(y_{35})+B_1(\op_3,\opshadow_4,x_{35},y_{35})}.
\end{align}
By comparison we get
\begin{equation}
B_1(\op_3,\op_4)=\shadow(\opshadow_1\opshadow_2[\op_4])^{-1}\d(\op_3,\opshadow_4)\vev{\op_3\op_3(\xy{4})}.
\end{equation}
In summary the bubble integral \eqref{eq:bubble1} mimicking the orthogonality relation \eqref{eq:cgorthogonality1} is
\begin{align}
B(\op_3,\op_4)
& =
\intt{dx_{1}dy_{1}dx_{2}dy_{2}} \vev{\op_{1}\op_{2}\op_{3}}\vev{\opshadow_{1}\opshadow_{2}\opshadow_{4}}\nn
\\
& =
\d(\op_3,\op_4)\d(x_{34},y_{34})+ \shadow(\opshadow_1\opshadow_2[\op_4])^{-1}\d(\op_3,\opshadow_4)\vev{\op_3\op_3(\xy{4})},
\label{eq:bubblelast}
\end{align}
where $\d(\op_3,\op_4)=8\pi^2\cN(\D_3,\xi_3)\d(r_{34})\d(s_{34})$.

\textbf{Incompleteness and projector.} The infinite-dimensional version of the completeness relation \eqref{eq:cgorthogonality2} should be
\begin{equation}
\intt{r^2dsdr}\intt{\dxy{0}}\vev{\op_1\op_2\opshadow_0}\vev{\op_0\opshadow_1(\xy{3})\opshadow_2(\xy{4})}\overset{?}{\sim} \d(x_{13})\d(y_{13})\d(x_{24})\d(y_{24}),
\label{eq:incomplete1}
\end{equation}
where we have relabeled the weight as $(\D_0=1+is,\xi_0=ir)$, and by \eqref{eq:inversePlancherel} the factor $r^2$ is proportional to the Plancherel measure of the principal series. However the set of Clebsch-Gordan kernels is an incomplete basis due to the following reason. Firstly, the Clebsch-Gordan kernel \eqref{eq:cgkernel1} corresponds to decomposing the tensor product of two tachyonic representations into another tachyonic one. But there should be massive and massless representations in this tensor product decomposition, since the sum of two spacelike momenta can be timelike or null. Secondly, according to \cite{ASENS_1972_4_5_3_459_0,ASENS_1973_4_6_1_103_0}, the Plancherel measure of the $3d$ Poincare group is 
\begin{equation}
c_1\inttt{r^2dr}{\Rpositive}\inttt{ds}{\R}+c_2\inttt{m^2dm}{\Rpositive}\sum_{j=-\inf}^{\inf},
\end{equation}
where $c_1,\, c_2$ are constants depending on the Haar measure, and the two terms count the contributions from tachyonic and massive representations respectively. Combining these two aspects and the orthogonality \eqref{eq:bubblelast}, we have
\begin{equation}
\d(x_{13})\d(y_{13})\d(x_{24})\d(y_{24})\sim\cP_{\text{t}}(\xy{1},\xy{2};\xy{3},\xy{4})+\cP_{\text{m}},
\end{equation}
where $\cP_{\text{t}}$ and $\cP_{\text{m}}$ are the projection operators of tachyonic and massive representations respectively, $\cP_{t,m}^2\sim\cP_{t,m},\, \cP_{t}\cdot \cP_{m}=0$, and
\begin{equation}
\cP_{t}(\xy{1},\xy{2};\xy{3},\xy{4})=\intt{r^2dsdr}\intt{\dxy{0}}\vev{\op_1\op_2\opshadow_0}\vev{\op_0\opshadow_1(\xy{3})\opshadow_2(\xy{4})}.
\label{eq:incomplete12}
\end{equation}

\section{Conformal Blocks and Partial Waves}

In relativistic CFTs, due to the convergence of OPE, the higher-point functions can be reduced to a sum of conformal blocks by applying the OPE relations repeatedly, and the coefficients are products of three-point coefficients. The conformal blocks are completely fixed by the conformal symmetry, depending on the external operators, the specific OPE channel, and the exchanged operators. 

This conformal block expansion can be regarded as an on-shell method, since the summation ranges over the physical Hilbert space. The correlation functions can also be expanded into an integral of the conformal partial waves over unphysical unitary principal series - this is the Euclidean inversion formula. Under suitable conditions, the block expansion is recovered from the inversion formula by a contour deformation argument.

In this section we develop the conformal block expansions for four different external singlet operators in Galilean CFT$_2$. We first calculate the conformal blocks of exchanged singlets and boost multiplets by solving the Casimir equations, then using the shadow formalism determine the conformal partial waves and establish the inversion formula. The previous results of singlet conformal blocks of the BMS algebra are in e.g. \cite{Bagchi:2017cpu,Hijano:2018nhq,Merbis:2019wgk,Ammon:2020wem}, see also the work on BMS torus blocks \cite{Bagchi:2020rwb}, and singlet conformal blocks with supersymmetric extensions \cite{Lodato:2018gyp}.

\textbf{Settings of conformal block expansion.} We firstly set up the conformal block expansion for four different external singlet operators. A priori, without the dynamical information of four-point functions $\vev{\op_1\op_2\dots }$, we do not know which kind of operators appearing in the OPE of $\op_1\times \op_2$. In GCFT$_2$, besides the singlets and the multiplets there can be other operators, e.g., the logarithmic multiplets\footnotemark{}. Starting from the simplest case, we assume that the exchanged operators are all singlets, and in the later section \ref{sec:blockproperties} we will add the boost multiplets into the conformal block expansion.

\footnotetext{Another tricky example is that there are staggered multiplets with respect to the BMS algebra in the BMS free scalar theory \cite{Hao:2021urq}, and when decomposing them into the  representations of global Galilean conformal symmetry, some conformal blocks forbidden by the global null state condition can be nonzero \cite{Chen:2022jhx}.}

The \schannel block expansion of a four-point function is
\begin{equation}
\wick{\vev{\c1 \op_1 \c1 \op_2 \c1 \op_3 \c1 \op_4}} =\sum_{n}\cD_{12n}\cD_{43n}\vev{\op_n(\xy{2})\op_n(\xy{3})}
=\sum_n p^{\schannel}_n \blockdressed^{\schannel}_n(\xy{i}),
\end{equation}
where $p^{\schannel}_n=c_{12n}c_{43n}$, and the conformal block with respect to the primary $\op_n$ is defined as
\begin{equation}
\label{eq:block4gcft}
\blockdressed^{\schannel}_n(\xy{i})=\cD_{12n}\cD_{43n}\vev{\op_n(\xy{2})\op_n(\xy{3})}.
\end{equation}
Similarly the \tchannel block expansion is,
\begin{equation}
\wick{\vev{\c1 \op_1 \c2 \op_2 \c2 \op_3 \c1 \op_4}} =\sum_{n}c_{14n}c_{23n}\cD_{14n}\cD_{23n}\vev{\op_n(\xy{3})\op_n(\xy{4})}=\sum_n p^{\tchannel}_n \blockdressed^{\tchannel}_n(\xy{i}).
\end{equation}
To further carry out calculations we introduce the stripped version of conformal blocks depending only on the cross ratios by factoring out the kinematical factors
\begingroup
\allowdisplaybreaks
\begin{align}
\blockdressed^{\schannel}_n(\xy{i})& =K^{\schannel}(\xy{i})\blockstripped^{\schannel}_{n}(\xy),
\label{eq:block1gcft}
\\
\blockdressed^{\tchannel}_n(\xy{i})& =K^{\tchannel}(\xy{i})\blockstripped^{\tchannel}_{n}(1-x,-y),
\end{align}
\endgroup
then the block expansion of the stripped four-point functions are
\begin{align}
\label{eq:blockexpansionstripped}
\fourpt^{\schannel}(\xy)& =\sum_n p_{n}^{\schannel} \blockstripped^{\schannel}_n(\xy),\\
\fourpt^{\tchannel}(1-x,-y)& =\sum_n p_{n}^{\tchannel} \blockstripped^{\tchannel}_n(1-x,-y).
\end{align}
% The $s-t$ crossing equation leads to
% \begin{eqnarray}
% \lefteqn{x^{-(\D_1+\D_2)}\exp[(\xi_1+\xi_2)\frac{y}{x}]
% \sum_n p_{n}^{\schannel} \blockstripped^{\schannel}_{n}(\xy)}\nn\\
% &&=
% (1-x)^{-(\D_2+\D_3)}\exp[(\xi_2+\xi_3)\frac{-y}{1-x}]
% \sum_n p_{n}^{\tchannel} \blockstripped^{\tchannel}_{n}(1-x,-y). 
% \label{eq:stcrossinggcft1} 
% \end{eqnarray}

\subsection{Conformal blocks from Casimir equations}

In this subsection we derive the Casimir equations of singlet and boost multiplet, then obtain the conformal blocks by solving the Casimir equations. 

\textbf{Conformal blocks of singlets.} In GCFT$_2$, the conformal blocks of exchanged singlet conformal families are the eigenfunctions of the Casimir differential operators. This originates from the fact that the Casimir elements \eqref{eq:casimirPoincare} of the Galilean conformal algebra act on the singlet $\rep{\D_0,\xi_0}\ni\ket{n}$ as scalars \eqref{eq:casimirsinglet}, $(C_{i}-\lm_{i})\ket{n}=0,\, i=1,2$. Notice that this is incorrect for boost multiplet and is insufficient for $\xi=0$ multiplet \cite{Chen:2022jhx}, where the Casimir equations must be modified appropriately.

In the appendix \ref{sec:derivingCasimirequations} we derive the Casimir equations of the stripped conformal block with exchanged  operator being singlet $\op_{0}\in\rep{\D_{0},\xi_{0}},\, \xi_{0}\in\Rnot$. The Casimir equations are
\begin{equation}
\ppair{\cC_i-\lm_{i}}\blockstripped^{\schannel}_0(x,k)=0,\qquad i=1,2,
\end{equation}
in which $(x,k)=(x,\frac{y}{x})$ are the slope coordinates, and the differential Casimir operators are
\begin{align}
\cC_1
& =
(1-x)\pdv[2]{k}
+(-\xi_{12}+\xi_{34}) x \pdv{k}
+\xi_{12} \xi_{34}x,
\\
\cC_2
& =
x k \pdv[2]{k}
+2x(x-1)\frac{\p^2}{\p x \p k}\nn
\\
& \peq 
+\ppair{2+(-\D_{12}+\D_{34})x+(\xi_{12}-\xi_{34})x k }\pdv{k}
+(\xi_{12}-\xi_{34}) x^2 \pdv{x}\nn
\\
& \peq 
+\ppair{\D_{12}\xi_{34}+\D_{34}\xi_{12}-\xi_{12}\xi_{34} k }x.
\end{align}
Then we solve the two Casimir equations in the appendix \ref{sec:solvingCasimirEquations}, and there are two independent solutions. By checking the \schannel OPE limit $x, k\to 0$, and redefining the normalization to ensure the exchanged primary operator contributes one: $\blockstripped_{\D_0,\xi_0}^{\schannel}\sim x^{\D_0} e^{-k \xi_0 }$, we find that the solution corresponding to the physical block is
\begingroup
\allowdisplaybreaks
\begin{align}
\blockstripped_{\D_0,\xi_0}^{\schannel}(x,k)
& =
\frac{N(\D_0,\xi_0)}{H(x)}
\exp\bpair{\frac{k}{1-x}\ppair{\half (\xi_{12}-\xi_{34})x - \xi_{0}H(x)}}
\nn
\\
& \peq
\cdot x^{\D_0}
\bpair{\xi_{0}^{2}-\half(\xi_{0}^{2}+\xi_{12}\xi_{34}) x + \xi_{0}^{2}H(x)}^{1-\D_0} 
\nn
\\
& \peq
\cdot \bpair{ \xi_{0}^{2}-\xi_{12}\xi_{34} + \half (2\xi_{0}^{2}- \xi_{12}^{2}- \xi_{34}^{2})x + (\xi_{12}-\xi_{34}) \xi_{0}H(x)}^{\half(\D_{12}-\D_{34})}
\nn
\\
& \peq 
\cdot \bpair{ \xi_{0}^{2}+\xi_{12}\xi_{34} -\half (\xi_{12}+\xi_{34})^2 x + (\xi_{12}+\xi_{34}) \xi_{0}H(x)}^{\half(\D_{12}+\D_{34})}
\label{eq:strippedblock}
\end{align}
\endgroup
in which
\begin{equation}
H(x)=\sqrt{1-(1+R_{12}R_{34})x +\frac{1}{4}(R_{12}+R_{34})^2 x^2},
\qquad 
R_{ij}\equiv\frac{\xi_{i}-\xi_{j}}{\xi_{0}},
\label{eq:functionHx}
\end{equation}
and the normalization factor is
\begin{equation*}
N(\D_0,\xi_0)=2^{\D_0-1} \xi_{0}^{2\D_{0}-3} \bpair{(\xi_0+\xi_{12})(\xi_0-\xi_{34})}^{-\half(\D_{12}-\D_{34})} \bpair{(\xi_0+\xi_{12})(\xi_0+\xi_{34})}^{-\half(\D_{12}+\D_{34})}.
\end{equation*}
The other solution is proportional to the shadow block $\blockstripped_{2-\D_0,-\xi_0}^{\schannel}(x,k)$, thus the two solutions respect the shadow symmetry $(\D_0,\xi_0)\to (2-\D_0,-\xi_0)$.

\textbf{Conformal blocks of boost multiplets.} The conformal blocks of exchanged boost multiplets are related to the ones of the singlets by a derivative relation. Following the logic of the previous subsection, we encounter the obstruction that the Casimir elements acting on the boost multiplets are not scalars, hence do not commute with the projection operators. As in the case of four identical external operators \cite{Chen:2020vvn}\footnote{See also the similar discussion of the logarithmic conformal blocks in LogCFT \cite{Hogervorst:2016itc}.}, for a rank-$r$ boost multiplet $\rep{\D_0,\xi_0,r}$, the following operators act as zero, $(C_{i}-\lm_{i})^{r}\ket{n}=0,\, i=1,2$. Denoting the conformal blocks of $\rep{\D_0,\xi_0,r}$ as 
\begin{equation}
\blockdressed^{\schannel}_{0,r}(\xy{i})=K^{\schannel}(\xy{i})\blockstripped^{\schannel}_{0,r}(x,k),
\end{equation}
the modified version of \eqref{eq:singletCasimir} is
\begin{equation}
\ppair{C^{(1+2)}_i-\lm_{i}}^{r}\blockdressed^{\schannel}_{0,r}=0,\qquad i=1,2.
\end{equation}
Then using the conjugation relation $\cC_{i} =(K^{\schannel})^{-1} C^{(1+2)}_i K^{\schannel}$, we get the Casimir equations of stripped conformal blocks
\begin{equation}
\ppair{\cC_i-\lm_{i}}^{r}\blockstripped^{\schannel}_{0,r}(x,k)=0,\qquad i=1,2,
\end{equation}
whose solution is a linear combination of $\xi_0$-derivative of the  singlet conformal block
\begin{equation}
\blockstripped^{\schannel}_{0,r}(x,k)=\sum_{a=0}^{r-1}p_{0,a}^{\schannel}
\pdv[a]{\xi_0}\blockstripped^{\schannel}_{0}(x,k).
\label{eq:multipletblock}
\end{equation}

By comparing the \schannel OPE limit, we relate the block coefficients $p_{0,a}^{\schannel}$ with the three-point coefficients. The \schannel OPE limit of the conformal block \eqref{eq:multipletblock} is
\begin{equation}
\blockstripped_{0,r}^{\schannel}(x,k)=
x^{\D_0} e^{-k \xi_0 } 
\sum_{a=0}^{r-1}p_{0,a}^{\schannel}
(-k)^{a}+  O(x^{\D_0+1}),
\label{eq:multipletOPElimit}
\end{equation}
while from the leading OPE \eqref{eq:opelimit1} the conformal block behaves as
\begin{equation}
\blockstripped_{0,r}^{\schannel}(x,k)\sim
x^{\D_0} e^{-k \xi_0 } 
\sum_{a=0}^{r-1}\sum_{b,b'=1}^{r}d_{120,b}d_{340,b'}\d_{a,r+1-b-b'}\frac{(-k)^{a}}{a!}.
\end{equation}
Hence matching the coefficients and using the relation \eqref{eq:opecoefficients1} we have
\begin{equation}
p_{0,a}^{\schannel}=\frac{1}{a!}\sum_{b=1}^{r}c_{120,b}c_{340,r+a+1-b}.
\label{eq:blocktothreepoint}
\end{equation}

\subsection{Analytic properties of conformal block expansion}\label{sec:blockproperties}
In this subsection we explore the analytic properties of the singlet and boost multiplet conformal blocks, and discuss the implications on the conformal block expansion.

\textbf{Analytic properties of singlet blocks.} As a first check of our calculation, taking $\D_{12}=\D_{34}=\xi_{12}=\xi_{34}=0$ the conformal block \eqref{eq:strippedblock} agrees with the result in the previous work \cite{Bagchi:2017cpu,Chen:2020vvn},
\begin{equation}
\blockstripped_{\D_0,\xi_0}^{\schannel}(x,k)=\frac{2^{-2+2\D_0}}{\sqrt{1-x}}e^{-\frac{k \xi_0}{\sqrt{1-x}}}(1+\sqrt{1-x})^{2-2\D_0}x^{\D_0}.
\end{equation}
Secondly near the \schannel OPE limit, the conformal block \eqref{eq:strippedblock} can be expanded into
\begin{equation}
\blockstripped_{\D_0,\xi_0}^{\schannel}(x,k)=x^{\D_0} e^{-k \xi_0 }\bpair{1+ a_0 x + a_1 k x + O(x^{2})} ,
\label{eq:singletOPElimit}
\end{equation}
where in the bracket it is a double Taylor series of $(x,k)$ counting the contribution of the descendants, and the next-to-leading coefficients are
\begin{align}
a_{1}& =-\frac{1}{2\xi_{0}}(\xi_{0}-\xi_{12})(\xi_{0}+\xi_{34}) ,\nn\\
a_{0}& =\frac{1}{2\xi_{0}^{2}} \bpair{\xi_{0}^{2}(\D_{0}-\D_{12}+\D_{34})-\xi_{0}(\D_{12}\xi_{34}+\D_{34}\xi_{12})+\D_{0}\xi_{12}\xi_{34}}.\nn
\end{align}
At first glance, it seems that the conformal block in GCFT$_2$ shares similar analytic structure as the one in CFT$_1$,
\begin{equation}
\blockstripped^{\schannel}_{\D_0}(x)=x^{\D_0}\Fba(\D_0-\D_{12},\D_0+\D_{34};2\D_0,x),
\end{equation}
as reviewed in appendix \ref{sec:cft1conformalblock}. The \schannel singularity in two cases is controlled by the power factor $x^{\D_0}$ and the rest part is analytic near $x=0$.

However there are two additional branch points in the conformal block \eqref{eq:strippedblock} at $x=x_{\pm}$, which are zeros of the function $H(x)$,
\begin{equation}
x_{\pm}=2\frac{1+R_{12}R_{34} \pm \sqrt{(1-R_{12}^2)(1-R_{34}^2)}}{(R_{12}+R_{34})^2},
\label{eq:rootsOffunctionHx}
\end{equation}
If $x_{\pm}\in (-1,1)$, the contributions from the descendants in \eqref{eq:singletOPElimit} grow too fast such that the \schannel convergent radius for each individual block is less than one, and the $s-t$ crossing equation can be invalid. 

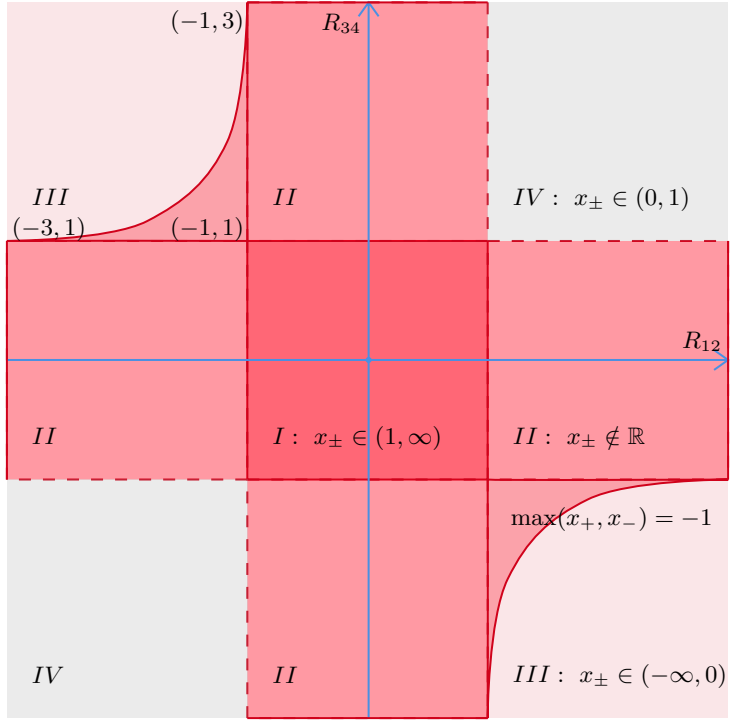
\begin{figure}[htbp]
\centering

\tikzset{every picture/.style={line width=0.75pt}} %set default line width to 0.75pt        

\begin{tikzpicture}[x=0.75pt,y=0.75pt,yscale=-1,xscale=1]
%uncomment if require: \path (0,409); %set diagram left start at 0, and has height of 409

%Shape: Rectangle [id:dp9380001671187128] 
\draw  [color={rgb, 255:red, 208; green, 2; blue, 27 }  ,draw opacity=1 ][fill={rgb, 255:red, 255; green, 2; blue, 27 }  ,fill opacity=0.6 ] (120,120) -- (240,120) -- (240,240) -- (120,240) -- cycle ;
%Flowchart: Process [id:dp5370051162252654] 
\draw  [color={rgb, 255:red, 208; green, 2; blue, 27 }  ,draw opacity=1 ][fill={rgb, 255:red, 255; green, 2; blue, 27 }  ,fill opacity=0.4 ][dash pattern={on 4.5pt off 4.5pt}] (120,0) -- (240,0) -- (240,120) -- (120,120) -- cycle ;
%Flowchart: Process [id:dp42214418448649194] 
\draw  [color={rgb, 255:red, 208; green, 2; blue, 27 }  ,draw opacity=1 ][fill={rgb, 255:red, 255; green, 2; blue, 27 }  ,fill opacity=0.4 ][dash pattern={on 4.5pt off 4.5pt}] (0,120) -- (120,120) -- (120,240) -- (0,240) -- cycle ;
%Flowchart: Process [id:dp6840300906704098] 
\draw  [color={rgb, 255:red, 208; green, 2; blue, 27 }  ,draw opacity=1 ][fill={rgb, 255:red, 255; green, 2; blue, 27 }  ,fill opacity=0.4 ][dash pattern={on 4.5pt off 4.5pt}] (240,120) -- (360,120) -- (360,240) -- (240,240) -- cycle ;
%Flowchart: Process [id:dp3119790017672559] 
\draw  [color={rgb, 255:red, 208; green, 2; blue, 27 }  ,draw opacity=1 ][fill={rgb, 255:red, 255; green, 2; blue, 27 }  ,fill opacity=0.4 ][dash pattern={on 4.5pt off 4.5pt}] (120,240) -- (240,240) -- (240,360) -- (120,360) -- cycle ;
%Flowchart: Process [id:dp7996805415336272] 
\draw  [draw opacity=0][fill={rgb, 255:red, 155; green, 155; blue, 155 }  ,fill opacity=0.2 ] (240,0) -- (360,0) -- (360,120) -- (240,120) -- cycle ;
%Flowchart: Process [id:dp4826832777103933] 
\draw  [draw opacity=0][fill={rgb, 255:red, 155; green, 155; blue, 155 }  ,fill opacity=0.2 ] (0,240) -- (120,240) -- (120,360) -- (0,360) -- cycle ;

%Shape: Rectangle [id:dp1939881067499547] 
\draw  [draw opacity=0][fill={rgb, 255:red, 208; green, 2; blue, 27 }  ,fill opacity=0.1 ] (0,0) -- (120,0) -- (120,120) -- (0,120) -- cycle ;
%Shape: Rectangle [id:dp20000228660759478] 
\draw  [draw opacity=0][fill={rgb, 255:red, 208; green, 2; blue, 27 }  ,fill opacity=0.1 ] (240,240) -- (360,240) -- (360,360) -- (240,360) -- cycle ;

%Shape: Axis 2D [id:dp0704613796264586] 
\draw [color={rgb, 255:red, 74; green, 144; blue, 226 }  ,draw opacity=1 ][line width=0.75]  (0,179.86) -- (360,179.86)(180.56,0) -- (180.56,360) (353,174.86) -- (360,179.86) -- (353,184.86) (175.56,7) -- (180.56,0) -- (185.56,7)  ;
%Shape: Circle [id:dp15704647295636898] 
\draw  [color={rgb, 255:red, 74; green, 144; blue, 226 }  ,draw opacity=1 ][fill={rgb, 255:red, 74; green, 144; blue, 226 }  ,fill opacity=1 ] (179.56,179.86) .. controls (179.56,179.31) and (180.01,178.86) .. (180.56,178.86) .. controls (181.11,178.86) and (181.56,179.31) .. (181.56,179.86) .. controls (181.56,180.41) and (181.11,180.86) .. (180.56,180.86) .. controls (180.01,180.86) and (179.56,180.41) .. (179.56,179.86) -- cycle ;
%Shape: Polygon Curved [id:ds03889829225660857] 
\draw  [color={rgb, 255:red, 208; green, 2; blue, 27 }  ,draw opacity=1 ][fill={rgb, 255:red, 255; green, 2; blue, 27 }  ,fill opacity=0.3 ] (290,250) .. controls (309.67,240.17) and (359.85,240) .. (360.09,240) .. controls (360.33,240) and (240,240.5) .. (240,240) .. controls (240,239.5) and (240.11,360) .. (240,360) .. controls (239.89,360) and (240.33,309.83) .. (250,290) .. controls (259.67,270.17) and (270.33,259.83) .. (290,250) -- cycle ;
%Shape: Polygon Curved [id:ds41154448300837787] 
\draw  [color={rgb, 255:red, 208; green, 2; blue, 27 }  ,draw opacity=1 ][fill={rgb, 255:red, 255; green, 2; blue, 27 }  ,fill opacity=0.3 ] (110,70) .. controls (119.22,50) and (120.05,0) .. (120.09,0) .. controls (120.14,0) and (119.86,120.22) .. (120.09,120) .. controls (120.33,119.78) and (0.11,120) .. (0,120) .. controls (-0.11,120) and (51.22,119.78) .. (70,110) .. controls (88.78,100.22) and (100.78,90) .. (110,70) -- cycle ;
%Straight Lines [id:da39316168672625906] 
\draw [color={rgb, 255:red, 208; green, 2; blue, 27 }  ,draw opacity=1 ]   (360,120) -- (360,240) ;
%Straight Lines [id:da07972997406999238] 
\draw [color={rgb, 255:red, 208; green, 2; blue, 27 }  ,draw opacity=1 ]   (0,120) -- (0,240) ;
%Straight Lines [id:da9874807786598865] 
\draw [color={rgb, 255:red, 208; green, 2; blue, 27 }  ,draw opacity=1 ]   (240,360) -- (120,360) ;
%Straight Lines [id:da9767848365600975] 
\draw [color={rgb, 255:red, 208; green, 2; blue, 27 }  ,draw opacity=1 ]   (240,0) -- (120,0) ;

% Text Node
\draw (251,211.4) node [anchor=north west][inner sep=0.75pt]  [font=\footnotesize]  {$II:\ x_{\pm } \notin \mathbb{R}$};
% Text Node
\draw (251,91.4) node [anchor=north west][inner sep=0.75pt]  [font=\footnotesize]  {$IV:\ x_{\pm } \in ( 0,1)$};
% Text Node
\draw (251,331.4) node [anchor=north west][inner sep=0.75pt]  [font=\footnotesize]  {$III:\ x_{\pm } \in ( -\infty ,0)$};
% Text Node
\draw (131,211.4) node [anchor=north west][inner sep=0.75pt]  [font=\footnotesize]  {$I:\ x_{\pm } \in ( 1,\infty )$};
% Text Node
\draw (167,10.5) node  [font=\footnotesize]  {$R_{34}$};
% Text Node
\draw (347,170.5) node  [font=\footnotesize]  {$R_{12}$};
% Text Node
\draw (250,252.4) node [anchor=north west][inner sep=0.75pt]  [font=\footnotesize]  {$\max( x_{+} ,x_{-}) =-1$};
% Text Node
\draw (80,2.4) node [anchor=north west][inner sep=0.75pt]  [font=\footnotesize]  {$( -1,3)$};
% Text Node
\draw (1,106.4) node [anchor=north west][inner sep=0.75pt]  [font=\footnotesize]  {$( -3,1)$};
% Text Node
\draw (80,106.4) node [anchor=north west][inner sep=0.75pt]  [font=\footnotesize]  {$( -1,1)$};
% Text Node
\draw (11,332.4) node [anchor=north west][inner sep=0.75pt]  [font=\footnotesize]  {$IV$};
% Text Node
\draw (11,92.4) node [anchor=north west][inner sep=0.75pt]  [font=\footnotesize]  {$III$};
% Text Node
\draw (131,332.4) node [anchor=north west][inner sep=0.75pt]  [font=\footnotesize]  {$II$};
% Text Node
\draw (11,212.4) node [anchor=north west][inner sep=0.75pt]  [font=\footnotesize]  {$II$};
% Text Node
\draw (131,92.4) node [anchor=north west][inner sep=0.75pt]  [font=\footnotesize]  {$II$};

\end{tikzpicture}

\caption{The $(R_{12},R_{34})$-plane. The different regions are separated by the lines $|R_{12}|=1,\,|R_{34}|=1$, and the ones labeled by the same Roman numerals share the same behaviours. The curve in region III is $\max(x_{+},x_{-})=-1$, i.e. $8 + R_{12}^2 + 6 R_{12} R_{34} + R_{34}^2=0$.}
\label{fig:Rregions}
\end{figure}

The relation between $(R_{12},R_{34})$ and $x_{\pm}$ is plotted in figure \ref{fig:Rregions}. The gray region IV is ruled out since the branch points enter into the $s-t$ crossing region $(0,1)\times \R$. The region III is divided by the curve $\max(x_{+},x_{-})=-1$, and outside the curve, the convergent radius of an individual conformal block is less than one. Ignoring this issue, the crossing equation still holds in $(0,1)\times \R$. In the region I and II, the conformal blocks behave similarly as those in CFT$_1$.

To illustrate these features, we consider the special case $\xi_{12}=\xi_{0}$, and $x_{\pm}=\frac{2}{1+R_{34}}$. When $R_{34}\leq 1$, the conformal block can be analytically continued from the one with $(R_{12},R_{34})=(0,0)$ to $(1,R_{34})$ along the curve $(0,0)\to(1,0)\to(1,R_{34})$ while keeping single-valued, and the result is 
\begin{equation}
\blockstripped_{\D_0,\xi_0}^{\schannel}(x,k)
=
e^{-k \xi _0} x^{\D_0} \left(1- \half(R_{34}+1) x\right)^{\D_{12}-\D_0}.
\end{equation}
When $R_{34}>1$, the two roots $x_{\pm}$ enter into $x\in(0,1)$, and the conformal block stops being single-valued. Notice that at $R_{34}=-1$ the conformal block is simply the leading factor $e^{-k \xi _0} x^{\D_0} $, i.e. the contributions of descendant operators are canceled with each other.

The second example is $\xi_{34}=\xi_{12},\, \D_{34}=\D_{12}$, and $x_{-}=1,\, x_{+}=\frac{1}{R_{12}^2}$. When $|R_{12}|\leq 1$, the conformal block can be analytically continued along the diagonal line $(0,0)\to (R_{12},R_{12})$,
\begin{align}
\blockstripped_{\D_0,\xi_0}^{\schannel}(x,k)
&=
\frac{2^{\Delta _0-1}\left(R_{12}+1\right)^{-2 \D_{12}}  }{H(x)}
x^{\Delta _0} 
\exp \frac{-k \xi_0 \sqrt{1-R_{12}^2 x}}{\sqrt{1-x}} 
\\
& \peq 
\cdot 
\bpair{1-\half(1+R_{12}^2)x+H(x)}^{1-\D_0}
\bpair{1+(1-2x)R_{12}^2+2R_{12}H(x)}^{\D_{12}}
\nn
\end{align}
where $H(x)=\sqrt{(1-x)(1-R_{12}^2x)}$. And when $|R_{12}|>1$ the conformal block loses its single-valuedness.

\textbf{Necessity of boost multiplet blocks.} Near the \schannel limit, the analytic behaviour of boost multiplet conformal block \eqref{eq:multipletOPElimit} is different from the singlet one \eqref{eq:singletOPElimit}: after taking apart of the factor $x^{\D_0} e^{-k \xi_0 }$, the power of $k$ in the singlet block cannot exceed the power of $x$, while in the multiplet block there is no such restriction. In concrete examples like MFT, the stripped four-point function divided by the factor $x^{\D_0} e^{-k \xi_0 }$ is analytic near $(x,k)=(0,0)$, and can be expanded as a double Taylor series. The contribution from the singlet blocks is not enough to match the double series of $(x,k)$, hence the boost multiplets must enter into the conformal block expansion in this situation.

Including these boost multiplets into the OPE, the \schannel block expansion of the stripped four-point functions \eqref{eq:blockexpansionstripped} are modified to
\begin{alignat}{4}
\label{eq:blockexpansion}
& \fourpt^{\schannel}(x,k)= 
& \,
& \sum_n p_{n}^{\schannel} \blockstripped^{\schannel}_{n}(x,k)\,+
& \,
& \sum_{n}\sum_{a=0}^{r_n-1}p_{n,a}^{\schannel}
\pdv[a]{\xi_n}\blockstripped^{\schannel}_{n}(x,k)\,+
& \,
& \dots 
\\[-1ex]
& 
& 
& \,\text{\footnotesize	singlets}
& 
& \,\text{\footnotesize	boost multiplets $r_n\geq 2$}
& 
& \,\text{\footnotesize	other operators}
\nn
\end{alignat}
Recovering the kinematical factor, the unstripped version of \eqref{eq:blockexpansion} is
\begin{equation}
\label{eq:blockexpansion2}
\vev{\op_1\op_2\op_3\op_4}= 
\sum_n p_{n}^{\schannel}\blockdressed^{\schannel}_{n}\,+
\sum_{n}\sum_{a=0}^{r_n-1}p_{n,a}^{\schannel}
\pdv[a]{\xi_n}\blockdressed^{\schannel}_{n}\,+ 
\dots .
\end{equation}

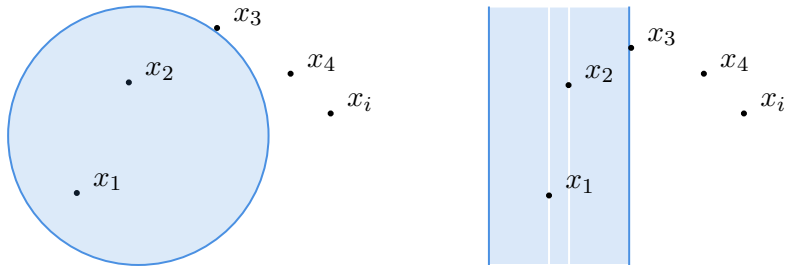
\begin{figure}[htbp]
\centering

\tikzset{every picture/.style={line width=0.75pt}} %set default line width to 0.75pt        

\begin{tikzpicture}[x=0.75pt,y=0.75pt,yscale=-1,xscale=1]

%Shape: Circle [id:dp11700218716495137] 
\draw  [color={rgb, 255:red, 0; green, 0; blue, 0 }  ,draw opacity=1 ][fill={rgb, 255:red, 0; green, 0; blue, 0 }  ,fill opacity=1 ] (43.29,103.79) .. controls (43.29,103.23) and (43.73,102.79) .. (44.29,102.79) .. controls (44.84,102.79) and (45.29,103.23) .. (45.29,103.79) .. controls (45.29,104.34) and (44.84,104.79) .. (44.29,104.79) .. controls (43.73,104.79) and (43.29,104.34) .. (43.29,103.79) -- cycle ;

%Shape: Circle [id:dp5829131366313034] 
\draw  [color={rgb, 255:red, 0; green, 0; blue, 0 }  ,draw opacity=1 ][fill={rgb, 255:red, 0; green, 0; blue, 0 }  ,fill opacity=1 ] (69.26,48.17) .. controls (69.26,47.62) and (69.7,47.17) .. (70.26,47.17) .. controls (70.81,47.17) and (71.26,47.62) .. (71.26,48.17) .. controls (71.26,48.72) and (70.81,49.17) .. (70.26,49.17) .. controls (69.7,49.17) and (69.26,48.72) .. (69.26,48.17) -- cycle ;

%Shape: Circle [id:dp5578761977912772] 
\draw  [color={rgb, 255:red, 74; green, 144; blue, 226 }  ,draw opacity=1 ][fill={rgb, 255:red, 74; green, 144; blue, 226 }  ,fill opacity=0.19 ] (10,75) .. controls (10,39.1) and (39.1,10) .. (75,10) .. controls (110.9,10) and (140,39.1) .. (140,75) .. controls (140,110.9) and (110.9,140) .. (75,140) .. controls (39.1,140) and (10,110.9) .. (10,75) -- cycle ;
%Shape: Circle [id:dp337375700131864] 
\draw  [color={rgb, 255:red, 0; green, 0; blue, 0 }  ,draw opacity=1 ][fill={rgb, 255:red, 0; green, 0; blue, 0 }  ,fill opacity=1 ] (150,43.79) .. controls (150,43.23) and (150.45,42.79) .. (151,42.79) .. controls (151.55,42.79) and (152,43.23) .. (152,43.79) .. controls (152,44.34) and (151.55,44.79) .. (151,44.79) .. controls (150.45,44.79) and (150,44.34) .. (150,43.79) -- cycle ;

%Shape: Circle [id:dp1551875747311855] 
\draw  [color={rgb, 255:red, 0; green, 0; blue, 0 }  ,draw opacity=1 ][fill={rgb, 255:red, 0; green, 0; blue, 0 }  ,fill opacity=1 ] (113.29,20.79) .. controls (113.29,20.23) and (113.73,19.79) .. (114.29,19.79) .. controls (114.84,19.79) and (115.29,20.23) .. (115.29,20.79) .. controls (115.29,21.34) and (114.84,21.79) .. (114.29,21.79) .. controls (113.73,21.79) and (113.29,21.34) .. (113.29,20.79) -- cycle ;

%Shape: Rectangle [id:dp5870642574292404] 
\draw  [color={rgb, 255:red, 255; green, 255; blue, 255 }  ,draw opacity=1 ][fill={rgb, 255:red, 74; green, 144; blue, 226 }  ,fill opacity=0.2 ] (250,10) -- (320,10) -- (320,140) -- (250,140) -- cycle ;
%Straight Lines [id:da3806222602422429] 
\draw [color={rgb, 255:red, 74; green, 144; blue, 226 }  ,draw opacity=1 ]   (250,10) -- (250,140) ;
%Straight Lines [id:da3304244560865306] 
\draw [color={rgb, 255:red, 74; green, 144; blue, 226 }  ,draw opacity=1 ]   (320,10) -- (320,140) ;
%Straight Lines [id:da25854758851591986] 
\draw [color={rgb, 255:red, 255; green, 255; blue, 255 }  ,draw opacity=1 ]   (280,10) -- (280,140) ;
%Straight Lines [id:da1412140487534026] 
\draw [color={rgb, 255:red, 255; green, 255; blue, 255 }  ,draw opacity=1 ]   (290,10) -- (290,140) ;
%Shape: Circle [id:dp4398836228808265] 
\draw  [color={rgb, 255:red, 0; green, 0; blue, 0 }  ,draw opacity=1 ][fill={rgb, 255:red, 0; green, 0; blue, 0 }  ,fill opacity=1 ] (279,105.04) .. controls (279,104.48) and (279.45,104.04) .. (280,104.04) .. controls (280.55,104.04) and (281,104.48) .. (281,105.04) .. controls (281,105.59) and (280.55,106.04) .. (280,106.04) .. controls (279.45,106.04) and (279,105.59) .. (279,105.04) -- cycle ;

%Shape: Circle [id:dp5214008477838914] 
\draw  [color={rgb, 255:red, 0; green, 0; blue, 0 }  ,draw opacity=1 ][fill={rgb, 255:red, 0; green, 0; blue, 0 }  ,fill opacity=1 ] (288.75,49.54) .. controls (288.75,48.98) and (289.2,48.54) .. (289.75,48.54) .. controls (290.3,48.54) and (290.75,48.98) .. (290.75,49.54) .. controls (290.75,50.09) and (290.3,50.54) .. (289.75,50.54) .. controls (289.2,50.54) and (288.75,50.09) .. (288.75,49.54) -- cycle ;

%Shape: Circle [id:dp7716525364013498] 
\draw  [color={rgb, 255:red, 0; green, 0; blue, 0 }  ,draw opacity=1 ][fill={rgb, 255:red, 0; green, 0; blue, 0 }  ,fill opacity=1 ] (320,30.79) .. controls (320,30.23) and (320.45,29.79) .. (321,29.79) .. controls (321.55,29.79) and (322,30.23) .. (322,30.79) .. controls (322,31.34) and (321.55,31.79) .. (321,31.79) .. controls (320.45,31.79) and (320,31.34) .. (320,30.79) -- cycle ;

%Shape: Circle [id:dp6449961997875444] 
\draw  [color={rgb, 255:red, 0; green, 0; blue, 0 }  ,draw opacity=1 ][fill={rgb, 255:red, 0; green, 0; blue, 0 }  ,fill opacity=1 ] (170,63.79) .. controls (170,63.23) and (170.45,62.79) .. (171,62.79) .. controls (171.55,62.79) and (172,63.23) .. (172,63.79) .. controls (172,64.34) and (171.55,64.79) .. (171,64.79) .. controls (170.45,64.79) and (170,64.34) .. (170,63.79) -- cycle ;

%Shape: Circle [id:dp5019176929819313] 
\draw  [color={rgb, 255:red, 0; green, 0; blue, 0 }  ,draw opacity=1 ][fill={rgb, 255:red, 0; green, 0; blue, 0 }  ,fill opacity=1 ] (356.29,43.79) .. controls (356.29,43.23) and (356.73,42.79) .. (357.29,42.79) .. controls (357.84,42.79) and (358.29,43.23) .. (358.29,43.79) .. controls (358.29,44.34) and (357.84,44.79) .. (357.29,44.79) .. controls (356.73,44.79) and (356.29,44.34) .. (356.29,43.79) -- cycle ;

%Shape: Circle [id:dp4345229846409837] 
\draw  [color={rgb, 255:red, 0; green, 0; blue, 0 }  ,draw opacity=1 ][fill={rgb, 255:red, 0; green, 0; blue, 0 }  ,fill opacity=1 ] (376.29,63.79) .. controls (376.29,63.23) and (376.73,62.79) .. (377.29,62.79) .. controls (377.84,62.79) and (378.29,63.23) .. (378.29,63.79) .. controls (378.29,64.34) and (377.84,64.79) .. (377.29,64.79) .. controls (376.73,64.79) and (376.29,64.34) .. (376.29,63.79) -- cycle ;

% Text Node
\draw (51,92.4) node [anchor=north west][inner sep=0.75pt]    {$x_{1}$};
% Text Node
\draw (76.97,36.79) node [anchor=north west][inner sep=0.75pt]    {$x_{2}$};
% Text Node
\draw (157.71,32.4) node [anchor=north west][inner sep=0.75pt]    {$x_{4}$};
% Text Node
\draw (121,9.4) node [anchor=north west][inner sep=0.75pt]    {$x_{3}$};
% Text Node
\draw (327.71,19.4) node [anchor=north west][inner sep=0.75pt]    {$x_{3}$};
% Text Node
\draw (296.46,38.15) node [anchor=north west][inner sep=0.75pt]    {$x_{2}$};
% Text Node
\draw (286.71,93.65) node [anchor=north west][inner sep=0.75pt]    {$x_{1}$};
% Text Node
\draw (177.71,52.4) node [anchor=north west][inner sep=0.75pt]    {$x_{i}$};
% Text Node
\draw (384,52.4) node [anchor=north west][inner sep=0.75pt]    {$x_{i}$};
% Text Node
\draw (364,32.4) node [anchor=north west][inner sep=0.75pt]    {$x_{4}$};

\end{tikzpicture}

\caption{OPE convergence: Euclidean CFT vs. GCFT$_2$. The left shows the OPE convergence region in compact CFTs, the right shows the one in GCFT$_2$ under suitable assumptions.}
\label{fig:opeconvergence}
\end{figure}

Assuming that the operator spectrum and conformal block coefficients $p_{0,a}^{\schannel}$ are well-controlled, like in the MFT, the $s-t$ crossing equation holds in the region $(0,1)\times \R$. This indicates that under suitable conditions, the OPE convergence region in GCFT$_2$ is a stripe as shown in figure \ref{fig:opeconvergence}, exhibiting the non-locality of $y$-direction.

\subsection{Conformal partial waves and blocks from shadow formalism}\label{sec:shadowpartialwaves}
In this subsection we derive the conformal partial waves and conformal blocks from the shadow formalism, and discuss their relations.

\textbf{Conformal partial waves.} The \schannel conformal partial wave $\CPWdressed_{\D_0,\xi_0}^{1234}(\xy{i})$ depends on the four external virtual operators $\op_i \in \urep{\D_i, \xi_i},\,\xi_i=\xi_0 R_i,\, i=1,2,3,4$ and the exchanged virtual operator $\op_0\in \urep{\D_0,\xi_0},\,\xi_0\neq 0$. The external indices will be omitted if no ambiguity. The conformal partial waves can be constructed as
\begin{equation}
\CPWdressed_{\D_0,\xi_0}(\xy{i})
=
\inttt{dx_0 dy_0}{\R^2} 
\vev{\op_1\op_2 \op_0(\xy{0})} 
\vev{\opshadow_0(\xy{0})\op_3\op_4}.
\label{eq:shadowCPW1}
\end{equation}
The stripped conformal partial waves $\CPWstripped_{\D_0,\xi_0}(\xy)$ are defined by factoring out the kinematical factor $K^{\schannel}$,
\begin{equation}
\CPWdressed_{\D_0,\xi_0}(\xy{i})
=K^{\schannel}(\xy{i})\CPWstripped_{\D_0,\xi_0}(x,k).
\end{equation}

Since the stripped conformal partial wave depends only on the cross ratios, we fix the gauge to the standard conformal frame, and find
\begin{equation}
\CPWstripped_{\D_0,\xi_0}(x,k)=\inttt{dx_0dy_0}{\R^2} F_x \exp (F_y),
\label{eq:cpwgcft2}
\end{equation}
in which
\begin{align}
F_x& = |x|^{\D_0} |x_{0}|^{-\D_{01,2}} |x-x_{0}|^{-\D_{02,1}} |1-x_{0}|^{\D_{04,3}-2},
\label{eq:cpwFx}
\\
F_y& = \xi_0 y_{0} \frac{(1+R_{34})x_{0}^{2}- (2+R_{12}x+R_{34}x) x_{0}+ (1+R_{12})x }{(x-x_{0})(1-x_{0})x_{0}}
+\xi_0 k \frac{x_0-R_{12}x}{x-x_0}.
\end{align}
The exponential factor is a pure phase due to the analytic continuation $\xi_0\in i\Rnot$, hence the integration of $y_0$ gives Dirac $\delta$-distributions of $x_0$
\begin{equation}
2\pi\d\bpair{\frac{|\xi_{0}|(1+R_{34})(x_{0}-x_{0,-})(x_{0}-x_{0,+})}{(x-x_{0})(1-x_{0})x_{0}}}
=\frac{2\pi|x_0(x-x_0)(1-x_0)|}{|\xi_{0}||1+R_{34}||x_{0,+}-x_{0,-}|}\bpair{\d(x_0-x_{0,+})+\d(x_0-x_{0,-})},
\end{equation}
where the two roots of $x_{0}$ are 
\begin{equation}
x_{0,\pm}=\frac{1+\half (R_{12}+R_{34}) x \pm H(x)}{1+R_{34}},
\label{eq:cpwRoots}
\end{equation}
and $H(x)$ is the same as \eqref{eq:functionH}. Substituting the $\delta$-distributions into the integrand, the resulting conformal partial wave is a combination of two conformal blocks with analytical continued weights
\begin{equation}
\CPWstripped_{\D_0,\xi_0}=
\shadow(\op_3\op_4[\opshadow_0])  \blockstripped_{\D_0,\xi_0}(x,k)
+
\shadow(\op_1\op_2[\op_0])  \blockstripped_{2-\D_0,-\xi_0}(x,k),
\label{eq:cpw}
\end{equation}
in which $\d(x_0-x_{0,+})$ contributes to the physical block $\blockstripped_{\D_0,\xi_0}(x,k)$ and the prefactors are simply the shadow coefficients. Notice that in \eqref{eq:cpw}  the power function parts should be understood as the absolute values since they come from \eqref{eq:cpwFx}. Recovering the kinematical factor we get the unstripped conformal partial wave
\begin{equation}
\CPWdressed_{\D_0,\xi_0}(\xy{i})=
\shadow(\op_3\op_4[\opshadow_0])  \blockdressed_{\D_0,\xi_0}(\xy{i})
+
\shadow(\op_1\op_2[\op_0])  \blockdressed_{2-\D_0,-\xi_0}(\xy{i}),
\label{eq:cpw2}
\end{equation}
and the shadow partial wave $\CPWdressed_{2-\D_0,-\xi_0}$ is proportional to $\CPWdressed_{\D_0,\xi_0}$,
\begin{equation}
\CPWdressed_{2-\D_0,-\xi_0}(\xy{i})=\frac{\shadow(\op_1\op_2[\opshadow_0])}{\shadow(\op_3\op_4[\opshadow_0])}\CPWdressed_{\D_0,\xi_0}(\xy{i}),
\label{eq:shadowcpw}
\end{equation}
by using the identity \eqref{eq:shadowCoefficientsToPlancherelMeasure}.

\textbf{Relation to conformal blocks.} The relation \eqref{eq:cpw2} between the conformal block and the conformal partial wave is similar to that in relativistic CFT \cite{Simmons-Duffin:2017nub,Karateev:2018oml}. However in our case, when the external operators are identical, the conformal partial wave constructed from the shadow formalism is not the same as the one from the spectral decomposition of the Casimir operators \cite{Chen:2020vvn}. This could be due to the fact that different boundary conditions at $x=1$ lead to different self-adjoint extensions of the Casimir operators, hence the eigenvalues and eigenfunctions are not the same. The analog of \eqref{eq:cpw} in relativistic CFT appears in the alpha space approach \cite{Hogervorst:2017sfd,Hogervorst:2017kbj,Rutter:2020vpw}, where the resulting stripped partial waves are only supported on $z\in(0,1)$.

The conformal partial wave \eqref{eq:cpw} is also not supported on the whole cross-ratio plane $\R^{2}$. When the $x_0$-roots \eqref{eq:cpwRoots} take complex values, the $\d$-distributions vanish and the corresponding partial wave also vanishes. This reality condition gives the support of conformal partial waves, $x\in(-\inf,x_{-})\cup (x_{+},\inf)$, where $x_{\pm}$ are the zeros of $H(x)$ \eqref{eq:rootsOffunctionHx}, and when $x_{\pm}\notin\R$, the support is $x\in\R$.

The relation \eqref{eq:cpw2} between conformal partial waves and blocks can be understood from the integral expression \eqref{eq:block4gcft} of conformal blocks. Assuming $x_1<x_2<x_3<x_4$ and $R_{12},R_{34}\in(-1,1)$, inserting the OPE block \eqref{eq:opeblockshadow1} and the normalization factor \eqref{eq:opeblocknormalization1} into \eqref{eq:block4gcft} we get,
\begingroup
\allowdisplaybreaks
\begin{align}
\blockdressed^{\schannel}_{\D_{0},\xi_{0}}(\xy{i})
&=\cD_{120}\cD_{430}\vev{\op_0(\xy{2})\op_0(\xy{3})}
\nn
\\
&=\inttt{dx'_0dy'_0}{I_{34}}\inttt{dx_0dy_0}{I_{12}}
\vev{\op_1 \op_2 \opshadow_0(\xy{0})}\vev{\op_0(\xy{0})\op_0(x'_{0},y'_{0})}
\nn
\\
&\peq \cdot 
\vev{\opshadow_0(x'_{0},y'_{0})\op_3\op_4} 
\shadow(\op_1\op_2[\opshadow_0])^{-1}\shadow(\op_4\op_3[\opshadow_0])^{-1}
\nn
\\
&=\shadow(\op_4\op_3[\opshadow_0])^{-1}\inttt{dx'_0dy'_0}{I_{34}} \vev{\op_1\op_2\opshadow_0(x'_{0},y'_{0})}
\vev{\opshadow_0(x'_{0},y'_{0})\op_3\op_4}, 
\label{eq:blocktheshadow}
\end{align}
\endgroup
where the integration region is $I_{ij}=(x_i,x_j)\times\R$. Under the assumption $R_{12}\in(-1,1)$, by \eqref{eq:shadowco1} and \eqref{eq:shadowco2}, the integration region $I_{12}$ can be extended to $\R^2$. Hence in the second line we can apply the integral expression of the shadow coefficients \eqref{eq:star2} and cancel the factor $\shadow(\op_1\op_2[\opshadow_0])^{-1}$. 

Since after gauge fixing the physical block comes from $\d(x_0-x_{0,+})$, to match the result \eqref{eq:blocktheshadow} with the first term in \eqref{eq:block4gcft} we must have the inequality,
\begin{equation}
R_{12},R_{34}\in(-1,1) \qand x\in(-\inf,x_{-})\cup (x_{+},\inf) \quad\implies\quad x_{0,+}\in(x_3,x_4)=(1,\inf),
\end{equation}
and this is indeed correct.

\subsection{Inversion formula}

In this subsection we discuss the orthogonality of partial waves, and then establish an inversion formula of four-point functions. In the following, the point $\xi=0$ should be excluded from the $\xi$-contour $\int^{i\inf}_{-i\inf}\frac{d\xi}{2\pi i}$ in the inversion formula, since there is no massless representation in the Plancherel measure of $ISO(2,1)$\footnote{The $\xi=0$ four-point functions are not in the Hilbert space of the $\xi\neq 0$ conformal partial waves, and should be expanded in a different basis, see \cite{Chen:2022jhx}.}. Alternatively, the two intervals $(-i\oo,0)$ and $(0,i\oo)$ can be joined by a small semicircle, and the integral along this semicircle vanishes since the Plancherel measure is proportional to $\xi^{2}d\xi$.

\textbf{Orthogonality.} The orthogonality of conformal partial waves can be derived from the bubble integral \eqref{eq:bubblelast} in the following way. Denoting $\CPWdressed_i$ as the unstripped conformal partial waves of exchanged virtual operators $\urep{\D_i,\xi_i}$, they admit a natural inner product, which is invariant under the Galilean conformal transformations,
\begin{align}
(\CPWdressed_1,\CPWdressed_2)
=\int\,\frac{\prod_{i=1}^{4} dx_{i}dy_{i}}{\vol{ISO(2,1)}}
\CPWdressed_{1}^{*}(\xy{i})\CPWdressed_{2}(\xy{i}),
\label{eq:CPWpairing}
\end{align}
where the infinite volume factor $\vol{ISO(2,1)}$ is to cancel the divergence of the integral. Noticing that under the complex conjugation $\vev{\op_i\dots }\Leftrightarrow\vev{\opshadow_i\dots }$, the inner product \eqref{eq:CPWpairing} can be evaluated as follows,
\begingroup
\allowdisplaybreaks
\begin{align}
(\CPWdressed_5,\CPWdressed_6)
& =\int\,\frac{\prod_{i=1}^{6} dx_{i}dy_{i}}{\vol{ISO(2,1)}}
\vev{\opshadow_1\opshadow_2\opshadow_5}\vev{\op_5\opshadow_3\opshadow_4}
\vev{\op_1\op_2\op_6}\vev{\opshadow_6\op_3\op_4}
\\
& =\int\,\frac{\prod_{i=3}^{6} dx_{i}dy_{i}}{\vol{ISO(2,1)}}
\vev{\opshadow_6\op_3\op_4}\vev{\op_5\opshadow_3\opshadow_4}
\int\,\prod_{i=1}^{2} dx_{i}dy_{i}
\vev{\op_1\op_2\op_6}\vev{\opshadow_1\opshadow_2\opshadow_5}\nn
\\
& =\int\,\frac{\prod_{i=3}^{6} dx_{i}dy_{i}}{\vol{ISO(2,1)}}
\vev{\opshadow_6\op_3\op_4}\vev{\op_5\opshadow_3\opshadow_4}
\big[
\d(\op_5,\op_6)\d(x_{56})\d(y_{56})
\nn
\\
&\peq 
+\, \shadow(\opshadow_1\opshadow_2[\op_5])^{-1}\d(\op_6,\opshadow_5)\vev{\op_6(\xy{5})\op_6}
\big]
\nn
\\
& =8\pi^2 A_{\text{3-pt}}\, \cN(\D_5,\xi_5)\bpair{\d(r_{56})\d(s_{56}) 
+ \frac{\shadow(\op_3\op_4[\op_5])}{\shadow(\op_1\op_2[\op_5])} \d(r_{5}+r_{6})\d(s_{5}+s_{6}) },
\nn
\end{align}
\endgroup
where in the third line we have used the bubble integral \eqref{eq:bubblelast}, and the integration of $(\xy{6})$ in the second term is done by the shadow coefficient \eqref{eq:star2}. In the last line the prefactor $A_{\text{3-pt}}$ is a three-point pairing
\begin{equation}
A_{\text{3-pt}}=\int\,\frac{\prod_{i=3}^{5} dx_{i}dy_{i}}{\vol{ISO(2,1)}}
\vev{\opshadow_5\op_3\op_4}\vev{\op_5\opshadow_3\opshadow_4}.
\label{eq:threepointpairing}
\end{equation}
This integral is an adjustable numerical factor, since the three points can be fixed by the conformal symmetry $ISO(2,1)$ and there are no integrals and residual symmetries left. We take $A_{\text{3-pt}}=\half$ by rescaling $\vol{ISO(2,1)}$, and the inner product \eqref{eq:CPWpairing} is
\begin{equation}
(\CPWdressed^{1234}_i,\CPWdressed^{1234}_{j})=4\pi^2\cN(\D_i,\xi_i)\bpair{\d(r_{ij})\d(s_{ij})+\frac{\shadow(\op_3\op_4[\op_i])}{\shadow(\op_1\op_2[\op_i])} \d(r_{i}+r_{j})\d(s_{i}+s_{j})}.
\end{equation}

\textbf{Inversion formula and inversion function.} Now we can decompose the four-point function $\vev{\op_1\op_2\op_3\op_4}$ by projecting onto the conformal partial waves,
\begin{equation}
I(\D,\xi):=(\Psi^{1234}_{\D,\xi},\vev{\op_1\op_2\op_3\op_4}).
\label{eq:inversionfunction}
\end{equation}
However as discussed in section \ref{sec:cgkernelO}, the set of conformal partial waves with respect to the unitary principal series is not a complete basis of the space of normalizable four-point functions $V=L^2((\R^2)^4_{x_i\neq x_j,y_i\neq y_j})$ with inner product \eqref{eq:CPWpairing}, and the corresponding projection operator $\cP$ acting on this space gives a subspace $\cP V$. We denote the projected four-point function by the subscript $\cP$.

The purpose of the Euclidean inversion formula is to diagonalize the four-point functions and to recover the OPE data. In practice, the information of the \schannel conformal block expansion is stored in the analytic structure near the \schannel OPE limit. If the projected subspace $\cP V$ contains four-point functions supported on a neighborhood of the \schannel limit, we can expect to read the conformal block expansion from the inversion function \eqref{eq:inversionfunction} by the contour deformation procedure.
To be concrete, we have the following inversion formula,
\begin{align}
\label{eq:inversionformula}
\vev{\op_1\op_2\op_3\op_4}_{\cP}
&=A_{\text{4-pt}}\int^{1+i\inf}_{1}\frac{d\D}{2\pi i}\int^{i\inf}_{-i\inf}\frac{d\xi}{2\pi i}\frac{I(\D,\xi)}{\cN(\D,\xi)}\Psi^{1234}_{\D,\xi}
\nn\\
&=\int^{1+i\inf}_{1-i\inf}\frac{d\D}{2\pi i}\int^{i\inf}_{-i\inf}\frac{d\xi}{2\pi i}\frac{I(\D,\xi)}{\cN(\D,\xi)}\shadow(\op_3\op_4[\opshadow])\blockdressed_{\D,\xi}\nn
\\
&=\int^{1+i\inf}_{1-i\inf}\frac{d\D}{2\pi i}\int^{i\inf}_{-i\inf}\frac{d\xi}{2\pi i}c(\D,\xi)\blockdressed_{\D,\xi}.
\end{align}
In the second line we have changed the variables $(\D,\xi)\to(2-\D,\xi)$ and used the identity
\begin{equation}
I(2-\D,-\xi)=\frac{\shadow(\op_1\op_2[\op])}{\shadow(\op_3\op_4[\op])}I(\D,\xi),
\end{equation}
which comes from \eqref{eq:shadowcpw}. The overall numerical factor $A_{\text{4-pt}}$ in \eqref{eq:inversionformula} is determined as follows: inserting the orthogonality relation \eqref{eq:inversionfunction} into \eqref{eq:inversionformula}, the conformal partial wave should be recovered and this gives $A_{\text{4-pt}}=(2A_{\text{3-pt}})^{-1}=1$. For convenience, in the last line we introduce
\begin{equation}
c(\D,\xi)=\frac{I(\D,\xi)}{\cN(\D,\xi)}\shadow(\op_3\op_4[\opshadow]),
\end{equation}
and call $c(\D,\xi)$ as the inversion function of the four-point function.

From \eqref{eq:multipletOPElimit} the conformal blocks decay exponentially in the right $\D$-plane and right $\xi$-plane when $0<x<1,\, k>0$, hence if the inversion function $c(\D,\xi)$ in \eqref{eq:inversionformula} are sub-exponential we can deform the contours of $(\D,\xi)$ from the unitary principal series to the right infinities and pick up the poles inside the right half-planes.  Writing the partial fraction decomposition of $c(\D,\xi)$ as
\begin{equation}
c(\D,\xi)=\sum_{n}\sum_{a=0}^{r_n-1}\frac{p^{\text{inversion}}_{n,a}}{(\D-\D_{n})(\xi-\xi_{n})^{a+1}}+\dots
\end{equation}
where the poles $(\D_{n},\xi_{n})$ are in the right half-planes, we get the conformal block expansion \eqref{eq:blockexpansion} for the projected four-point functions,
\begin{equation}
\vev{\op_1\op_2\op_3\op_4}_{\cP}= 
\sum_{n}\sum_{a=0}^{r_n-1}p_{n,a}
\pdv[a]{\xi_n}\blockdressed_{n}(x,k), \qand p_{n,a}=\frac{1}{a!}p^{\text{inversion}}_{n,a}.
\end{equation}

\textbf{Projector.} Finally we discuss the projection operator. Inserting the inversion function \eqref{eq:inversionfunction} into the inversion formula \eqref{eq:inversionformula} and we get the projection operator
\begin{align}
\label{eq:projector1}
\vev{\op_1\op_2\op_3\op_4}_{\cP}=
\intt{\frac{\prod_{i=1}^{4} dx'_{i}dy'_{i}}{\vol{ISO(2,1)}}}
\vev{\op'_1\op'_2\op'_3\op'_4}\cdot\cP^{1'2'3'4'}_{1234}(x'_{i},y'_{i};\xy{i}),
\end{align}
where $\op'_{i}=\op_{i}(x'_{i},y'_{i})$ and the kernel of $\cP$ is
\begin{align}
\cP^{1'2'3'4'}_{1234}
&=
\inttt{\frac{dr_{0}ds_{0}}{4\pi^{2} \cN(\D_{0},\xi_{0})}}{\R\times \Rpositive}
\ppair{\CPWdressed^{1234}_{\D_{0},\xi_{0}}(x'_{i},y'_{i})}^*\CPWdressed^{1234}_{\D_{0},\xi_{0}}(x_{i},y_{i})\\
&=
\inttt{\frac{dr_{0}ds_{0}}{4\pi^{2} \cN(\D_{0},\xi_{0})}}{\R\times \Rpositive}
\intt{dx_{0}dy_{0}dx'_{0}dy'_{0}}
\vev{\opshadow'_1\opshadow'_2\opshadow'_{0}}\vev{\op'_0\opshadow'_3\opshadow'_4}
\vev{\op_1\op_2\op_0}\vev{\opshadow_0\op_3\op_4}.
\nn
\end{align}

In the appendix \ref{sec:completeness1} we show that in the case of four identical external operators, the projector is proportional to 
\begin{equation}
\cP^{1'2'3'4'}_{1234}\sim \th(1-x_{12,34})\d(x_{12,34}-x'_{12,34})\d(k_{12,34}-k'_{12,34}),
\end{equation}
where $x_{12,34},\, k_{12,34}$ are the cross ratios of four points. Hence for the stripped four-point function $\fourpt(x,k)$, $\cP$'s action gives $\fourpt(x,k)_{\cP}=\fourpt(x,k),\, x<1$. Furthermore, as discussed in section \ref{sec:shadowpartialwaves} the zeros of $H(x)$ are $x_{-}=1,x_{+}=\inf$ and the stripped conformal partial waves are supported and orthogonal on $(-\inf,1)$. Combining these two aspects, the projected subspace $\cP V$ contains normalizable four-point functions on the region $x_{12,34}<1$, and the set of conformal partial waves is an orthogonal and complete basis of $\cP V$.

Besides, in \cite{Chen:2020vvn} the conformal partial waves form a complete basis on the interval $(0,1)$ by the alpha space method. And under the symmetry of four-point functions the interval $(0,1)$ is mapped to $(-\inf,0)$, hence they are complete and orthogonal on $(-\inf,1)$. 
This shows the equivalence of shadow formalism and the spectral decomposition of Casimir operators.

In the case of $\vev{\op_1\op_2\op_2\op_1}$-type four-point functions, $\xi_{34}=-\xi_{12}$, the set of conformal partial waves runs over the anti-diagonal line in figure \ref{fig:Rregions}. 
The zeros of $H(x)$ are $\set{\frac{\xi^2}{\xi^2-\xi_{12}^2},\inf}$ as discussed in section \ref{sec:shadowpartialwaves}, 
and if $|\xi|>|\xi_{12}|$ or $|\xi|<|\xi_{12}|$, the conformal partial wave is single-valued with support $(-\inf,\frac{\xi^2}{\xi^2-\xi_{12}^2})\supset (-\inf,1)$ or $(\frac{\xi^2}{\xi^2-\xi_{12}^2},\inf)\supset (0,\inf)$ respectively.
Although the subspace $\cP V$ does not admit a simple description, the supports of conformal partial waves always cover the interval $(0,1)$, and normalizable four-point functions in the crossing region $(0,1)\times\RR$ can be expanded in $\cP V$ without loss of information.
Therefore, we can expect that the \schannel conformal block expansion is captured by the projected four-point functions.
In the next section \ref{sec:MFT1} we will show that this is correct by the example of mean field theory.

% In the case of $\vev{\op_1\op_2\op_2\op_1}$-type four-point functions, $\xi_{34}=-\xi_{12}$, the set of conformal partial waves runs over the anti-diagonal line in figure \ref{fig:Rregions}. The subspace $\cP V$ does not admit a simple description due to the following reason. 
% The zeros of $H(x)$ are $\set{\frac{\xi^2}{\xi^2-\xi_{12}^2},\inf}$. 
% If $|\xi|>|\xi_{12}|$, the conformal partial wave is single-valued with support $(-\inf,\frac{\xi^2}{\xi^2-\xi_{12}^2})\supset (-\inf,1)$ as discussed in section \ref{sec:shadowpartialwaves}.
% If $|\xi|<|\xi_{12}|$, the conformal partial wave is single-valued with support $(\frac{\xi^2}{\xi^2-\xi_{12}^2},\inf)\supset (0,\inf)$. 
% Nevertheless, the interval $(0,1)$ is always included in the support of conformal partial wave, and we can expect that the correct \schannel conformal block expansion is captured by the projected four-point functions. In the next section \ref{sec:MFT1} we show that this is correct by the example of mean field theory.

\section{Applications and Generalizations of Shadow Formalism}

In this section, we discuss a few applications and generalizations of shadow formalism in GCFT. The first one is to reconsider the decomposition of the four-point functions in Galilean mean field theory, of which a special case has been studied in \cite{Chen:2020vvn}. The second application is to construct Lagrangian of Galilean MFT. We manage to find a series of bilocal actions, corresponding to the Galilean MFT, with the help of the kernel of the shadow transform. 

\subsection{Decomposition of four-point functions in mean field theory}\label{sec:MFT1}

The mean field theory (MFT), or the generalized free theory, is defined as that all its correlation functions are the Wick contractions of the two-point functions. Regarding the field theory as a stochastic process, the MFT is equivalent to the Gaussian process, and the two-point function is called the covariance function. The MFT provides a simple example of CFT when the two-point function is conformally covariant.

In the relativistic case, the MFT is the leading contribution of a large-$N$ CFT, and corresponds to the free theory in AdS for a holographic CFT \cite{Heemskerk:2009pn}. It also gives the leading contribution at large spin in the context of analytic bootstrap \cite{Komargodski:2012ek,Fitzpatrick:2012yx,Alday:2016njk}. Finally it relates to the long-range models. The MFT with a fundamental scalar $\f$, $\D_{\f}\neq \frac{d-2}{2}$ admits an unusual Lagrangian description: the kinematical term contains the fractional Laplacian and is nonlocal, see e.g. \cite{Rajabpour:2011qr}. Deformed by a relevant quartic interaction, it flows to the long-range Ising model for a window of the parameters \cite{Paulos:2015jfa}.

In GCFT$_2$, the MFT \cite{Chen:2020vvn} is one of the few concrete models besides the BMS free scalar \cite{Hao:2021urq} and free fermion \cite{Bagchi:2017cte}. In this subsection we consider the MFT containing two different bosonic singlets $\f_i\in \rep{\D_i,\xi_i}$. The two point functions are
\begin{align}
\vev{\f_1\f_1}& =|x_{12}|^{-2\D_1}e^{2\xi_1 k_{12}},
\\
\vev{\f_2\f_2}& =|x_{12}|^{-2\D_2}e^{2\xi_2 k_{12}},
\\
\vev{\f_1\f_2}& =0.
\end{align}
The four-point function of $\f_1,\,\f_2$ we are interested in is
\begin{equation}
\vev{\f_1\f_2\f_2\f_1}=\vev{\f_1(\xy{1})\f_1(\xy{4})}\vev{\f_2(\xy{2})\vev{\f_2(\xy{3})}}.
\label{eq:fourpoint1}
\end{equation}
The \tchannel $\f_1\times \f_1 \to \f_2 \times \f_2$ OPE is trivial, and there is only one exchanged operator: the identity operator $\id$. The \schannel $\f_1\times \f_2 \to \f_1 \times \f_2$ OPE is expected to exchange double trace operators, schematically $\norder{\f_1\p^{n}\f_2}$. 

In the following we review the operator construction method \cite{Chen:2020vvn}. Then using the shadow formalism and the inversion formula we decompose the four-point function \eqref{eq:fourpoint1} and obtain the conformal block expansion. The derivation of the conformal block expansion from the inversion formula necessitates a dispersion-like relation on the $\xi$-plane.

\textbf{Method of operator construction.} The leading and next-to-leading terms in the conformal block expansion was calculated by operator construction in \cite{Chen:2020vvn}. The leading exchanged conformal family is the singlet with primary operator $\op_{\text{L}}=\norder{\f_1\f_2}$. At the next-to-leading order, there are four composite operators
\begin{equation}
\set{
\norder{\p_{x}\f_1\f_2},\, \norder{\p_{y}\f_1\f_2},\, \norder{\f_1\p_{x}\f_2},\, \norder{\f_1\p_{y}\f_2}
}.
\end{equation}
Diagonalizing them by $L_{0},\, M_{0}$, two operators are descendants of $\norder{\f_1\f_2}$, and the rest two constitute the primaries $\op_{\text{NL}}^{a},\, a=1,2$ in a rank-$2$ boost multiplet,
\begin{align}
\op_{\text{NL}}^{1}
&=
\frac{\Delta_{2} \xi_{1}^{2}-\Delta_{1} \xi_{2}\left(2 \xi_{1}+\xi_{2}\right)}{2 \xi_{1}\left(\xi_{1}+\xi_{2}\right)}
\norder{\p_{y}\f_1\f_2}
-
\frac{\Delta_{1} \xi_{2}^{2}-\Delta_{2} \xi_{1}\left(\xi_{1}+2 \xi_{2}\right)}{2 \xi_{2}\left(\xi_{1}+\xi_{2}\right)}
\norder{\f_1\p_{y}\f_2}
\nn
\\
&\peq
-
\xi_{2}\norder{\p_{x}\f_1\f_2}
+
\xi_{1}\norder{\f_1\p_{x}\f_2},
\nn
\\
\op_{\text{NL}}^{2}
&=\xi_{2}\norder{\p_{y}\f_1\f_2}-\xi_{1}\norder{\f_1\p_{y}\f_2}.
\end{align}
Here the two-point functions of $\op_{\text{NL}}^{a}$ are not normalized to the standard form \eqref{eq:twopointrankr} and the overall factor is $d=2\xi_1\xi_2(\xi_1+\xi_2)$. Accordingly the block coefficients \eqref{eq:blocktothreepoint} should be divided by $d$. The three-point coefficients in $\vev{\f_1\f_2\op_{\text{NL}}^{a}}$ are
\begin{equation}
c_{\f_1\f_2}{}_{\op_{\text{NL}}^{1}}=-2\xi_1\xi_2,
\hspace{3ex}
c_{\f_1\f_2}{}_{\op_{\text{NL}}^{2}}=-\frac{\Delta_{2} \xi_{1}^{2}+\Delta_{1} \xi_{2}^{2}}{\xi_{1}+\xi_{2}}.
\end{equation}
Then the contributions of these two conformal families to the conformal block expansion are
\begin{equation}
\vev{\f_1\f_2\f_2\f_1}=p_{0}\blockdressed_{\D_1+\D_2,\xi_1+\xi_2}+ \bpair{p_{1,0}\blockdressed_{\D_1+\D_2+1,\xi_1+\xi_2}+p_{1,1}\blockdressed_{\D_1+\D_2+1,\xi_1+\xi_2}}+\dots
\end{equation}
in which the block coefficients are
\begin{equation}
p_{0}=1,
\hs{3ex} 
p_{1,0}=\frac{2\left(\Delta_{2} \xi_{1}^{2}+\Delta_{1} \xi_{2}^{2}\right)}{\left(\xi_{1}+\xi_{2}\right)^{2}}, 
\hs{3ex} 
p_{1,1}=\frac{2 \xi_{1} \xi_{2}}{\xi_{1}+\xi_{2}}.
\label{eq:opconstruction}
\end{equation}
By the operator counting technique, there should be a rank-$(n+1)$ boost multiplet in each order of the conformal block expansion. In the following we derive the conformal block expansion from the inversion formula and justify the result of operator construction.

\textbf{Inversion function.} Following the calculation in relativistic CFT, we firstly analytically continue the external weights $(\D_i,\xi_i)$ onto the unitary principal series, then the inversion function is
\begin{align}
\peq I^{\text{u.p.s.}}(\D,\xi)
& =(\CPWdressed^{1221}_{\D,\xi},\vev{\f_1\f_2\f_2\f_1})
\\
& =\intt{\frac{\prod_{i=0}^{4} dx_{i}dy_{i}}{\vol{ISO(2,1)}}}
\vev{\fshadow_1\fshadow_2\opshadow}
\vev{\op\fshadow_2(\xy{3})\fshadow_1(\xy{4})}
\vev{\f_1\f_1(\xy{4})}\vev{\f_2\f_2(\xy{3})}
\nn\\
& =\intt{\frac{\prod_{i=0}^{2} dx_{i}dy_{i}}{\vol{ISO(2,1)}}}
\vev{\fshadow_1\fshadow_2\opshadow}
\vev{\op\f_2\f_1}
\shadow(\op\fshadow_2[\fshadow_1])
\shadow(\op\f_1[\fshadow_2])
\nn
\\
& =
A_{\text{3-pt}}
\shadow(\op\fshadow_2[\fshadow_1])
\shadow(\op\f_1[\fshadow_2]),
\nn
\end{align}
where the exchanged virtual operator $\op_{\D,\xi}$ is located at $(\xy{0})$. In the third line we have used the result of the shadow coefficients \eqref{eq:star2}, and in the last line the remaining three-point pairing is a numerical constant $A_{\text{3-pt}}$ \eqref{eq:threepointpairing}. Then the inversion function is
\begin{align}
c^{\text{u.p.s.}}(\D,\xi)
&=A_{\text{3-pt}}\cN^{-1}(\D,\xi)\shadow(\op\fshadow_2[\fshadow_1])
\shadow(\op\f_1[\fshadow_2])
\shadow(\f_1\f_2[\opshadow])
\nn
\\
&=
\pi F(\xi,\xi_1,\xi_2) \left(\xi-\xi_{1}-\xi_{2}\right)^{-1-\D+\D_{1}+\D_{2}}
\end{align}
where,
\begin{align}
F(\xi,\xi_1,\xi_2)
&=
2^{5-2 \D-2 \D_{1}-2 \D_{2}} 
\xi^{3-2 \D} 
\xi_{1}^{1-2 \D_{1}}
\xi_{2}^{1-2 \D_{2}}
\left(\xi+\xi_{1}+\xi_{2}\right)^{-3+\D+\D_{1}+\D_{2}}
\nn\\
&
\peq \cdot\left(\xi-\xi_{12}\right)^{-1+\D-\D_{12}}
\left(\xi+\xi_{12}\right)^{-1+\D+\D_{12}}.
\end{align}

\begin{figure}[htbp]
\centering

\tikzset{every picture/.style={line width=0.75pt}} %set default line width to 0.75pt        

\begin{tikzpicture}[x=0.75pt,y=0.75pt,yscale=-1,xscale=1]
%uncomment if require: \path (0,539); %set diagram left start at 0, and has height of 539

%Straight Lines [id:da4725378698652545] 
\draw [color={rgb, 255:red, 74; green, 144; blue, 226 }  ,draw opacity=1 ]   (200,52) -- (200,190) ;
\draw [shift={(200,50)}, rotate = 90] [color={rgb, 255:red, 74; green, 144; blue, 226 }  ,draw opacity=1 ][line width=0.75]    (10.93,-3.29) .. controls (6.95,-1.4) and (3.31,-0.3) .. (0,0) .. controls (3.31,0.3) and (6.95,1.4) .. (10.93,3.29)   ;
%Straight Lines [id:da8799919688205711] 
\draw [color={rgb, 255:red, 74; green, 144; blue, 226 }  ,draw opacity=1 ]   (300,190) -- (199.8,190) -- (100,190) ;
%Shape: Circle [id:dp020337296903600866] 
\draw  [color={rgb, 255:red, 74; green, 144; blue, 226 }  ,draw opacity=1 ][fill={rgb, 255:red, 255; green, 255; blue, 255 }  ,fill opacity=1 ] (197.5,190) .. controls (197.5,188.62) and (198.62,187.5) .. (200,187.5) .. controls (201.38,187.5) and (202.5,188.62) .. (202.5,190) .. controls (202.5,191.38) and (201.38,192.5) .. (200,192.5) .. controls (198.62,192.5) and (197.5,191.38) .. (197.5,190) -- cycle ;
%Straight Lines [id:da8891350047234592] 
\draw    (300,190) .. controls (301.67,188.33) and (303.33,188.33) .. (305,190) .. controls (306.67,191.67) and (308.33,191.67) .. (310,190) .. controls (311.67,188.33) and (313.33,188.33) .. (315,190) .. controls (316.67,191.67) and (318.33,191.67) .. (320,190) .. controls (321.67,188.33) and (323.33,188.33) .. (325,190) .. controls (326.67,191.67) and (328.33,191.67) .. (330,190) .. controls (331.67,188.33) and (333.33,188.33) .. (335,190) .. controls (336.67,191.67) and (338.33,191.67) .. (340,190) -- (340,190) ;
\draw [shift={(300,190)}, rotate = 0] [color={rgb, 255:red, 0; green, 0; blue, 0 }  ][fill={rgb, 255:red, 0; green, 0; blue, 0 }  ][line width=0.75]      (0, 0) circle [x radius= 1.34, y radius= 1.34]   ;
%Straight Lines [id:da23876263129335284] 
\draw    (249.9,190) .. controls (252.14,190.74) and (252.89,192.23) .. (252.14,194.47) .. controls (251.39,196.71) and (252.14,198.2) .. (254.38,198.94) .. controls (256.62,199.68) and (257.37,201.17) .. (256.62,203.41) .. controls (255.87,205.65) and (256.62,207.14) .. (258.86,207.88) .. controls (261.1,208.62) and (261.85,210.11) .. (261.1,212.35) .. controls (260.35,214.59) and (261.1,216.08) .. (263.34,216.82) .. controls (265.58,217.56) and (266.33,219.05) .. (265.58,221.29) .. controls (264.83,223.53) and (265.58,225.02) .. (267.82,225.76) .. controls (270.06,226.5) and (270.81,227.99) .. (270.06,230.23) .. controls (269.31,232.47) and (270.06,233.96) .. (272.3,234.7) .. controls (274.54,235.44) and (275.29,236.93) .. (274.54,239.17) .. controls (273.79,241.41) and (274.54,242.9) .. (276.78,243.64) .. controls (279.02,244.38) and (279.77,245.87) .. (279.02,248.11) .. controls (278.27,250.35) and (279.02,251.84) .. (281.26,252.58) .. controls (283.49,253.33) and (284.24,254.83) .. (283.49,257.06) .. controls (282.74,259.3) and (283.49,260.79) .. (285.73,261.53) .. controls (287.97,262.27) and (288.72,263.76) .. (287.97,266) .. controls (287.22,268.24) and (287.97,269.73) .. (290.21,270.47) .. controls (292.45,271.21) and (293.2,272.7) .. (292.45,274.94) .. controls (291.7,277.18) and (292.45,278.67) .. (294.69,279.41) .. controls (296.93,280.15) and (297.68,281.64) .. (296.93,283.88) .. controls (296.18,286.12) and (296.93,287.61) .. (299.17,288.35) -- (300,290) -- (300,290) ;
\draw [shift={(249.9,190)}, rotate = 63.39] [color={rgb, 255:red, 0; green, 0; blue, 0 }  ][fill={rgb, 255:red, 0; green, 0; blue, 0 }  ][line width=0.75]      (0, 0) circle [x radius= 1.34, y radius= 1.34]   ;
%Straight Lines [id:da4962407092402592] 
\draw    (149.9,190) .. controls (147.67,189.25) and (146.92,187.76) .. (147.67,185.53) .. controls (148.42,183.3) and (147.67,181.8) .. (145.44,181.05) .. controls (143.2,180.31) and (142.45,178.82) .. (143.2,176.58) .. controls (143.95,174.35) and (143.2,172.85) .. (140.97,172.1) .. controls (138.74,171.35) and (137.99,169.86) .. (138.74,167.63) .. controls (139.49,165.4) and (138.74,163.91) .. (136.51,163.16) .. controls (134.27,162.41) and (133.52,160.92) .. (134.27,158.68) .. controls (135.02,156.45) and (134.27,154.96) .. (132.04,154.21) .. controls (129.81,153.46) and (129.06,151.96) .. (129.81,149.73) .. controls (130.56,147.5) and (129.81,146.01) .. (127.58,145.26) .. controls (125.34,144.52) and (124.59,143.03) .. (125.34,140.79) .. controls (126.09,138.56) and (125.34,137.06) .. (123.11,136.31) .. controls (120.88,135.56) and (120.13,134.07) .. (120.88,131.84) .. controls (121.63,129.61) and (120.88,128.12) .. (118.65,127.37) .. controls (116.41,126.62) and (115.66,125.13) .. (116.41,122.89) .. controls (117.16,120.66) and (116.41,119.17) .. (114.18,118.42) .. controls (111.95,117.67) and (111.2,116.17) .. (111.95,113.94) .. controls (112.7,111.71) and (111.95,110.22) .. (109.72,109.47) .. controls (107.48,108.73) and (106.73,107.24) .. (107.48,105) .. controls (108.23,102.77) and (107.48,101.27) .. (105.25,100.52) .. controls (103.02,99.77) and (102.27,98.28) .. (103.02,96.05) .. controls (103.77,93.82) and (103.02,92.32) .. (100.79,91.57) -- (100,90) -- (100,90) ;
\draw [shift={(149.9,190)}, rotate = 243.48] [color={rgb, 255:red, 0; green, 0; blue, 0 }  ][fill={rgb, 255:red, 0; green, 0; blue, 0 }  ][line width=0.75]      (0, 0) circle [x radius= 1.34, y radius= 1.34]   ;
%Curve Lines [id:da6626125632716071] 
\draw [color={rgb, 255:red, 155; green, 155; blue, 155 }  ,draw opacity=1 ] [dash pattern={on 4.5pt off 4.5pt}]  (200,114) .. controls (234,154.75) and (270,224) .. (310,194) ;
%Curve Lines [id:da018224577633068062] 
\draw [color={rgb, 255:red, 155; green, 155; blue, 155 }  ,draw opacity=1 ] [dash pattern={on 4.5pt off 4.5pt}]  (200,114) .. controls (234.25,114.38) and (290.25,142.38) .. (310,188) ;
%Straight Lines [id:da243771717015558] 
\draw    (100,190) .. controls (98.33,191.67) and (96.67,191.67) .. (95,190) .. controls (93.33,188.33) and (91.67,188.33) .. (90,190) .. controls (88.33,191.67) and (86.67,191.67) .. (85,190) .. controls (83.33,188.33) and (81.67,188.33) .. (80,190) .. controls (78.33,191.67) and (76.67,191.67) .. (75,190) .. controls (73.33,188.33) and (71.67,188.33) .. (70,190) .. controls (68.33,191.67) and (66.67,191.67) .. (65,190) .. controls (63.33,188.33) and (61.67,188.33) .. (60,190) -- (60,190) ;
\draw [shift={(100,190)}, rotate = 180] [color={rgb, 255:red, 0; green, 0; blue, 0 }  ][fill={rgb, 255:red, 0; green, 0; blue, 0 }  ][line width=0.75]      (0, 0) circle [x radius= 1.34, y radius= 1.34]   ;
%Straight Lines [id:da02066994380143461] 
\draw    (200,190) .. controls (201.67,191.67) and (201.67,193.33) .. (200,195) .. controls (198.33,196.67) and (198.33,198.33) .. (200,200) .. controls (201.67,201.67) and (201.67,203.33) .. (200,205) .. controls (198.33,206.67) and (198.33,208.33) .. (200,210) .. controls (201.67,211.67) and (201.67,213.33) .. (200,215) .. controls (198.33,216.67) and (198.33,218.33) .. (200,220) .. controls (201.67,221.67) and (201.67,223.33) .. (200,225) .. controls (198.33,226.67) and (198.33,228.33) .. (200,230) .. controls (201.67,231.67) and (201.67,233.33) .. (200,235) .. controls (198.33,236.67) and (198.33,238.33) .. (200,240) .. controls (201.67,241.67) and (201.67,243.33) .. (200,245) .. controls (198.33,246.67) and (198.33,248.33) .. (200,250) .. controls (201.67,251.67) and (201.67,253.33) .. (200,255) .. controls (198.33,256.67) and (198.33,258.33) .. (200,260) .. controls (201.67,261.67) and (201.67,263.33) .. (200,265) .. controls (198.33,266.67) and (198.33,268.33) .. (200,270) .. controls (201.67,271.67) and (201.67,273.33) .. (200,275) .. controls (198.33,276.67) and (198.33,278.33) .. (200,280) .. controls (201.67,281.67) and (201.67,283.33) .. (200,285) .. controls (198.33,286.67) and (198.33,288.33) .. (200,290) .. controls (201.67,291.67) and (201.67,293.33) .. (200,295) .. controls (198.33,296.67) and (198.33,298.33) .. (200,300) .. controls (201.67,301.67) and (201.67,303.33) .. (200,305) .. controls (198.33,306.67) and (198.33,308.33) .. (200,310) .. controls (201.67,311.67) and (201.67,313.33) .. (200,315) .. controls (198.33,316.67) and (198.33,318.33) .. (200,320) .. controls (201.67,321.67) and (201.67,323.33) .. (200,325) .. controls (198.33,326.67) and (198.33,328.33) .. (200,330) -- (200,330) ;
\draw [shift={(200,190)}, rotate = 90] [color={rgb, 255:red, 0; green, 0; blue, 0 }  ][fill={rgb, 255:red, 0; green, 0; blue, 0 }  ][line width=0.75]      (0, 0) circle [x radius= 1.34, y radius= 1.34]   ;

% Text Node
\draw (180,58) node  [font=\footnotesize]  {$\Im \xi $};
% Text Node
\draw (359,188) node  [font=\footnotesize]  {$\Re \xi $};
% Text Node
\draw (324,209.5) node  [font=\footnotesize]  {$\xi _{1} +\xi _{2}$};
% Text Node
\draw (242.5,210.5) node  [font=\footnotesize]  {$\xi _{12}$};
% Text Node
\draw (162.5,210.5) node  [font=\footnotesize]  {$\xi _{21}$};
% Text Node
\draw (104,210.5) node  [font=\footnotesize]  {$-\xi _{1} -\xi _{2}$};

\end{tikzpicture}

\caption{The analytic structure of the inversion function of MFT for $\xi_2>\xi_1>0$. There are five branch points at $\xi =\pm \xi_1 \pm \xi_2,0$ drawn as the wavy lines. The physical cut corresponding to the double-trace operators is anchored at $\xi=\xi_1+\xi_2$. The dashed lines are different analytic continuations to the imaginary axis, and from this ambiguity we read off the discontinuity along the cut $\xi=\xi_1+\xi_2$.}
\label{fig:branchcut}
\end{figure}
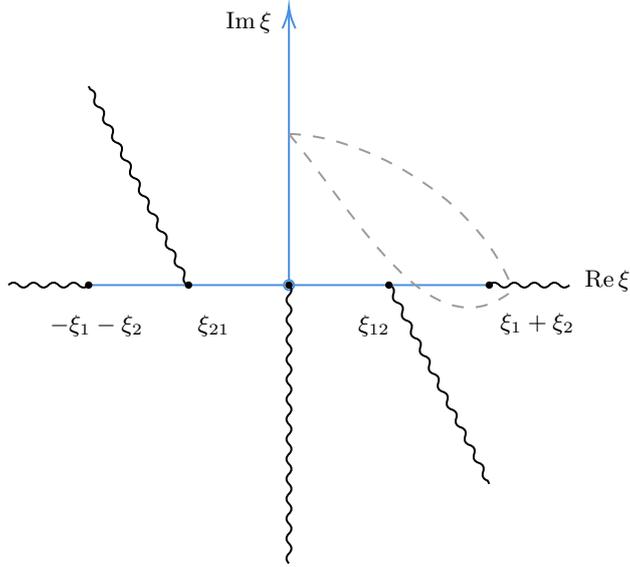

\textbf{Dispersion relation.} Notice that $c^{\text{u.p.s.}}(\D,\xi)$ has no poles of $(\D,\xi)$, and instead there are five branch cuts anchored at $\xi =\pm \xi_1 \pm \xi_2,0$ shown in figure \ref{fig:branchcut}. This phenomenon is expected as explained in \cite{Chen:2020vvn}. In the case of four identical external operators, the physical inversion function $c(\D,\xi)$ has two branch cuts, and $c^{\text{u.p.s.}}(\D,\xi)$ with respect to virtual external operators is the imaginary part of $c(\D,\xi)$ along the physical cut, due to the ambiguity of analytic continuation of weights. The prototypical example of this ambiguity is
\begin{equation}
\intt{dt}\frac{f(t)}{t\pm i\e}=\PV\intt{dt} \frac{f(t)}{t} \mp i\pi f(0).
\end{equation}
The inversion function $c^{\text{u.p.s.}}(\D,\xi)$ plays the role of $-\pi f(0)$ and $c(\D,\xi)$ is multi-valued. For the preceding reason we propose the dispersion relation
\begin{equation}
\disc_{\text{phy. cut}} c(\D,\xi)= \lim_{\e\to 0}\bpair{c(\D,\xi+\e)-c(\D,\xi-\e)}
=
2i\,  c^{\text{u.p.s.}}(\D,\xi),
\label{eq:dispersion}
\end{equation}
and then
\begin{equation}
\disc_{\text{phy. cut}} I(\D,\xi)
=
2i\,  I^{\text{u.p.s.}}(\D,\xi).
\end{equation}
The real part is recovered by the Kramers–Kronig relation\footnote{It is also known as the Hilbert transform, and will be reviewed in appendix \ref{sec:distributions}.},
\begin{equation}
\Re c(\D,\xi)=\frac{1}{\pi}\PV\inttt{d\xi'}{\R} \frac{1}{\xi'-\xi}c^{\text{u.p.s.}}(\D,\xi').
\end{equation}

A natural question is that whether the inversion function can be continued onto the unitary principal series. This requires the external operators of the input four-point functions can be continued according to the scheme $\xi_i\to e^{i\th}\xi_i$ shown in figure \ref{fig:analyticcontinuation1}. In other words we need a family of four-point functions $G_4(\lm)$ depending on $\lm$ analytically. In relativistic CFT this is rare since the unitary theories are expected to be isolated points in the space of CFTs. In GCFT$_2$ we show in the appendix \ref{sec:outer} that the family $G_4(\lm)$ exists at least for real $\lm$, due to the existence of outer-automorphism of the Galilean conformal algebra. We leave this question for further study and return to the discussion on MFT.

\textbf{Inversion formula.} After inserting the inversion function into the inversion formula \eqref{eq:inversionformula}, we enclose the $\xi$-contour along the physical branch cut,
\begin{equation}
\vev{\f_1\f_2\f_2\f_1}_{\cP}
=
\int^{1+i\inf}_{1-i\inf}\frac{d\D}{2\pi i}
\int^{+\inf}_{\xi_1+\xi_2}\frac{d\xi}{2\pi i}\disc c(\D,\xi)\blockdressed_{\D,\xi},
\end{equation}
then using the relation \eqref{eq:dispersion} we get
\begin{align}
\rhs
&=\int^{1+i\inf}_{1-i\inf}\frac{d\D}{2\pi i}
\int^{+\inf}_{\xi_1+\xi_2}\,\frac{d\xi}{\pi}\, c^{\text{u.p.s.}}(\D,\xi)\blockdressed_{\D,\xi}\nn
\\
&=
\int^{1+i\inf}_{1-i\inf}\frac{d\D}{2\pi i}
\int^{+\inf}_{\xi_1+\xi_2}\,d\xi\,\left(\xi-\xi_{1}-\xi_{2}\right)^{-1-\D+\D_{1}+\D_{2}}F(\xi,\xi_1,\xi_2)\blockdressed_{\D,\xi}.
\label{eq:inversion11}
\end{align}

As discussed in the appendix \ref{sec:distributions}, the power factor $\left(\xi-\xi_{1}-\xi_{2}\right)^{-1-\D+\D_{1}+\D_{2}}$ as tempered distribution possesses simple poles at
\begin{equation}
\D_n=\D_{1}+\D_{2}+n, \hspace{3ex} n=0,1,2,\dots .
\label{eq:doubletracepoles}
\end{equation}
The remaining factors are analytic at $\xi=\xi_1+\xi_2$ and the Taylor expansion is
\begin{equation}
F(\xi,\xi_1,\xi_2)\blockdressed_{\D,\xi}=\sum_{n=0}^{\inf} a_n(\D)\left(\xi-\xi_{1}-\xi_{2}\right)^n.
\end{equation}
Then separating the $\xi$-integration of \eqref{eq:inversion11} into two parts\footnote{Choosing different separations does not affect the analytic structure and give the same result. The contributions come from the singularity $\xi=\xi_1+\xi_2$.}, we find that the part on the interval $(\xi_1+\xi_2+1,\inf)$ contributes no $\D$-poles and hence can be dropped when deforming the $\D$-contour,
\begingroup
\allowdisplaybreaks
\begin{align}
\rhs
% &=
% \int^{1+i\inf}_{1-i\inf}\frac{d\D}{2\pi i}
% \int^{+\inf}_{\xi_1+\xi_2}\sum_{n=0}^{\inf} a_n\,d\xi\,\left(\xi-\xi_{1}-\xi_{2}\right)^{-1-\D+\D_{1}+\D_{2}+n}
% \\
&=
\int^{1+i\inf}_{1-i\inf}\frac{d\D}{2\pi i}
\bpair{
\ppair{\int^{+\inf}_{\xi_1+\xi_2+1}d\xi + \int^{\xi_1+\xi_2+1}_{\xi_1+\xi_2}d\xi}
\sum_{n=0}^{\inf} \left(\xi-\xi_{1}-\xi_{2}\right)^{-1-\D+\D_{1}+\D_{2}+n} 
}
\nn
\\
&=
\int^{1+i\inf}_{1-i\inf}\frac{d\D}{2\pi i}
\sum_{n=0}^{\inf}
\int^{\xi_1+\xi_2+1}_{\xi_1+\xi_2}d\xi \left(\xi-\xi_{1}-\xi_{2}\right)^{-1-\D+\D_{1}+\D_{2}+n}a_n(\D)
\nn
\\
&=
-\int^{1+i\inf}_{1-i \inf}\frac{d\D}{2\pi i}
\sum_{n=0}^{\inf}\frac{a_n(\D)}{\D-\D_{1}-\D_{2}-n}
\nn
\\
&=
\sum_{n=0}^{\inf}a_n(\D_{1}+\D_{2}+n),
\label{eq:doubletrace1}
\end{align}
\endgroup
where in the last line we have deformed the $\D$-contour and pick up the double-trace poles \eqref{eq:doubletracepoles}, and the minus sign is canceled due to the clock-wise order of the contour.

Each term in the summation \eqref{eq:doubletrace1} fits into a conformal block of rank-$(n+1)$ boost multiplet with double-trace weight $(\D,\xi)=(\D_1+\D_2+n,\xi_1+\xi_1)$
\begin{equation}
a_n(\D_{1}+\D_{2}+n)=\sum_{a=0}^{n}p_{n,a}\p_{\xi}^{a}\blockdressed_{\D_{1}+\D_{2}+n,\xi_1+\xi_2},
\end{equation}
where the block coefficients are
\begin{equation}
\label{eq:block}
p_{n,a}=\frac{1}{n!}\binom{n}{a}\p_{\xi}^{n-a}F(\xi_{1}+\xi_{2},\xi_{1},\xi_{2})\eval_{\D=\D_1+\D_2+n},
\end{equation}
and they match with \eqref{eq:opconstruction}.

In summary we have the conformal block expansion of the projected four-point function
\begin{equation}
\label{eq:projectedfourpoint}
\vev{\f_1\f_2\f_2\f_1}_{\cP}
=\sum_{n=0}^{\inf}\sum_{a=0}^{n}p_{n,a}\p_{\xi}^{a}\blockdressed_{\D_{1}+\D_{2}+n,\xi_1+\xi_2},
\end{equation}
with the block coefficients \eqref{eq:block}. 
In the crossing region $(0,1)\times\RR$, the projected four-point function \eqref{eq:projectedfourpoint} converges rapidly and can be re-expanded as a double Taylor series of $(x,k)$ in the \schannel limit. The expansion matches with that of the four-point function \eqref{eq:fourpoint1} order by order. Hence \eqref{eq:projectedfourpoint} should equal to \eqref{eq:fourpoint1} in the crossing region. In this way we derive the conformal block expansion of the four-point function and confirm the validity of the inversion formula.

% We check that the partial sum of $n\leq 5$ divided by the four-point function \eqref{eq:fourpoint1} is $1+O(x^6)$, hence $\vev{\f_1\f_2\f_2\f_1}_{\cP}$ should equal to the four-point function in the region $(0,1)\times \R$, confirming the validity of the inversion formula.

\subsection{Bilocal actions of mean field theory}\label{sec:bilocalsinglet}

There are few GCFT$_2$ models with concrete actions. One of them that has been thoroughly discussed in
the literature is the BMS free scalar model \cite{Hao:2021urq} with the action on the plane
\begin{equation}
S=\half \intt{dxdy}\f \p_{y}^2 \f.
\label{eq:actionBMSfreescalar}
\end{equation}
As a worldsheet theory, this model describes the tensionless limit of the free bosonic string, see e.g. \cite{Schild:1976vq,Isberg:1993av,Bagchi:2015nca} and the ambitwistor strings \cite{Mason:2013sva,Casali:2016atr,Casali:2017zkz}. This action can also be realized as a $\sqrt{T\bar{T}}$ deformation of the relativistic free scalar \cite{Rodriguez:2021tcz}. The higher dimensional Carrollian analog of \eqref{eq:actionBMSfreescalar} was discussed in \cite{Bagchi:2019xfx,Chen:2021xkw,Bagchi:2022emh}.

In this subsection we explore the Lagrangian description of Galilean MFT. Inspired by the MFT in AdS/CFT \cite{Witten:1998qj} and the long-range models in statistical physics, see e.g. \cite{Rajabpour:2011qr}, we find a series of bilocal actions labeled by $(\D,\xi)$ corresponding to the Galilean MFT. Moreover at the special value $\xi=0$ we get additional actions labeled by $(\D_1,\D_2)$, one of which gives the BMS free scalar \eqref{eq:actionBMSfreescalar}. Starting from the ansatz of bilocal action
\begin{equation}
S=\intt{dx_1dy_1dx_2dy_2}\f_1(\xy{1})K(x_{12},y_{12})\f_2(\xy{2})
\label{eq:actionbilocal1}
\end{equation}
with two scalars transforming as $\f_i\in \rep{\D_i,\xi_i}$, and imposing the Galilean conformal invariance as in table \ref{table:global_gca} on the action, we get four equations of $K$, which are related to the two-point Ward identities \eqref{eq:wardsinglet} by the shadow replacement $(\D_1,\xi_1,\D_2,\xi_2)\to(2-\D_1,-\xi_1,2-\D_2,-\xi_2)$.
% \begin{align}
% \label{eq:bilocal}
% & \ppair{x_{12}\p_{y_1}+\xi_{1}+\xi_{2}}K(\xy{12})=0,\\
% & \ppair{x_{12}\p_{x_1}+y_{12}\p_{y_1}+4-\D_1-\D_2}K(\xy{12})=0,\nn\\
% & \bpair{(x_{1}^{2}-x_{2}^{2})\p_{y_1} + 2\xi_1 x_1 + 2\xi_2 x_2}K(\xy{12})=0,\nn\\
% & \bpair{(x_{1}^{2}-x_{2}^{2})\p_{x_1} + 2(x_{1}y_{1}-x_{2}y_{2})\p_{y_1} +2(2x_1+2x_2-\D_1 x_1 -\D_2 x_2 + \xi_1 y_1 +\xi_2 y_2 )}K(\xy{12})=0.\nn
% \end{align}

Mathematically speaking, the discussion in section \ref{sec:derivationshadow} is on the intertwining maps, and the action \eqref{eq:actionbilocal1} is an intertwining bilinear form introduced by Bruhat \cite{bruhat1956representations}, see also chapter 3 of \cite{gel2016generalized}. Intuitively, the finite dimensional analogs for two representations $\cV_i\ni V_i$ are
\begin{equation}
    V_{1}^{a}=K^{a}_{b}V_{2}^{b},\qquad (V_1,V_2)_{K}=K_{ab}V_{1}^{a}V_{2}^{b}.
\end{equation}
For unitary representations the two concepts are essentially the same. We can use the positive-definite inner product $(V_1,U_1)=g_{ab} V_{1}^{a} U_{1}^{b}$ to raise and lower the indices, then the intertwining map and bilinear form are related by $K_{ac}=g_{ab} K^{b}_{c}$. For non-unitary representations they are not necessarily equivalent.

\textbf{Bilocal actions.} By solving the differential equations of $K$, we can read the bilocal actions. We find the following distributional solutions for $K$ in two different cases.

Case 1: $\xi_1+\xi_2\neq 0$. The last two equations in \eqref{eq:wardsinglet} force the scalars to be identical, $\xi_1=\xi_2,\,\D_1=\D_2$, and the solutions are exactly the kernel of the shadow transform \eqref{eq:shadowgcft2}, 
\begin{equation}
K(x_{12},y_{12})=e^{-2\xi_1\frac{y_{12}}{x_{12}}}\bpair{c_1|x_{12}|^{2\D_1-4}+c_2 \sign(x_{12})|x_{12}|^{2\D_1-4}},
\label{eq:actiontypical}
\end{equation}
where the first one corresponds to bosonic statistics and the second to fermionic statistics. At special values $\xi_1=0,\, \D_1=\frac{3-n}{2},\,n\in \Zgeq$, one of the power-type distributions should be regularized to the $n$-th derivative of the $\d$-distribution $\d^{(n)}(x_{12})$, and the action is local with respect to $x$.

Case 2: $\xi_1+\xi_2= 0$. Besides the previous power-law solutions, the first two equations in \eqref{eq:wardsinglet} admit an additional solution
\begin{equation}
K(x_{12},y_{12})= \d(x_{12})\bpair{d_1|y_{12}|^{\D_1+\D_2-3}+d_2 \sign(y_{12})|y_{12}|^{\D_1+\D_2-3}}.
\label{eq:actionanomalous}
\end{equation}
The third equation gives no constraint and the last equation sets $\xi_1=\xi_2=0$. Notice that in this case the scalars are not compulsory to be equal. At special values $\D_1+\D_2=2-n,\,n\in \Zgeq$, one of the power-type distributions in the additional solutions \eqref{eq:actionanomalous} should be regularized to the $n$-th derivative of the $\d$-distribution $\d^{(n)}(y_{12})$, and the action turns out to be local and BMS-invariant,
\begin{equation}
S_n=\intt{dxdy}\f_1(\xy)\p_{y}^{n}\f_2(\xy).
\end{equation}

For example, in the case $\xi_1=\xi_2=0,\, \D_1+\D_2=0$ there are totally four actions as follows,
\begin{align}
S& =\intt{dx_1dy_1dx_2dy_2}\f(\xy{1})\f(\xy{2})x_{12}^{-4},
\label{eq:actionbad1}
\\
S& =\intt{dxdy_1dy_2}\f(x,y_1)\p_{x}^{3}\f(x,y_2),
\label{eq:actionbad2}
\\
S& =\intt{dxdy_1dy_2}\f_1(x,y_1)\f_2(x,y_2)y_{12}^{-3},
\\
S& =\intt{dxdy}\f_1(x,y)\p_{y}^{2}\f_2(x,y),
\end{align}
and when the two fields are identical, the last one is the BMS free scalar model \eqref{eq:actionBMSfreescalar}. Another example is that if $\f_1=\f_2=\psi,\, \D_1=\half,\, \xi_1=\xi_2=0$ and setting $\psi$ to be Grassmann-valued, we get the fermionic part of the homogeneous tensionless superstring \cite{Bagchi:2017cte},
\begin{equation}
S=\intt{dxdy}\psi(\xy)\p_{y}\psi(\xy).
\end{equation}

\textbf{Path integral quantization.} These actions \eqref{eq:actiontypical} and \eqref{eq:actionanomalous} are free and the correlation functions can be evaluated from the path integral. Assuming the two scalars are identically bosonic and normalizing the action by a factor $\half$, the partition function with the source is
\begin{equation}
Z[J]=\intt{D\f}e^{-S+\intt{dxdy}J\f}.
\end{equation}
Substituting $\f(\xy{1})$ with $ \f(\xy{1})+\intt{dx_2dy_2}K^{-1}(x_{12},y_{12})J(\xy{2})$, where $K^{-1}$ is the inverse of $K$, the partition function is solved as
\begin{equation}
Z[J]=Z[0]\intt{D\f}e^{-\half \intt{dx_1dy_1dx_2dy_2}J(\xy{1})K^{-1}(x_{12},y_{12})J(\xy{2})},
\end{equation}
then the two-point function is $\vev{\f\f}=-K^{-1}$ and the higher-point functions are the Wick contractions of $\vev{\f\f}$. 

For the first series of actions \eqref{eq:actiontypical} and $\xi\in\Rnot$, we need to start from the imaginary Carrollian or Galilean time at the beginning as in section \ref{sec:derivationshadow}, and the integral equation of the inverse kernel is
\begin{equation}
\int dx_0dy_0\,e^{-2\xi\frac{y_{10}}{x_{10}}}|x_{10}|^{2\D-4}K^{-1}(x_{02},y_{02})=\d(x_{12},y_{12}).
\end{equation}
Using \eqref{eq:doubleshadow} the two-point function is
\begin{equation}
\vev{\f(\xy{1})\f(\xy{2})}=c\, e^{2\xi\frac{y_{12}}{x_{12}}}|x_{12}|^{-2\D},
\end{equation}
where $c$ is an unimportant constant, and we get back to the scalar MFT in section \ref{sec:MFT1}.

Notice that when $\xi=0$ the first series of actions \eqref{eq:actiontypical},
\begin{equation}
S=\intt{dx_1dx_2 dy_1dy_2} \f(\xy{1})|x_{12}|^{2\D-4}\f(\xy{2}),
\end{equation}
possess gauge redundancies. Any field configuration $\f(x,y)$ satisfying
\begin{equation}
\intt{dy}\f(x,y)=:\Phi(x),
\end{equation}
gives the same contribution $e^{-\intt{dx_1dx_2}\Phi(x_1)|x_{12}|^{2\D-4}\Phi(x_2)}$ to the path integral. In other words, the bilocal kernel $K(x_{12},y_{12})=|x_{12}|^{2\D-4}$ is highly degenerate and has no inverse unless we first mod out the zero-modes, and the two-point function should satisfy
\begin{equation}
\intt{dx_0}|x_{10}|^{2\D-4}\vev{\Phi(x_0)\Phi(x_2)}=-\d(x_{12}),
\end{equation}
instead of the naive one,
\begin{equation}
\intt{dx_0 dy_0}|x_{10}|^{2\D-4}\vev{\f(\xy{0})\f(\xy{2})}=-\d(x_{12})\d(y_{12}).
\end{equation}
Therefore the theory is equivalent to the MFT in CFT$_1$. The action is 
\begin{equation}
S=\intt{dx_1dx_2}\Phi(x_1)|x_{12}|^{2\D-4}\Phi(x_2),
\end{equation}
with $\D_{\Phi}=\D-1$ and the two-point function is
\begin{equation}
\vev{\Phi(x_1)\Phi(x_2)}=c\, |x_{12}|^{2-2\D}.
\end{equation}

The second series of actions \eqref{eq:actionanomalous} are local with respect to $x$,
\begin{equation}
S=\intt{dx dy_1dy_2} \f(x,y_1)|y_{12}|^{2\D-3}\f(x,y_2).
\end{equation}
The two-point function satisfies the following integral equation,
\begin{equation}
\intt{dy_0}|y_{01}|^{2\D-3}\vev{\f(x_1,y_0)\f(\xy{2})}=-\d(x_{12})\d(y_{12}).
\end{equation}
and the solution is,
\begin{equation}
\vev{\f(\xy{1})\f(\xy{2})}=c\, \d(x_{12})|y_{12}|^{1-2\D}.
\end{equation}
Notice that if using the path integral method to quantize the BMS free scalar model \eqref{eq:actionBMSfreescalar}, we will get the correlation functions with respect to the trivial vacuum instead of the highest weight vacuum \cite{Hao:2021urq}.

\subsection{Localization of shadow integrals}

In this subsection we discuss the localization of the shadow integrals. In the previous sections we notice that the integrals in the shadow formalism are localized to Dirac $\d$-distributions. The reason is similar to that of ambitwistor strings \cite{Mason:2013sva,Casali:2016atr,Casali:2017zkz}. We call the involved integrals as ``shadow integrals" with the following form
\begin{align}
\cA(x_{e},y_{e},\D_{n},\xi_{n})
& =\intt{\prod_{i\in I}{dx_i dy_i}}\prod\vev{\op\op}\vev{\op\op\op}\cdots\nn\\
& =\intt{\prod_{i\in I}{dx_i dy_i}}F_x(x_{n},\D_{n}) e^{F_y(x_{n},y_{n},\xi_{n})},
\label{eq:shadowIntegral}
\end{align}
where all the operators are virtual singlets $\op_n\in \urep{\D_n,\xi_n}$, and $F_x$ is the collection of power functions and $F_y$ is the collection of exponential factors. Borrowing the terminology of Feynman diagram, we label the integrated positions by $i\in I$ and call them ``internal", $(x_{i},y_{i}),\,i\in I$, and label the remaining positions by $e\in E$ and call them ``external", $(x_{e},y_{e}),\,e\in E$. Since the exponential factors are bilinear with respect to $(y_n,\xi_n)$, we have
\begin{equation}
F_y=i A_0(x_n,y_e,r_n)+i\sum_{i\in I}y_i A_{i}(x_n,r_n),
\end{equation}
and the $y_i$-integration gives a $\d$-distribution localized on (the real points of) the algebraic variety
\begin{equation}
\cV[\cA]=\set{A_{i}(x_n,r_n)=\sum_{a\neq b}\frac{f_{ab}(r_n)}{x_{ab}}=0:i\in I}\subset \R^{2 |N|}\backslash\set{x_{ab}\neq 0:a,b\in N,a\neq b}
\label{eq:shadowVariety}
\end{equation}
where $f_{ab}$'s are linear functions of $r_n$. Then the integral \eqref{eq:shadowIntegral} simplifies to
\begin{equation}
\cA(x_{e},y_{e},\D_{n},\xi_{n})=(2\pi)^{|I|}\intt{\prod_{i\in I} dx_i \d(A_i)}F_x(x_{n},\D_{n}) e^{i A_0(x_n,y_e,r_n)}.
\label{eq:umbraIntegral}
\end{equation}

For convenience we refer to the variety $\cV:=\cV[\cA]$ associated with the shadow integral $\cA$ as ``shadow variety". The shadow variety controls the behaviour of the shadow integral. We firstly summarize the properties of the shadow varieties appearing in the previous sections, and then briefly discuss the higher-point conformal blocks.
\begin{itemize}
\item Four-point conformal partial waves and blocks: $\cV$ can be solved as the equations of $x_{i},\,i\in I$, and each root corresponds to a block in the partial wave. The shadow variety $\cV$ may not have real points and then the integral vanishes. 
\begin{alignat}{2}
&\cA:\hspace{3ex}
&&
\intt{dx_0 dy_0} 
\vev{\op_1\op_2 \op_0} 
\vev{\opshadow_0\op_3\op_4},
\\
&\cV:\hspace{3ex} 
&&\frac{r_{01,2}}{x_{01}}+\frac{r_{02,1}}{x_{02}}-\frac{r_{04,3}}{x_{03}}-\frac{r_{03,4}}{x_{04}}=0.
\end{alignat}
\item Bubble diagram: $\cV$ has two irreducible components, and the dimension of the intersection between $\cV$ with $\set{r_i=\text{const}:i\in I}$ is not zero, then there are nontrivial integrals of $x_i$ survived in \eqref{eq:umbraIntegral}.
\begin{alignat}{2}
&\cA:\hspace{3ex}
&&
\intt{\dxy{1}\dxy{2}} 
\vev{\op_1\op_2 \op_3} 
\vev{\opshadow_1\opshadow_2 \opshadow_4},
\\
&\cV:\hspace{3ex} 
&& \{\frac{r_{34}}{x_{12}}-\frac{r_{13,2}}{x_{13}}+\frac{r_{14,2}}{x_{14}}=0,\, \frac{r_{34}}{x_{12}}+\frac{r_{23,1}}{x_{23}}-\frac{r_{24,1}}{x_{24}}=0 \}
\end{alignat}
\item OPE block: the factor $F_x$ in the shadow integral depends on $y_{ab}$, then the remaining integral \eqref{eq:umbraIntegral} contains derivatives of $\d$-distribution on $\cV$.
\begin{alignat}{2}
&\cA:\hspace{3ex}
&&
\intt{\dxy{0}} 
\vev{\op_1\op_2 \opshadow_0} \op_{0},
\\
&\cV:\hspace{3ex} 
&& \frac{r_{01,2}}{x_{02}}+\frac{r_{02,1}}{x_{01}}=0.
\end{alignat}
\end{itemize}

\textbf{Higher-point conformal blocks.}
We briefly discuss the calculation of five-point conformal blocks via the shadow formalism. For the five-point there is only one type of OPE, named as the comb channel. The conformal partial wave with respect to exchanged singlets $\op_{a},\, \op_{b}$ can be constructed as
\begin{equation}
\CPWdressed_{a,b}(\xy{i})
=\intt{\dxy{a}\dxy{b}}
\vev{\op_1\op_2\op_{a}}
\vev{\opshadow_{a}\op_{3}\op_{b}}
\vev{\opshadow_{b}\op_{4}\op_{5}},
\label{eq:fivepoint}
\end{equation}
and the shadow variety is 
\begin{equation}
\cV=\{
\frac{r_{a1,2}}{x_{a1}}+\frac{r_{a2,1}}{x_{a2}}-\frac{r_{ab,3}}{x_{a3}}-\frac{r_{a3,b}}{x_{ab}}=0,\, 
\frac{r_{b4,5}}{x_{b5}}+\frac{r_{b5,4}}{x_{b4}}-\frac{r_{ab,3}}{x_{b3}}-\frac{r_{a3,b}}{x_{ab}}=0
\}.
\end{equation}
The conformal partial wave contains four conformal blocks due to the shadow symmetry, implying that $\cV$ as equations of $x_{a},\, x_{b}$ should have four roots, each of which contributes to a conformal block. After eliminating $x_{b}$ we indeed get a fourth order equation of $x_{a}$. However since the power factors are of the form $\prod_{i=0}^{5}|F_{i}|^{\D_{i}}$, after inserting the solutions of $x_{a},\, x_{b}$, the result is unlikely to be simplified unless $\D_{i}\in \Z$. We leave this possibility for further study.

For six-point conformal partial waves, there are two types of OPE, named as the comb channel and the snowflake channel. In both cases the shadow varieties $\cV$ are defined by three equations. As equations of $x_{i},\, i\in I$, we check numerically that $\cV$ has eight different roots, as expected from the shadow symmetry.

\section{Conclusion and Discussions}
In this work, we studied the shadow formalism for two-dimensional Galilean conformal field theory. As $2d$ Galilean conformal group is isomorphic to $3d$ Poincare group, we are allowed to use the Wigner-Mackey classification of $3d$ Poincare group to identify the unitary principal series representations and then to construct the shadow transform for GCFT$_2$. Using the shadow transform, we computed the OPE blocks, and discussed the Clebsch-Gordan kernels and the shadow coefficients. Moreover, we studied the conformal blocks and conformal partial waves in the framework of shadow formalism. 

Furthermore we investigated several applications of shadow formalism, including the revisit of the decomposition of four-point function in Galilean MFT and the construction of a series of bilocal actions of Galilean MFT. In the revisit of the four-point function in MFT, we proposed a new inversion formula, due to different form of conformal partial waves from the one in \cite{Chen:2020vvn}. The resulting inversion function should be treated carefully and led to the correct conformal block expansion of four-point function. In constructing the bilocal actions of Galilean MFT, we used the intertwining bilinear forms, which obey the Ward identities. In special cases, these actions reduce to those of the BMS free scalar and the homogeneous tensionless fermionic string.

In our study, we came across a kind of integrals, which we called shadow integrals. The shadow integral can be reduced to the integral over an algebraic variety, due to the localization in the $y_i$ integration. This remarkable property makes analytic bootstrap in GCFT$_2$ feasible.

There are several future directions:

\textbf{More on shadow formalism.} In this work we mainly focused on the shadow formalism related to singlet representations, and the exchanged boost multiplets were dealt separately by the method of Casimir equations. Firstly it would be interesting to extend the current approach to boost multiplets in GCFT$_2$ and similarly to logarithmic multiplets in LogCFT. This requires a way to bypass the unitarity of the ``Euclidean" shadow transform. Secondly, when $\xi=0$ operators are involved in the shadow formalism there are technical difficulties to be settled. For example, the $\xi=0$ shadow transform is not unique due to the existence of different types of solutions of the Ward identities. Besides the shadow formalism, the $\xi=0$ subsector are found related to the celestial CFT recently \cite{Bagchi:2022emh}, and one may try to explore the role of the $\xi\neq 0$ subsector played in the celestial holography.

\textbf{GCFT$_2$ with/as defect.} In \cite{Setare:2011hc}, the lower-point correlation functions of boundary GCFT$_2$ were determined. Interestingly there are two types of boundaries with different residual symmetries, and in result the coincidence of $2d$ Galilean and Carrollian conformal symmetries can be distinguished. One could adopt the shadow formalism to the analytic studies of boundary and crosscap GCFT$_2$, and even interfaces between Galilean, Carrollian and relativistic CFTs. On the other side, the GCFT$_2$ itself can be regarded as the null defect in Lorentzian CFT$_3$, which deserves to be explored further.

\textbf{Deformations of MFT.} Constructing interacting GCFT$_2$ is an important task, and the numerical Galilean conformal bootstrap is obstructed by the lack of unitarity in GCFT$_2$. 
The problem is that the highest weight representations contain negative-norm states except for the $\xi=0$ singlets, and if the spectrum contains only $\xi=0$ singlets, the theory reduces to a CFT$_1$. Hence a prior there is no positivity constraint on the bootstrap equations. 
The unitary representations in the Wigner classification do not correspond to local operators, but are more interesting and can be related to the celestial/Carrollian correspondence, since the analogs in CFTs - the principal series representations have already appeared in the discussions of dS bootstrap and celestial CFTs.
In many aspects, the GCFT$_2$ in this paper is more similar to the logarithmic CFT. It would be interesting to add relevant interactions to the Galilean MFT and to investigate whether there are nontrivial fixed points, which will further serve as the target of the numerical Galilean conformal bootstrap.

\section*{Acknowledgments}
We would like to thank Peng-Xiang Hao and Zhe-Fei Yu for the participation in the early stage of this project. We are grateful to P. Hao, J. Qiao, Z. Yu, Y. Zheng for many valuable discussions. 
We would like to thank L. Apolo, B. Czech, C. Chang, J. Lu, W. Song, J. Wu, G. Yang, W. Yang, X. Zhou for their stimulating questions in the workshops and seminars. The work is in part supported by NSFC Grant No. 11735001.

\vspace{1cm}
\appendix
\renewcommand{\appendixname}{Appendix~\Alph{section}}

\section{Conventions and Notations}\label{sec:app1}

\textbf{Notations in CFT$_1$ and GCFT$_2$:}
\begin{alignat}{2}
    x_{12}& =x_1-x_2, 
    & \hs
    k_{12}& =\frac{y_{12}}{x_{12}},
    \\
    \D_{12}& =\D_1-\D_2, 
    &  
    \xi_{12}& =\xi_1-\xi_2,
    \\
    x_{12,3}& =\frac{x_{13}x_{23}}{x_{12}},   
    & 
    k_{12,3}& =\frac{y_{13}}{x_{13}}+\frac{y_{23}}{x_{23}}-\frac{y_{12}}{x_{12}},
    \\
    \D_{12,3}& =\D_1+\D_2-\D_3, & \hs \xi_{12,3}& =\xi_1+\xi_2-\xi_3,\\
    x_{12,34}& =x=\frac{x_{12}x_{34}}{x_{13}x_{24}},  & \hs 
    k_{12,34}& =\frac{y}{x}=\frac{y_{12}}{x_{12}}+\frac{y_{34}}{x_{34}}-\frac{y_{13}}{x_{13}}-\frac{y_{24}}{x_{24}},
    \\
    R_{i}& =\frac{\xi_{i}}{\xi_{0}},
    & 
    R_{ij}& =\frac{\xi_{i}-\xi_{j}}{\xi_{0}}.
\end{alignat}

\textbf{Four-point configuration.} The permutation group $S_4$ acts transitively on the cross ratios of four points $(\xy{i}),\,i=1,2,3,4$, and the stabilizer subgroup is the Klein four-group
\begin{equation}
V_4=\set{(1),(12)(34),(13)(24),(14)(23)}\simeq \Z_2\times \Z_2,
\end{equation}
that is, $g\cdot x_{12,34}=x_{12,34},\,g\cdot k_{12,34}=k_{12,34},\,\forall g\in V_4$. And effectively the action by $S_4/V_4 \simeq S_3$ is shown in the table \ref{tab:fourpointcrossratio}.
\begin{table}[htbp]
    \centering
    \tablevspace{1.6}
    \begin{tabular}{l|CCCCCC}
        $S_3$ & (1) & (12)& (13) & (23) & (123) & (132)  \\
        \hline
        action & \frac{12\cdot 34}{13\cdot 24} & \frac{21\cdot 34}{14\cdot 23}& \frac{14\cdot 23}{13\cdot 24}& \frac{13\cdot 24}{12\cdot 34}& \frac{14\cdot 23}{21\cdot 34}& \frac{13\cdot 24}{14\cdot 23}\\
        channel& s& & t& u& & \\
        $x$ & x   & -\frac{x}{1-x}& 1-x& \frac{1}{x}& -\frac{1-x}{x}& \frac{1}{1-x} \\
        $k$ & \frac{y}{x} & \frac{y}{x(1-x)} & -\frac{y}{1-x} & -\frac{y}{x} & -\frac{y}{x(1-x)} & \frac{y}{1-x}\\
        $y$ & y& -\frac{y}{(1-x)^2}& -y & -\frac{y}{x^2} & \frac{y}{x^2}& \frac{y}{(1-x)^2}\\
    \end{tabular}
    \caption{The four-point cross ratios. The last three rows of cross ratios can be directly obtained by setting the four points to the conformal frame: $\set{0,x,1,\inf}$ in CFT$_1$ and $\set{(0,0),(x,y),(1,0),(\inf,0)}$ in GCFT$_2$.}
    \label{tab:fourpointcrossratio}
\end{table}

\textbf{$3d$ Lorentzian/Euclidean conformal algebra.} In $\R^{2,1}$ or $\R^{3}$ we choose the coordinates $(x^{0},x^{1},x^{2})$ and signatures $(\pm 1,1,1)$. The generators of Lorentzian conformal algebra $\solie(3,2)$ or its Euclidean partner $\solie(4,1)$ are $M_{ab},D,P_{a},K_{a}$, $a=0,1,2$ and the corresponding vector fields are
\begin{equation}
p_{a}=\p_{a}, \hspace{3ex} 
d=x^{a}\p_{a}, \hspace{3ex} 
k_{a}=2x_{a}x^{b}\p_{b}-x^2\p_{a}, \hspace{3ex} m_{ab}=-x_{a}\p_{b}+x_{b}\p_{a}.
\end{equation}
The commutation relations are,
\begin{align}
& [M_{ab},M_{cd}]=g_{ad}M_{bc}+g_{bc}M_{ad}-g_{ac}M_{bd}-g_{bd}M_{ac},\nn\\
& [M_{ab},P_{c}]=-g_{ac}P_{b}+g_{bc}P_{a}, \hs [M_{ab},K_{c}]=-g_{ac}K_{b}+g_{bc}K_{a},\nn\\
& [D,P_{a}]=P_{a},\hs
 [D,K_{a}]=-K_{a},\hs
 [K_{a},P_{b}]=2g_{ab}D-2M_{ab},
\end{align}
and the default conjugation relation of the generators is anti-Hermitian, $Q^{\dagger}=-Q$. The BPZ conjugation $D^{\dagger}=D,\,P^{\dagger}=K$ can be derived by switching between the N-S quantization and the radial quantization.

\subsection{Regularization of tempered distributions}\label{sec:distributions}

In the main text the calculations are undertaken in the framework of tempered distributions, and we have come across the regularization of power-type distributions. In this subsection we provide a mild introduction to the regularization and normalization of distributions by examples, following \cite{gel2014properties}. For simplicity the functions and distributions are on $\R$.

\textbf{Regularization.} A tempered distribution $\f\in \cS'(\R)$ acting on the rapidly decreasing test function $f\in \cS(\R)$ can be formally written as an integral $\f(f)=\intt{d x}\f(x) f(x)$ with kernel $\f(x)$. One can imagine $f$ as a Gaussian wave-packet and $\f$ as some sharp observable. When the kernel $\f(x)$ is a function with singularities, e.g. $\frac{1}{x-y}$, the integral is convergent only for a small class $V$ of test functions $f\in V\subset \cS(\R)$. Then subtracting off all the divergent terms of the integral means extending the domain of $\f$ from $V$ to $\cS(\R)$. This procedure is called regularization of distributions. 

For example, the regularization of $\f(x)=\frac{1}{x-y}$ can be chosen as 
\begin{equation}
\PV\intt{dx}\frac{f(x)}{x-y}=\lim_{\e\to 0}\inttt{dx}{|x-y|>\e}\frac{f(x)}{x-y}.
\end{equation}
This is the Hilbert transform, also the shadow transform of $\D=\half$ in CFT$_1$ as reviewed in the appendix \ref{sec:cft1shadow}. It is a unitary operator on $L^2(\R)$\footnote{The Hilbert space $L^2(\R)$ is canonically embedded into $\cS'(\R)$: $\cS(\R)\to L^2(\R)\to \cS'(\R)$, i.e. an example of rigged Hilbert space.} with two eigenspaces $H^{2}_{+}(\R)\oplus H^{2}_{-}(\R)$, and as a result the (fermionic) principal series representation with $\D=\half$ of $SL(2,\R)$ is reducible.

The extension is usually not unique. We are interested in the case that a family of distributions $\f_{a}(x)$ depends on parameter $a$ analytically. Then the analyticity of $a$ helps us pick out a unique regularization of $\f_{a}(x)$. For example, the power-type distribution 
\begin{equation}
(x_{+}^{a},f(x))=\inttt{dx}{(0,\inf)}x^a f(x),
\end{equation}
is convergent if $\Re a> -1$ and hence is analytic with respect to $a$. If $\Re a\leq -1$ the integral is divergent and acquires regularization at $x=0$. Inserting a real-analytic test function $f(x)=\sum_{n=0}^{\inf} \frac{f^{(n)}(0)}{n!} x^n$ and interchanging the order of integration and summation, we have 
\begin{align}
(x_{+}^{a},f(x))
& =\inttt{dx}{(1,\inf)}x^a f(x) + \sum_{n=0}^{\inf} \frac{f^{(n)}(0)}{n!}\inttt{dx}{(0,1)}x^{a+n}\nn\\
& =\inttt{dx}{(1,\inf)}x^a f(x) + \sum_{n=0}^{\inf} \frac{f^{(n)}(0)}{n!}\frac{1}{a+n+1},
\end{align}
The first term is well-controlled due to the rapid decay of $f(x)$ and is irrelevant to our discussion. The second term implies that $(x_{+}^{a},f(x))$ as a function of $a$ is meromorphic in $\C$, with simple poles at $a=-1,-2,\dots$ and residues $\Res_{a=-n} (x_{+}^{a},f(x)) = \frac{f^{(n-1)}(0)}{(n-1)!}$. Then stripping off the test function we find $x_{+}^{a}$ is meromorphic with respect to $a\in \C$, with simple poles at $a=-1,-2,\dots $ and residues,
\begin{equation} 
\Res_{a=-n} x_{+}^{a} = \frac{(-1)^{n-1}}{(n-1)!}\d^{(n-1)}(0).
\end{equation}
Similarly $|x|^{a},\, |x|^{a}\sign(x)$ and $x_{-}^{a}=\th(-x)|x|^{a}$ are all meromorphic with respect to $a\in \C$. We summarize their analytic structure in the table \ref{tab:distributions2}.

\begin{table}[htpb]
    \centering
    \renewcommand{\arraystretch}{1.6}
    \begin{tabular}{L| L| L L}
    \text{distributions}     & \text{poles}       & \text{residues} & \text{at}\\
    \hline
    x_{+}^{a}     & -1,-2,-3,\dots         & \frac{(-1)^{n-1}}{(n-1)!}\d^{(n-1)}(0)& a=-n\\

    x_{-}^{a}   & -1,-2,-3,\dots          &\frac{1}{(n-1)!}\d^{(n-1)}(0) & a=-n\\

    |x|^{a}       & -1,-3,-5,\dots       & \frac{2}{(2n)!}\d^{(2n)}(0) & a=-2n-1\\
    
    |x|^{a}\sign(x) & -2,-4,-6\dots     & \frac{-2}{(2n-1)!}\d^{(2n-1)}(0) & a=-2n\\
    \hline
    \hline
    \text{normalized ver.} & \text{removable poles}                     & \text{values}& \text{at}\\
    \hline
    \frac{1}{\G(a+1)}x_{+}^{a}   &-1,-2,-3,\dots  & \d^{(n-1)}(x) & a=-n\\
    
    \frac{1}{\G(a+1)}x_{-}^{a}   &-1,-2,-3,\dots    & (-1)^{n-1}\d^{(n-1)}(x)& a=-n\\
    
    \frac{1}{\G\left(\frac{a+1}{2}\right)}|x|^{a}& -1,-3,-5,\dots   & \frac{(-1)^n n!}{(2n)!}\d^{(2n)}(x) & a=-2n-1\\
    
    \frac{1}{\G\left(\frac{a+2}{2}\right)}|x|^{a} \sign(x) & -2,-4,-6\dots  & \frac{(-1)^n (n-1)!}{(2n-1)!}\d^{(2n-1)}(x)& a=-2n\\
    \hline
    (x+i\e)^a& -1,-2,-3,\dots& x^{-n}-i\pi \frac{(-1)^{n-1}}{(n-1)!} \d^{(n-1)}(x)& a=-n\\
    
    (x-i\e)^a& -1,-2,-3,\dots& x^{-n}+i\pi \frac{(-1)^{n-1}}{(n-1)!} \d^{(n-1)}(x)& a=-n\\
    \end{tabular}
    \caption{Homogeneous distributions on $\R$.}\label{tab:distributions2}
\end{table}

\textbf{Normalization of distributions.} We can cancel the simple poles of $x_{+}^{a}$ by a suitable Gamma function, and the normalized distribution $\frac{x_{+}^{a}}{\G(a+1)}$ is holomorphic with respect to $a$. Then the values at the removable poles are,
\begin{equation}
\frac{x_{+}^{a}}{\G(a+1)}\eval_{a=-n}=\d^{(n-1)}(x).
\end{equation}

Two other useful homogeneous distributions derived from boundary values of meromorphic functions are
\begin{equation}
    (x+i\e)^a =x_{+}^{a}+e^{ia\pi}x_{-}^{a},
    \hs 
    (x-i\e)^a =x_{+}^{a}+e^{-ia\pi}x_{-}^{a},
\end{equation}
and the poles are canceled at $a=-1,-2,\dots $ due to the factor $e^{ia\pi}=(-1)^n$, hence they are holomorphic functions with respect to $a$. At the removed poles the values are
\begin{align}
    (x+i\e)^{-n}& =x^{-n}-i\pi \frac{(-1)^{n-1}}{(n-1)!} \d^{(n-1)}(x),\nn\\
    (x-i\e)^{-n}& =x^{-n}+i\pi \frac{(-1)^{n-1}}{(n-1)!} \d^{(n-1)}(x),
\end{align}
in which the distributions $x^{-n}$ are understood as $|x|^{-2n}$ and $|x|^{-2n-1}\sign(x)$ respectively.

\textbf{Example.} As an example we provide the calculation of \eqref{eq:cft1doubleshadow} as follows,
\begin{align}
    K(\shadow^2,x_1,x_2)
    & =\lim_{\e\to 0}\inttt{dx_0}{\R} |x_{02}|^{-2(\D-\e)} |x_{01}|^{2\D-2}\nn\\
    & = \lim_{\e\to 0}\pi\frac{\G(\D-\half)}{\G(1-\D)}\frac{\G(\half-\D)}{\G(\D)}\cdot \frac{|x_{12}|^{-1+2\e}}{\G(\e)}\nn\\
    & =\frac{2\pi \tan\pi\D}{2\D-1}\d(x_{12}),
    \label{eq:klt1}
\end{align}
where in the first line the integral is regularized by shifting the weight slightly, in the second line the $1d$ KLT integral \cite{Murugan:2017eto} is used, and in the last line the $\d$-distribution comes from regularization of the distribution $|x|^{\a}$. Notice that the KLT integral 
\begin{equation}
\inttt{dx_0}{\R} |x_{01}|^{-2\D_1}|x_{02}|^{-2\D_2}=\pi^{\half}\frac{\G(\half-\D_1)}{\G(\D_1)}\frac{\G(\half-\D_2)}{\G(\D_2)}\frac{\G(\D_1+\D_2-\half)}{\G(1-\D_1-\D_2)}|x_{12}|^{1-2\D_1-2\D_2},
\end{equation} 
is equivalent to the star-triangle relation by a special conformal transformation.

\section{Kinematics and shadow formalism of CFT\secmath{_1}}\label{sec:cft1}

The $1d$ CFT appears in the SKY model, conformal defect and lightcone limit of higher dimensional CFT. 
The discussion of GCFT$_2$ is similar to that of CFT$_1$ in many aspects.
In this appendix we review the kinematics of CFT$_1$, see e.g. \cite{howe1992non,Maldacena:2016hyu,Hogervorst:2017sfd,Bulycheva:2017uqj,Kitaev:2017hnr,Sun:2021thf}, including correlation functions, the shadow transform, OPE blocks and the conformal block expansion.

\subsection{Local operators and correlation functions}
The $1d$ Euclidean\footnotemark{} conformal group is $SO(2,1)$ with the generators $L_{n},\, n=\pm 1, 0$ obeying the commutation relations $[L_{n},L_{m}]=(n-m)L_{n+m}$, and to discuss fermionic representations, it should be replaced by the double cover $SL(2,\R)$. For simplicity we focus on the bosonic case. The conformal transformations on $\R^{1}$ are the fractional linear transformations as shown in the table \ref{table:cft1}. The primary operators at $x=0$ are defined by
\begin{equation}
[L_0,\op]=\D \op,\hs{3ex} [L_{-1},\op]=\p_{x}\op, \hs{3ex}[L_1,\op]=0,
\end{equation}
and the descendant operators are $\p^{n}_{x}\op$. The primary operators together with their descendants form a highest weight representation. When $\D> 0$ they are unitary irreducible representations of the Lorentzian conformal group $\wave{SL}(2,\R)$, named as the discrete series representations. The infinitesimal transformations of the primary operators $\op_{\D}(x)$ are
\begin{equation}
[L_n,\op(x)]=(x^{n+1}\p_x +(n+1)\D x^{n})\op(x),
\end{equation}
where $n=\pm1,0$, and the finite transformations are
\begin{equation}
    U(f)\cdot \op(x)=|f'|^{\D}\op(x'),
    \label{eq:globaltransformationscft1}
\end{equation}
where $x'=f(x)$ are the conformal transformations.

\footnotetext{The $1d$ Lorentzian conformal group acting on the Lorentzian cylinder is the universal cover $\wave{SL}(2,\R)$. The covering map of these related groups are summarized as 
\begin{equation*}
\wave{SL}(2,\R)
\overset{\Z}{\longrightarrow}
SL(2,\R)\simeq SU(1,1)\simeq Sp(1,\R) 
\overset{\Z_2}{\longrightarrow} 
PSL(2,\R)\simeq SO(2,1).
\end{equation*}
The $1d$ Euclidean conformal algebra is related to the Lorentzian one by the NS-quantization and Wick rotation,
\begin{equation*}
L_{0,E}=\frac{i}{2}(L_{-1,L}+L_{1,L}), \hs L_{\pm 1,E}=-\frac{i}{2}(L_{-1,L}-L_{1,L})\mp L_{0,L}.
\end{equation*}
}

\begin{table}[htpb]
    \centering
    \renewcommand{\arraystretch}{2.4}
    \begin{tabular}{l L | L | L}
    name                & \text{charge}  & \text{vector field}   & \text{finite transformation}   \\
    \hline
    translation     & L_{-1}        & \p_x                 & x'=x+a\\

    dilation            & L_{0}         & x\p_x       & x'=\lm\,x \\

    SCT             & L_{1}         & x^2\p_x  & x'=x/(1-\mu x)\\
    
    inversion       & I& & x'=-1/x
    \end{tabular}
    \caption{The generators of $1d$ conformal group. The last line is the inversion which is useful to check conformal covariance.}
    \label{table:cft1}
\end{table}

The two-point functions of primary operators $\op_{i}:=\op_{\D_i}$ are
\begin{equation}
\vev{\op_1 \op_2}=\d_{12}|x_{12}|^{-2\D}, \hs \D=\D_1=\D_2,
\end{equation}
and the three-point functions are
\begin{equation}
\vev{\op_1 \op_2 \op_3}=c_{123}|x_{12}|^{-\D_{12,3}}|x_{23}|^{-\D_{23,1}}|x_{31}|^{-\D_{31,2}},
\end{equation}
where $c_{123}$ is the three-point coefficient. The four-point function of $\op_{i}$ could be written as a product of the stripped four-point function $\fourpt^{\schannel}(x)$ and a kinematical factor $K^{\schannel}(x_i)$
\begin{equation}
\vev{\op_1 \op_2 \op_3 \op_4}= 
K^{\schannel}(x_i)
\fourpt^{\schannel}(x).
\label{eq:schannel1}
\end{equation}
We choose the \schannel $\op_1\times \op_2\to \op_3\times \op_4$ kinematical factor as
\begin{equation}
K^{\schannel}(x_i)=
|x_{12}|^{-(\D_1+\D_2)}|x_{34}|^{-(\D_3+\D_4)} 
\abs{\frac{x_{24}}{x_{14}}}^{\D_{12}}
\abs{\frac{x_{14}}{x_{13}}}^{\D_{34}},
\label{eq:kinematical1}
\end{equation}
then read the \tchannel $\op_2\times \op_3\to \op_1\times \op_4$ kinematical factor by the permutation $(13)$
\begin{equation}
\vev{\op_1 \op_2 \op_3 \op_4}= 
|x_{23}|^{-(\D_2+\D_3)}|x_{14}|^{-(\D_1+\D_4)} 
\abs{\frac{x_{24}}{x_{34}}}^{\D_{32}}
\abs{\frac{x_{34}}{x_{13}}}^{\D_{14}}
\fourpt^{\tchannel}(1-x).
\label{eq:tchannel1}
\end{equation}
The $s-t$ crossing equation from \eqref{eq:schannel1} and \eqref{eq:tchannel1} is 
\begin{equation}
x^{-(\D_1+\D_2)}\fourpt^{\schannel}(x)=(1-x)^{-(\D_2+\D_3)}\fourpt^{\tchannel}(1-x)\qas 0<x<1.
\end{equation}

\subsection{Unitary principal series and shadow transform}\label{sec:cft1shadow}
The unitary irreducible representations of the Euclidean conformal group $SO(2,1)$ are classified into three classes: unitary principal series, discrete series and complementary series.

The Euclidean shadow transform is an intertwining map between two unitary principal series representations of the Euclidean conformal group. The unitary principal series representation $\urep{\D}$ of $SO(2,1)$ is defined as follows: the representation space is $L^2(\R)\ni f(x)$, with the inner product
\begin{equation}
    (f_1,f_2)=\int_{\R}dx\,f_1^*(x)f_2(x),
\end{equation}
and the group action is the same as the one on the primary operators \eqref{eq:globaltransformationscft1} but with complex weight $(\D=\half+is),\, s\in\Rnot$,
\begin{equation}
    U(f)\cdot \op(x)=|f'|^{\D}\op(x').
\end{equation}
Given a unitary principal series representation $\urep{\D=\half+is},\,s\in \Rneq$, we denote the associated shadow representation as $\urep{\wave{\D}=1-\D}$ and an operator transforming in $\urep{\wave{\D}}$ as $\opshadow$. The shadow transform $\shadow$
\begin{equation}
\shadow{\op}(x)
=\int_{\R}d x_0\,\vev{\opshadow(x)\opshadow(x_0)}\op(x_0) 
=\int_{\R}d x_0\, |x-x_0|^{2\D-2}\op(x_0)
\end{equation}
is an intertwining map between two representations
\begin{equation}
\shadow: \urep{\D} \to \urep{1-\D}.
\end{equation}
If the representations $\urep{\D}$ and $\urep{1-\D}$ are both irreducible, by the Schur lemma, $\shadow$ is an isomorphism. To check this we apply the shadow transform twice $\shadow^2: \urep{\D} \to \urep{\D}$,
\begin{equation}
\int dx_1\, K(\shadow^2,x_1,x_2)\op(x_1)
=\int dx_0 dx_1\,\vev{\op(x_2)\op(x_0)}\vev{\opshadow(x_0)\opshadow(x_1)}\op(x_1).
\end{equation}
In the case that $\shadow$ is an isomorphism, the kernel of $\shadow^2$ should be proportional to identity, $K(\shadow^2)=\cN(\D) \delta(x_{12})$. This kernel $K(\shadow^2)$ can be evaluated via the KLT integrals and the prefactor is
\begin{equation}
\cN(\D)=\frac{2\pi \tan\pi\D}{2\D-1}.
\label{eq:cft1doubleshadow}
\end{equation}
When $\D=\half+is,\, s\in\Rnot$, the factor $\cN(\D)$ is finite and nonzero, hence the shadow transform is an isomorphism indeed.

At the position of the poles and the zeros of $\cN(\D)$, the operators $\shadow$ and $\shadow^{-1}$ are not isomorphisms. For example, when $\shadow$ is not injective, the kernel subspace $\ker \shadow$ is a sub-representation, and the maximal quotient $\urep{\D}/\ker \shadow{\urep{\D}}$ is the discrete series representation, see e.g. chapter 7 of \cite{gel2016generalized}.

\subsection{OPE blocks}\label{sec:opeblockcft1}
The OPE relation of two primary operators is
\begin{equation}
\op_1(x_1)\op_2(x_2)=  \sum_{k} c_{12}^{k} \cD_{12k}(x_{12},\p_2) \op_{k}(x_2),
\label{eq:ope}
\end{equation}
and the bilocal operator $\cD_{12k}\op_k(x_2)$, capturing the resummation of the derivative operators, is called the OPE block \cite{Ferrara:1973vz},
\begin{equation}
\cD_{12k}(x_{12},\p_{2})\op_k(x_2)
 =x_{12}^{-\D_{12,k}}\Faa(\D_{k1,2}, 2\D_k ; x_{12} \p_{2})\op_k(x_2)
\label{eq:opeblock0}
\end{equation}
Here our convention follows the one used in \cite{Fortin:2020zxw}, and is slightly different from the one in \cite{Czech:2016xec}, which is
\begin{equation}
\op_1(x_1)\op_2(x_2)= x_{12}^{-\D_1-\D_2} \sum_{k} c_{12}^{k} \cB_{12k}(x_1,x_2).
\end{equation}
The relation between two conventions is
\begin{equation}
\cD_{12k}(x_{12},\p_2)\op_k(x_2) =x_{12}^{-\D_1-\D_2} \cB_{12k}(x_1,x_2).
\end{equation}

Notice that the technically safe way of phrasing the OPE relation \eqref{eq:ope} is to introduce the vacuum OPE \cite{Mack:1976pa},
\begin{equation}
\cR\, \op_1(x_1)\op_2(x_2)\vac= \sum_{k} c_{12}^{k} \cD_{12k}(x_{12},\p_2) \op_{k}(x_2)\vac
\end{equation}
where $\cR$ is the operator ordering with respect to a specific quantization scheme.

There are a few equivalent methods of computing the OPE block: the first is to use the compatibility of the OPE relation and the three-point functions, the second is to apply the recursion relation by imposing the symmetries on the OPE relation, and the third is to use the shadow formalism. We first recall the method of the recursion relation and then discuss the OPE blocks from the shadow formalism.

\textbf{OPE blocks from recursion relations.} Shifting the vacuum OPE to the origin and denoting the descendants as $\ket{\op_k,n}:=L_{-1}^{n}\ket{\op_k}$, we have 
\begin{equation}
\op_1(x)\ket{\op_2}= \sum_{k} c_{12}^{k} \sum_{n=0}^{\inf} a_{n}(\op_k) x^{-\D_{12,k}+n} \ket{\op_{k},n}.
\end{equation}
The $L_{1}$'s action gives
\begin{align}
L_{1}\op_1(x)\ket{\op_2}
& =
[L_{1},\op_1(x)]\ket{\op_2}+\op_1(x) L_{1}\ket{\op_2}\nn\\
& =
(x^2\p +2\D_1 x)\op_1(x)\ket{\op_2}\nn\\
& =
\sum_{k} c_{12}^{k} \sum_{n=0}^{\inf} a_{n}(\op_k) (\D_{1k,2}+n)  x^{-\D_{12,k}+n+1} \ket{\op_{k},n},
\end{align}
and by using $L_1\ket{\op_k,n}=n(2\D_k+n-1)\ket{\op_k,n-1}$ we have 
\begin{equation}
L_{1}\op_1(x)\ket{\op_2}=
\sum_{k} c_{12}^{k} \sum_{n=0}^{\inf} a_{n}(\op_k) x^{-\D_{12,k}+n} n(2\D_k+n-1)\ket{\op_k,n-1}.
\end{equation}
Due to the orthogonality of the descendants we get the recursion relation of $a_n$
\begin{equation}
a_{n}(\op_k)=\frac{(\D_{1k,2}+n-1)}{n(2\D_k+n-1)}a_{n-1}(\op_k), \qand a_0(\op_k)=1,
\end{equation}
and the solution
\begin{equation}
a_n(\op_k)=\frac{(\D_{1k,2})_n}{(2\D_k)_n n!}
\end{equation}
matches with the Taylor coefficient of the hypergeometric function in \eqref{eq:opeblock0}.

\textbf{OPE blocks from shadow formalism.} The operator product can be expanded at any other point $x_0\in [x_2,x_1],\,x_2<x_1$, and the translation invariance of the OPE block \eqref{eq:opeblock0} is not manifest. For example, to switch from $x_2$ to $x_1$ we may use the Kummer identity $\Faa(a,b;x)=\Faa(b-a,b;-x)e^{x}$,
\begin{align}
\cD_{12k}(x_{12},\p_2)\op_k(x_2)\vac
& =x_{12}^{-\D_{12,k}}\Faa(\D_{2k,1}, 2\D_k ; -x_{12} \p_{2})e^{x_{12} \p_{2}}\op_k(x_2)\vac\nn\\
& =\cD_{12k}(x_{21},\p_1)\op_k(x_1)\vac.
\end{align}
To make the translational invariance apparent, we can average over the expansion point $x_0$ by a weight function $f_{12k}$,
\begin{equation}
\cD_{12k}(x_{12},\p_2)\op_k(x_2)\vac=\inttt{dx_0}{I}f_{12k}(x_1,x_2,x_0)\op_k(x_0)\vac.
\label{eq:smearing1}
\end{equation}
The left hand side transforms as $\op_1(x_1)\op_2(x_2)\vac$, hence $f_{12k}\sim\vev{\op_1(x_1)\op_2(x_2)\opshadow_{k}(x_0)}$.
If insisting on the Euclidean region, there are extra unphysical contributions to the integral due to the shadow symmetry $\D\to1-\D$, and to single out the correct terms we need to introduce the projector of monodromy by hand \cite{SimmonsDuffin:2012uy}. 

If Wick-rotating to the Lorentzian region, the three-point structure $\vev{\op_1(x_1)\op_2(x_2)\opshadow_{k}(x_0)}$ admits different analytic continuations, e.g. Wightman function and time-ordered function. For the time-ordered three-point structure, the OPE block \eqref{eq:smearing1} contains shadow terms as the Euclidean case. The Wightman three-point structure is a more suitable choice and there are no shadow terms \cite{Ferrara:1972ay}. Equivalently, we can modify the integration region to eliminate the shadow terms. The weight function should vanish in the region where the OPE loses convergence, e.g., hitting other operators or becoming timelike with respect to the external operators. Hence in the two dimensional case, the integral domain is the causal diamond $\Diamond_{12}$ associated with spacelike pair of points $\set{x_1,x_2}$ \cite{Czech:2016xec}. 

Comparing with the Taylor expansion \eqref{eq:opeblock0} of the OPE blocks, the integral expression \eqref{eq:smearing1} corresponds to changing the basis of a conformal family. The original basis is  $\set{\ket{\op},L_{-1}\ket{\op},\dots }$, and the new basis is $\set{\op(x)\vac:x\in\R^{1}}$. They are related by the vacuum mode expansion,
\begin{equation}
\op(x)\vac=\sum_{n=0}^{\inf}\frac{x^{n}}{n!}\p^n \op(0)\vac
\end{equation}
We can check the equivalence of the two approaches by evaluating the integral \eqref{eq:smearing1} directly,
\begin{equation}
\cD_{12k}(x_{12},\p_2)\op_k(x_2)=N_{12k} \inttt{d x_0}{I}\vev{\op_1(x_1) \op_2(x_2) \opshadow_k(x_0)} \op_k(x_0)
\end{equation}
where $N_{12k}$ is some normalization factor to be determined. In the CFT$_1$ case, the causal diamond degenerates to the interval $x_0\in (x_2,x_1)$ and we have
\begin{align}
    \lhs
    & =N_{12k} x_{12}^{-\D_{12,4}}\inttt{d x_0}{(x_2,x_1)}x_{02}^{-\D_{24,1}}x_{10}^{-\D_{41,2}} \op_k(x_0)\nn\\
    & =N_{12k} x_{12}^{-\D_{12,4}}\inttt{d x_0}{(x_2,x_1)}x_{02}^{-\D_{24,1}}x_{10}^{-\D_{41,2}} \sum_{n=0}^{\inf}\frac{x_{02}^{n}}{n!}\p^n \op_k(x_2)\nn\\
    & =N_{12k} \sum_{n=0}^{\inf}\p^n \op_k(x_2) I_n,
    \label{eq:vacuumope} 
\end{align}
where $\D_4=1-\D_k$ and the coefficient integrals are
\begin{equation}
    I_n
    =\frac{1}{n!}x_{12}^{-\D_{12,4}} \inttt{d x_0}{(x_2,x_1)}x_{02}^{-\D_{24,1}+n}x_{10}^{-\D_{41,2}}
    = \frac{1}{n!} B(\D_{k1,2}+n,\D_{k2,1})x_{12}^{-\D_{12,k}+n},
\end{equation}
then by choosing $N_{12k}=B(\D_{k1,2},\D_{k2,1})^{-1}$ we get back to the previous result,
\begin{align}
    \cD_{12k}(x_{12},\p_2)\op_k(x_2)
    & =N_{12k}x_{12}^{-\D_{12,k}} \sum_{n=0}^{\inf}\p^n \op(x_2) x_{12}^{n}\frac{1}{n!} B(\D_{k1,2}+n,\D_{k2,1})\nn\\
    & =x_{12}^{-\D_{12,k}} \Faa(\D_{k1,2}, 2\D_k ; x_{12} \p_{2}) \op(x_2).
\end{align}

\subsection{Conformal blocks}\label{sec:cft1conformalblock}
Using the OPE relations repeatedly the higher-point functions can be decomposed into
a sum of the conformal blocks multiplied by the OPE coefficients. The conformal blocks
capture the contributions from the exchanged conformal families. This procedure has been explicitly done in the CFT$_1$ and CFT$_2$ case \cite{Fortin:2020zxw} for arbitrary higher-point. There are a few efficient ways of calculating four-point conformal blocks: solving the Casimir equations \cite{Dolan:2011dv}, the recursion relations with respect to $\D$ \cite{Penedones2016}, and the shadow formalism \cite{Osborn_2012,Karateev:2018oml}. We briefly illustrate the method of Casimir equation in CFT$_1$.

\textbf{Settings of conformal block expansion.} The \schannel conformal block expansion of four-point function is
\begin{equation}
\wick{\vev{\c1 \op_1 \c1 \op_2 \c1 \op_3 \c1 \op_4}} =\sum_{n}c_{12n}c_{43n}\cD_{12n}(x_{12},\p_2)\cD_{43n}(x_{43},\p_3)\vev{\op_n(x_2)\op_n(x_3)}=\sum_n p^{\schannel}_n \blockdressed^{\schannel}_n(x_i),
\end{equation}
where the (unstripped) conformal block with respect to primary $\op_n$ is defined as
\begin{equation}
\label{eq:block4}
\blockdressed^{\schannel}_n(x_i)=\cD_{12n}(x_{12},\p_2)\cD_{43n}(x_{43},\p_3)\vev{\op_n(x_2)\op_n(x_3)},
\end{equation}
and $p^{\schannel}_n=c_{12n}c_{43n}$. To further carry out calculations we need to introduce the stripped version of conformal blocks depending only on the cross ratios by factoring out the kinematical factors
\begin{equation}
\blockdressed^{\schannel}_n(x_i)=K^{\schannel}(x_i)\blockstripped^{\schannel}_{n}(x),
\label{eq:block1}
\end{equation}
then the block expansion of the stripped four-point function is
\begin{equation}
\fourpt^{\schannel}(x)=\sum_n p^{\schannel}_n \blockstripped^{\schannel}_{n}(x).
\end{equation}

\textbf{Conformal blocks from Casimir equation.} Inserting a complete basis into the four-point functions in the radial quantization $x_4>x_3>x_2>x_1$, we get
\begin{equation}
\vev{\op_1 \op_2 \op_3 \op_4}=\bra{0} \op_4 \op_3 \sum |\op_0|\, \op_2 \op_1\vac, 
\end{equation}
where $|\op_0|$ is the projection operator of the conformal family $\rep{\D_0}$
\begin{equation}
|\op_0|=\sum_{n,m\in \rep{\D_0}}G^{-1}{}_{n,m}\ket{n}\bra{m},
\end{equation}
and $G_{n,m}=\bra{n}\ket{m},\, n,m\in \rep{\D_0}$ is the Gramian matrix. Then the conformal block can be written as a summation over the matrix elements $\bra{n}\op_1 \op_2 \vac$,
\begin{equation}
\label{eq:cft1block5}
\blockdressed^{\schannel}_0(\xy{i})=\bra{0} \op_4(x_4) \op_3(x_3) |\op_0| \op_2(x_2) \op_1(x_1)\vac.
\end{equation}

The Casimir differential operator is the representation of the abstract Casimir element of the conformal algebra when acting on the matrix elements $\bra{n}\op_1 \op_2 \vac,\, n\in \rep{\D_0}$. For element $X$ in the conformal algebra $\sllie(2,\R)$, the corresponding Ward identity is
\begin{equation}
\label{eq:Ward1}
\bra{n} X \op_1 \op_2 \vac
=
(X^{(1)}+X^{(2)})\bra{n}\op_1(x_1)\op_2(x_2)\vac,
\end{equation}
where the differential operators are $X^{(i)}\vev{\dots \op_i(x_i)\dots }=\vev{\dots [X,\op_i(x_i)]\dots }$. The Casimir element of $\sllie(2,\R)$ is $C=L_0(L_0-1)-L_{-1}L_1$. Hence by repeatedly using \eqref{eq:Ward1}, and the fact that $C$ acts on $\rep{\D_0}$ as a scalar $\D_0 (\D_0-1)$, we find
\begin{equation}
\label{eq:Casimir5}
C^{(12)}\bra{n}\op_1 \op_2 \vac
=\D_0 (\D_0-1)\bra{n}\op_1 \op_2 \vac,
\end{equation}
where the differential operator $C^{(1+2)}$ is
\begin{equation}
C^{(1+2)}=-x_{12}^2 \p_1\p_2 -2x_{12}\ppair{\D_1 \p_2-\D_2 \p_1} +(\D_1 + \D_2)(\D_1 + \D_2-1).
\end{equation}
Inserting \eqref{eq:Casimir5} into the expression of conformal blocks, we get the Casimir equation
\begin{equation}
\ppair{C^{(1+2)}-\D_0 (\D_0-1)}\blockdressed^{\schannel}_{\D_0}(x_i)=0.
\end{equation}
Then plugging the definition of stripped conformal blocks into this equation, we get the Casimir equation
\begin{equation}
\bpair{x^2(1-x)\p^2_x +(\D_{12}-\D_{34}-1)x^2\p_x + \D_{12}\D_{34}x -\D_0 (\D_0-1)}\blockstripped^{\schannel}_{\D_0}(x)=0,
\end{equation}
which has two independent solutions: the physical block
\begin{equation}
\blockstripped^{\schannel}_{\D_0}(x)=x^{\D_0}\Fba(\D_0-\D_{12},\D_0+\D_{34};2\D_0,x),
\label{eq:cft1block}
\end{equation}
and the shadow block $\blockstripped^{\schannel}_{1-\D_0}(x)$.

\section{GCFT\secmath{_2} as Null Defect of Lorentzian CFT\secmath{_3}}

In this appendix we establish the relation between GCFT$_2$ and null defect in Lorentzian CFT$_3$ at the kinematical level, see also the discussion in higher dimensions \cite{Chen:2021xkw} and a related discussion in \cite{Figueroa-OFarrill:2021sxz}. We firstly recall the ideas of conformal defect, see e.g. \cite{Cardy:1984bb,McAvity:1995zd,Liendo:2012hy,Gaiotto:2013nva,Gliozzi:2015qsa,Billo:2016cpy,Gadde:2016fbj} and the analytic studies in \cite{Rastelli:2017ecj,Hogervorst:2017kbj,Mazac:2018biw,Kaviraj:2018tfd}, then discuss the residual symmetry of null defects in the first subsection, the outer automorphism of Carrollian conformal algebra in the second subsection. Notice that this defect picture rules out the infinite-dimensional BMS symmetry, and the result does not contradict to the symmetry enhancement argument in \cite{Chen:2019hbj}, since the latter relies on the existence of local conserved currents, which a defect theory may not have.

Denoting the conformal group of the bulk\footnote{Not in the holographic sense.} CFT as $G$ and the one of the defect CFT as $G_{d}\subset G$, all the conformal defects connected by conformal transformations in $G$ can be regarded as points in the homogeneous space $G/G_{d}$, and the broken symmetries transforming one defect to another describe the motions in this homogeneous space. The defect can be regarded as a non-local object added to the original CFT spectrum, and besides the bulk-local operators describing local excitations in the bulk, there are defect-local operators describing local excitations on the defect. The dynamical information is captured by the bulk-bulk OPE, the bulk-defect OPE, the defect-defect OPE and the defect expansion, the last of which is a generalization of the Ishibashi states in $2d$ Virasoro CFT \cite{Cardy:1986gw,Ishibashi:1988kg,Cardy:1989ir}.

Among the defects of various codimensions discussed in the literature, a trivial example is the point-like defect containing isolated points. For the one-point case, $G_{d}$ is the one-point stabilizer subgroup. If we insist that the state $\ket{\defect}$ describing the defect is $G_{d}$-invariant, in unitary CFT after modding out null states, $\ket{\defect}$ is translational-invariant and hence $\ket{\defect}=\vac$. Next considering the defect-local operators transforming in a non-trivial representation of $G_{d}$, together with their partners in the homogeneous space $G/G_{d}=\R^{d}_{c}$, where $\R^{d}_{c}$ is the conformal compactification of $\R^{d}$, we return to the construction of local operators. 

For a Lorentzian CFT$_3$ with conformal group $SO(3,2)$, there are three types of $\codim-1$ conformal defects: timelike defect with residual symmetry $SO(2,2)$, spacelike defect with $SO(3,1)$ and null defect with $SO(1,1)\ltimes ISO(2,1)$, and the defect local operators should compose a Lorentzian CFT$_2$, a Euclidean CFT$_2$ and a special type of Carrollian CFT$_2$ respectively. 

The residual conformal group of null defect $G_{d}=SO(1,1)\ltimes SO(2,1)$ is exactly the same as one-point stabilizer subgroup of local operators, and $G/G_{d}=\R^{2,1}_{c}$, indicating a relation between null defects and points in $\R^{2,1}_{c}$.
Apparently we can associate the null-cone $(x-X)\cdot (x-X)=0$ centered at $X\in \R^{2,1}$ with the point $X$ itself. To see this map from the null defect to the point preserving the symmetry, we consider the set of hyperboloids and hyperplanes in $\R^{2,1}$,
\begin{equation}
\M=\set{X^{+}x\cdot x+2X\cdot x+X^{-}=0 \,:\,X\in \R^{2,1} ,(X^{+},X^{-})\in\R^{1,1}}\simeq \R^{3,2}.
\end{equation}
In this set, the elements satisfying $X\cdot X-X^{+}X^{-}<0,>0,=0$ correspond to the timelike, the spacelike and the null defects respectively. Remarkably the conformal transformations of $SO(3,2)$ act linearly on $\M$.
Now the null-cone $X\cdot X-X^{+}X^{-}=0$ in $\M$ can naturally be identified with the embedding space of $\R^{2,1}$, i.e. $\R^{2,1}_{c}$.
In this way we establish a 1-1 symmetry-preserving correspondence between the null defects and the points in the conformal compactification of $\R^{2,1}$,
\begin{equation}
\text{null defect of }\R^{2,1}_{c} \hspace{3ex} \Longleftrightarrow \hspace{3ex} \text{point in }\R^{2,1}_{c},
\end{equation}
in which the null-planes correspond to the points at the conformal boundary of $\R^{2,1}$, and the conformal boundary itself corresponds to the spacelike infinity\footnote{Notice that this is not the asymptotic infinity. In the conformal compactification of $\R^{2,1}$, the timelike infinities and the spacelike infinity are identified.}. And as a byproduct we re-derive the embedding space formalism.

On the other side, the residual symmetries $ISO(2,1)$ act on the defect exactly as the Carrollian conformal transformations, hence we can describe the defect-local operators by Carrollian CFT$_2$. The extra $SO(1,1)\subset G_{d}$ can be identified with the non-trivial outer automorphism of the Carrollian conformal group $ISO(2,1)$. Despite that a Carrollian CFT may not possess this enhanced symmetry $SO(1,1)$, we can still learn the properties of the Carrollian CFT from the defect viewpoint.

\subsection{Null-plane and null-cone defects}\label{sec:nulldefect1}
In this subsection we discuss three typical configurations of null defects: the null-plane, the lightcone and the conformal boundary. The settings of Lorentzian CFT$_3$ are summarized in appendix \ref{sec:app1}.

\textbf{Null-plane.} For the defect located at $x^{0}-x^{1}=0$, we use the lightcone coordinates, $y'=x^{0}-x^{1},\,y=x^{0}+x^{1},\,x=x^{2}$. Obviously the residual symmetry of the defect surface contains $x$-translation, $y$-translation and dilatation, and we can identify them with the generators of Carrollian conformal algebra, $L_{-1} =P_{2},\,M_{-1} =\half(P_{0}+P_{1}),\,L_{0} =D$. With the null vector $(1,1,0)$, the null rotation along the $x-y$ plane,
\begin{equation}
\exp \half v(M_{20}-M_{12})=\bmat{1+\frac{1}{8}v^2 &-\frac{1}{8}v^2 &\half v \\\frac{1}{8} v^2 &1-\frac{1}{8}v^2 &\half v \\\half v& -\half v& 1}
\end{equation}
preserves the defect as well. It acts as $x\to x,\,y\to y+vx$, hence suggesting that $M_{0}=-\half(M_{20}-M_{12})$.

To get the rest two generators, noticing that the action of the inversion $I: x^{0}\to \frac{x^{0}}{x\cdot x},\,x^{1}\to \frac{x^{1}}{x\cdot x},\,x^{2}\to \frac{-x^{2}}{x\cdot x}$ preserves the defect and acts as the inversion $x\to -\frac{1}{x},\,y\to \frac{y}{x^2}$ on the defect. Hence we can identify the SCTs as
\begin{equation}
L_{1}=K_{2}, \hs M_{1} =-\half(K_{0}+K_{1}).
\end{equation}
One can check that the six generators form a subalgebra $\isolie(2,1)\subset \solie(3,2)$. We summarize the identification of the generators in the following,
\begin{alignat}{3}
L_{-1}& = P_{2},          & \hs            L_{0}& = D,           & \hs            L_{1}& = K_{2},       \nn\\
M_{-1}& = \half(P_{0}+P_{1}),        & \hs            M_{0}& = -\half(M_{20}-M_{12}),            & \hs            M_{1}& = -\half(K_{0}+K_{1}).
\end{alignat}
The last residual symmetry is $D_{b}:=-M_{01}$ acting on the null-plane as $x\to x,\,y\to e^{-t}y$, and its commutation relations with other generators are
\begin{equation}
[D_{b},L_{n}]=0,     \hs  [D_{b},M_{n}]=-M_{n}.
\end{equation}

\textbf{Lightcone.} Another type of null defect locates at the lightcone $x\cdot x=0$. Apparently the Lorentz transformations $M_{ab}$ and the three SCTs, which reduce to
\begin{equation}
K: x^{a}\to \frac{x^{a}-t^{a}x^2}{1-2t\cdot x+t^2 x^2} \overset{x^2=0}{\longrightarrow}\frac{x^{a}}{1-2t\cdot x},
\end{equation}
preserve the lightcone, and should be identified with $L$'s and $M$'s in $\isolie(2,1)$ respectively. 

The correct parameterization manifesting the Carrollian conformal symmetry in table \ref{table:global_gca} turns out to be
\begin{equation}
(x^{0},x^{1},x^{2})=(\frac{1+x^2}{c y},\frac{1-x^2}{c y},\frac{2x}{c y}), \hs (\xy)\in \R^2,
\label{eq:lightcone1}
\end{equation}
where $c$ is an arbitrary non-vanishing constant. 

The $x$-dependence in \eqref{eq:lightcone1} is inspired by the N-S quantization and the embedding space formalism of CFT$_1$. The Cayley tranform $z=\frac{1+ix}{1-ix}=\frac{1-x^2}{1+x^2}+i \frac{2x}{1+x^2}$ maps a CFT$_1$ on the real line $x\in \R^{1}$ to a CFT$_1$ on the unit circle $\th=2\arctan x \in S^{1}$, and from the embedding space viewpoint, this corresponds to choosing non-compact vs. compact slicing of the projective lightcone. 

In our case the null direction of the lightcone is physical, and the parameterization such that $x$ transforms as linear fractional transformation is
\begin{equation}
(x^{0},x^{1},x^{2})=F(\xy)(1,\frac{1-x^2}{1+x^2},\frac{2x}{1+x^2}).
\end{equation}
By comparing the actions of six residual symmetries on the defect with the Carrollian conformal transformations on the plane, we get a set of over-constrained equations of $F(\xy)$ with a unique solution $F(\xy)=\frac{(1+x^2)}{c y}$. Choosing $c=4$ we get the subalgebra inclusion,
\begin{alignat}{3}
L_{-1}& = -\ppair{M_{20}+M_{12}},          & \hs            L_{0}& = -M_{01} ,          & \hs            L_{1}& = M_{20}-M_{12},          
\nn\\
M_{-1}& = K_{0}-K_{1},          & \hs            M_{0}& = - K_{2},            & \hs            M_{1}& = K_{0}+K_{1}.
\end{alignat}
The bulk dilatation $D_{b}:=D$ acts on the null-cone as $x\to x,\,y\to e^{-t}y$, and its commutation relations are the same as before.

\textbf{Conformal boundary.} Notice that we can exchange the translations and SCTs in $\R^{2,1}$ by the inversion $I$, and the lightcone is mapped to the conformal boundary of $\R^{2,1}_{c}$. Then the subalgebra inclusion is
\begin{alignat}{3}
L_{-1}& = -\ppair{M_{20}+M_{12}},          & \hs            L_{0}& = -M_{01} ,          & \hs            L_{1}& = M_{20}-M_{12},          
\nn\\
M_{-1}& = P_{0}-P_{1},          & \hs            M_{0}& = -P_{2},            & \hs            M_{1}& = P_{0}+P_{1},
\label{eq:inf}
\end{alignat}
and $D_{b}:=-D$. In the picture of embedding space of $\R^{2,1}$, $(X^{+},X^{-},X)=(1,x^2,x^{a})$, the lightcone defect is $X^{-}=0$ and the conformal boundary is $X^{+}=0$.

\subsection{Bulk dilatation as outer automorphism of Carrollian CFT}\label{sec:outer}

In relativistic CFT$_d$, the semi-simplicity of the (complexified) conformal algebra $\mathfrak{so}(d+2,\C)$ implies that the outer automorphism group is finite. In Carrollian CFT$_2$, the outer automorphism group of both the global and local Carrollian conformal algebras is the multiplicative group, $\Out(\mathfrak{cca}_2)=\Rneq$. The connected component of identity in the group is generated by the bulk dilatation $D_{b}$, since there is no intrinsic scale in $\R^{2,1}$ mimicking the AdS radius. 

From the null-cone defect picture, the bulk dilatation $D_{b}$ simply rescale the Carrollian time $y$. The infinitesimal action of $D_{b}$ is
\begin{alignat}{2}
    [D_b,L_n]&=0,            & \hs    [D_b,c_L]&=0,\nn\\
    [D_b,M_n]&=-M_n,         &       [D_b,c_M]&=-c_M.
\end{alignat}
Defining the flowed generators as $Q(t)=U(t)QU(-t),\,U(t)=e^{t D_{b}}$, then we have 
\begin{alignat}{2}
    L_n(t)&= L_n,          & \hs   c_L(t)&= c_L,\nn\\
    M_n(t)&=\frac{1}{\lm} M_n,         &       c_M(t)&=\frac{1}{\lm} c_M.
\end{alignat}
where $\lm=e^{t}$. Defining the flowed highest weight state as $\ket{\D,\xi}_t=U_t\ket{\D,\xi}$ by $M_0(t)\ket{\D,\xi}_t=\xi \ket{\D,\xi}_t$, we find the boost charge 
\begin{equation}
M_0 \ket{\D,\xi}_t  = \lm U_t M_0 \ket{\D,\xi}=\lm\xi \ket{\D,\xi}_t.
\end{equation}
Hence $D_{b}$ sends $\rep{\D,\xi,r}$ to a series of inequivalent ones $\rep{\D,\lm\xi,r}$. Either this series of operators is in the same theory, or there is a continuous series of theories labeled by $\lm$. If the theory is invariant under the outer automorphism, either there are only $\xi=0$ operators or the $\xi$-spectrum is continuous, and for a BMS field theory with this invariance, the central charge $c_{M}$ is forced to vanish, $c_{M}=0$.

Besides the multiplicative outer automorphism, there is a discrete automorphism. The discrete one is not in the identity component of the outer automorphism group, and it flips the sign of supertranslations
\begin{alignat}{2}
    L_n'&= L_n,          & \hs   c_L'&= c_L,
    \nn\\
    M_n'&=- M_n,         &       c_M'&=- c_M. 
\end{alignat}

\section{Miscellaneous Calculations}
\subsection{Calculation of the OPE blocks}\label{sec:calculationOPEblocks}

In this appendix, we present some technical details in computing the OPE blocks. Inspired by the discussion in the appendix \ref{sec:opeblockcft1}, we choose the integral region as $I= (x_2,x_1)\times \R, \, x_1>x_2$, then the integral \eqref{eq:opeblockshadow1} can be evaluated as,
\begin{align}
    &\peq\cD_{123}\op_3(\xy{2})
    \nn\\
    & =
    N_{123}
    x_{12}^{-\D_{12,4}}e^{\xi_{12,4}k_{12}}
    \inttt{d x_0y_0}{I}
    x_{02}^{-\D_{42,1}}x_{10}^{-\D_{41,2}} 
    e^{\xi_{42,1}k_{02}+\xi_{41,2}k_{10}}
    \op_3(\xy{0})
    \nn\\
    & =
    N_{123}
    x_{12}^{-\D_{12,4}}e^{\xi_{12,4}k_{12}}
    \inttt{d x_0y_0}{I}
    x_{02}^{-\D_{42,1}}x_{10}^{-\D_{41,2}} 
    e^{\xi_{42,1}k_{02}+\xi_{41,2}k_{10}}
    \sum_{n,m=0}^{\inf}
    \frac{x_{02}^{n}}{n!}\frac{y_{02}^{m}}{m!}
    \p_{x_{2}}^n\p_{y_{2}}^m \op_3(\xy{2})
    \nn\\
    & =
    x_{12}^{-\D_{12,3}}e^{\xi_{12,3}k_{12}}
    \sum_{n,m=0}^{\inf}\p_{x_{2}}^n\p_{y_{2}}^m \op_3(\xy{2}) 
    I_{n,m}, 
    \label{eq:In1}
\end{align}
where $\D_4=2-\D_3,\,\xi_4=-\xi_3$. The coefficient integrals are
\begin{align}
I_{n,m}
& =
N_{123}
\frac{x_{12}^{-\D_{12,4}}e^{\xi_{12,4}k_{12}}}{x_{12}^{-\D_{12,3}}e^{\xi_{12,3}k_{12}}}
\inttt{d x_0y_0}{I}
x_{02}^{-\D_{42,1}}x_{10}^{-\D_{41,2}} 
e^{\xi_{42,1}k_{02}+\xi_{41,2}k_{10}}
\frac{x_{02}^{n}}{n!}\frac{y_{02}^{m}}{m!}
\label{eq:In}
\\
& =
N_{123}
\frac{1}{n!m!}
x_{12}^{-2\D_3+2}e^{2\xi_3 k_{12}}
\inttt{d x_0}{I}
x_{02}^{-\D_{42,1}+n}x_{10}^{-\D_{41,2}} 
\exp\bpair{\frac{\xi_{42,1}(-y_2)}{x_0-x_2}+\frac{\xi_{41,2}(y_1)}{x_1-x_0}}
I'_{m},\nn
\end{align}
where the $y_0$-integral is
\begin{equation}
I'_{m}
=
\inttt{d y_0}{\R}
y_{02}^{m}
\exp\bpair{\ppair{\frac{\xi_{42,1}}{x_0-x_2}-\frac{\xi_{41,2}}{x_1-x_0}}y_0}.
\end{equation}
If \eqref{eq:opeblockgcft1} is the correct ansatz for the OPE block, the integral \eqref{eq:In} should be a homogeneous polynomial of degree $(n+m)$ with respect to $x_{12}$ and $y_{12}$
\begin{equation}
I_{n,m}(x_{12},y_{12})=\sum_{k=0}^{n+m}a_{n,m,k}(\D_{12},\D_3,\xi_{12},\xi_3)x_{12}^{k}y_{12}^{n+m-k}.
\end{equation}
Let us look the first coefficient integral, which is related to the normalization factor and the shadow coefficient. The related $y_0$-integral is
\begin{equation}
I'_{0}
=
\inttt{d y_0}{\R}
\exp\bpair{\ppair{\frac{\xi_{42,1}}{x_0-x_2}-\frac{\xi_{41,2}}{x_1-x_0}}y_0}.
\end{equation}
For virtual operators, the $\xi$'s are purely imaginary, hence this integral is proportional to a $\delta$-distribution
\begin{equation}
I'_{0}
=
2\pi \d\bpair{\frac{2 \xi_3(x_0-X)}{(x_0 - x_2) (x_1 - x_0)}}
=
\frac{\pi}{|\xi_3|}(x_0 - x_2) (x_1 - x_0) \d(x_0-X),
\end{equation}
where the special point is
\begin{equation}
X=\frac{1}{2}(x_1+x_2)+\frac{R}{2} (x_1-x_2), \qand R:=\frac{\xi_1-\xi_2}{\xi_3}.
\end{equation}
The interesting thing is that the constraint $X\in (x_2,x_1)$ is equivalent to
\begin{equation}
-1<R<1.
\end{equation}
Thus, the first integral is
\begin{equation}
\label{eq:opeblocknormalization}
I_{0,0}
=
N_{123}^{-1}
% =
% 2^{2-2\D_3}\frac{\pi}{|\xi_3|} \xi_3^{2-2\D_3}\xi_{312}^{-1+\D_{312}}\xi_{321}^{-1+\D_{321}}.
=
2^{2-2\D_3}\frac{\pi}{|\xi_3|}(1+R)^{-1+\D_{312}}(1-R)^{-1+\D_{321}}.
\end{equation}

For other coefficient integrals, the computation is straightforward. The higher $y_0$-integrals give rise to the derivatives of $\d(x)$
\begin{equation}
I'_m
=
\frac{\pi}{|\xi_3|}(2\xi_3)^{-m} (x_{02}x_{10})^{m+1} \d^{(m)}(x_0-X) \exp(\frac{2 \xi_3(x_0-X)y_2}{(x_0 - x_2) (x_1 - x_0)}),
\end{equation}
where we have used the analytic continuation $\xi_n=ir_n\in i\R$. Inserting this into the expression of $I_{n,m}$ \eqref{eq:In}, the integration with respect to $x_0$ turns into the derivatives of order $m$
\begin{align}
\label{eq:inm}
I_{n,m}(x_{12},y_{12})
& =
I_{0,0}^{-1}
\frac{(-1)^{m}}{n!m!}
\frac{\pi}{|\xi_3|}(2\xi_3)^{-m} 
x_{12}^{-2\D_3+2}e^{2\xi_3 k_{12}}
\nn
\\
& \peq 
\cdot \dv[m]{x_0}\eval_{x_0=X}
x_{02}^{\D_{31,2}+n+m-1}x_{10}^{\D_{32,1}+m-1} 
\exp(\frac{-\xi_{321} y_{12}}{x_{10}}).
\end{align}
For example, at the level $L=n+m=1$, there are two descendants $\set{\p_x\op,\p_y\op}$, with $2\times (L+1)$ coefficients
\begin{align}
\cD_{123}& \supset I_{1,0}(\xy)\p_x+I_{0,1}(\xy)\p_y\nn\\
& \supset \frac{1+R}{2} x \p_{x}+\ppair{-\frac{\D_{12}-\D_3 R}{2\xi_3}x+\frac{1+R}{2}y}\p_y.
\end{align}
We list the coefficients at level $2$ in table \ref{table:anmk23}. In the table, the unlisted coefficient $a_{0,2,2}$ is
\begin{equation}
a_{0,2,2}
=
\frac{1}{16 \xi _3^2} \ppair{2 \D_{12}^2+\left(\D_3+1\right) \left(\left(2 \D_3+1\right) R^2-1\right)-2 \D_{12} \left(2 \D_3 R+R\right)}.
\end{equation}

\begin{table}[t]
\begin{center}
\renewcommand{\arraystretch}{2}
    \begin{tabular}{L| L L L L}
    L=2 & x^2& xy& y^2& \\
    \hline
    \p_x^2 & \frac{1}{8} (R+1)^2 & 0 & 0 & \\
    \p_x\p_y &  \frac{(R+1) \left(-2 \D_{12}+2 \D_3 R+R-1\right)}{8 \xi _3} & \frac{1}{4} (R+1)^2 & 0 & \\
    \p_{y}^{2} & a_{0,2,2}
    & \frac{(R+1) \left(-2 \D_{12}+2 \D_3 R+R-1\right)}{8 \xi _3} 
    & \frac{1}{8} (R+1)^2 & \\
    \end{tabular}
    \caption{The OPE block coefficients at level 2.
    \label{table:anmk23}}
    \end{center}
\end{table}

Along the diagonal direction the OPE block coefficients are the same up to a binomial coefficient, and this holds at the higher levels. This inspires us to make the simpler ansatz for the OPE block
\begin{equation}
\label{eq:opeblockgcft2}
\cD_{123}(x_{12},y_{12},\p_{x_2},\p_{y_2})=x_{12}^{-\D_{123}}e^{\xi_{123}\frac{y_{12}}{x_{12}}} \sum_{n,m} a_{n,m} (x_2\p_{x_2}+y_2\p_{y_2})^{n} (x_2\p_{y_2})^{m}, 
\end{equation}
in which the derivatives should be understood as acting on $\op_3$ directly, and the coefficients are related to the previous ones by
\begin{equation}
a_{n,m}=a_{n,m,n+m} \qiff I_{n,m}(x,0)=x^{n+m}a_{n,m}.
\end{equation}
Equivalently there is a recursion relation for $I_{n,m}$,
\begin{equation}
\label{eq:recursiongcft1}
x\p_y I_{n,m}(\xy)=(n+1)I_{n+1,m-1}(\xy) \qif m\geq 1,\,n\geq 0,
\end{equation}
which can be checked straightforwardly using \eqref{eq:inm}. Actually this is originated from the Ward identity with respect to $M_0$.

To calculate the leading coefficient $a_{n,m}$, from \eqref{eq:inm} we read
\begin{equation}
\label{eq:opeblockgcft4}
I_{n,m}(x_{12},0)=
I_{0,0}^{-1}
\frac{(-1)^{m}}{n!m!}
\frac{\pi}{|\xi_3|}(2\xi_3)^{-m} 
x_{12}^{-2\D_3+2}
\dv[m]{x_0}\eval_{x_0=X}
x_{02}^{\D_{31,2}+n+m-1}x_{10}^{\D_{32,1}+m-1}. 
\end{equation}
Here the derivative term can be expressed as a Jacobi polynomial. To be concrete, setting $z=R+2\frac{x_0-X}{x_{12}}$, then $x_{02}=\half (1+z) x_{12},\,x_{10}=\half (1-z) x_{12}$, and
\begin{align}
& 
\peq \dv[m]{x_0}\eval_{x_0=X}
x_{02}^{\D_{31,2}+n+m-1}x_{10}^{\D_{32,1}+m-1} \nn\\
& =
\ppair{\frac{x_{12}}{2}}^{2\D_{3}+n+m-2}\dv[m]{}{z} \eval_{z=R}\bpair{(1-z)^{\D_{321}-1}(1+z)^{\D_{312}+n-1}(1-z^2)^{m}}\\
& =
\ppair{\frac{x_{12}}{2}}^{2\D_{3}+n+m-2}
(-1)^{m}2^{m}m!
(1-R)^{\D_{321}-1}(1+R)^{\D_{312}+n-1}
P^{(\D_{321}-1,\D_{312}+n-1)}_{m}(R).\nn
\end{align}
Substituting it into \eqref{eq:opeblockgcft4}, we get
\begin{equation}
I_{n,m}(x_{12},0)
=x_{12}^{n + m}\cdot 
\frac{2^{-n - m} \xi_3^{-m}}{n!}   (1 + R)^{n}   P^{(\D_{321}-1,\D_{312}+n-1)}_{m}(R).
\end{equation}
Hence the closed form of the OPE block is
\begin{align}
\cD_{123}(\xy,\p_{x},\p_{y})
&=x^{-\D_{12,3}}e^{\xi_{12,3}\frac{y}{x}} 
\sum_{n,m} \frac{ (2\xi_3)^{-m}}{n!}
\ppair{\frac{1 + R}{2}}^{n} 
P^{(\D_{32,1}-1,\D_{31,2}+n-1)}_{m}(R)\nn
\\
& \peq\cdot
(x\p_{x}+y\p_{y})^{n} (x\p_{y})^{m}.
\end{align}
In the above, we have used the definition and the Rodrigues' formula for the Jacobi polynomials
\begin{align}
P_{n}^{(a, b)}(z)
& =
\frac{(a+1)_{n}}{n !}\Fba(-n, 1+a+b+n ; a+1 ; \half(1-z))\\
& =
\frac{(-1)^{n}}{2^{n} n !}(1-z)^{-a}(1+z)^{-b} \dv[n]{z}\bpair{(1-z)^{a}(1+z)^{b}(1-z^{2})^{n}}
\end{align}

\subsection{Relation between shadow coefficient and OPE block normalization}\label{sec:relationshadowcoefficientandopeblock}
Consider the three-point structure $\vev{\op_1\op_2\shadow{\opshadow_3}}$, using \eqref{eq:star1} and then inserting the OPE block \eqref{eq:opeblockshadow1} into the three-point function $\vev{\op_1 \op_2 \op_3}$, we get
\begin{align}
\vev{\op_1\op_2\shadow{\opshadow_3}(\xy{4})}
& =\shadow(\op_1\op_2[\opshadow_3])\vev{\op_1\op_2\op_3(\xy{4})}\nn\\
& =\shadow(\op_1\op_2[\opshadow_3])N_{123}\int_I \vev{\op_1\op_2\opshadow_3}\vev{\op_3\op_3(\xy{4})}\label{eq:shadowco11}\\
& =\int_{\R^2} \vev{\op_1\op_2\opshadow_3}\vev{\op_3\op_3(\xy{4})},\label{eq:shadowco21}
\end{align}
where the last line comes from expanding the definition of $\shadow{\opshadow_3}$ in $\vev{\op_1\op_2\shadow{\opshadow_3}}$. 

Notice that the integral regions in \eqref{eq:shadowco11} and \eqref{eq:shadowco21} are not the same. Assuming $x_1>x_2$, the OPE convergence condition implicitly used in \eqref{eq:shadowco11} is $x_4\notin (x_2,x_1),\,R\in(-1,1)$, and the OPE block integration is localized on $\set{X'}\times \R$, where
\begin{equation}
X'=\half(x_1+x_2)+\frac{R}{2}(x_1-x_2) \in (x_2,x_1).
\end{equation}
We can re-parametrize $x_4$ as
\begin{equation}
   x_4=\half(x_1+x_2)+\frac{R_0}{2}(x_1-x_2), 
\end{equation}
then the former conditions are summarized as $[R,1,R_0,-1]$, where $[\dots ]$ denotes the cyclic order on the real projective line $\mathbb{RP}^{1}$. On the other hand, from \eqref{eq:localizationpoint1} the shadow coefficient integration \eqref{eq:shadowco21} is localized on
\begin{equation}
X = \frac{-(1-R)(1+R_0)x_1+(1-R_0)(1+R)x_2}{2(R-R_0)},
\end{equation}
and the relations are
\begin{align}
& X\in(x_2,x_1)\implies [R,1,R_0,-1] \qor [R_0,1,R,-1],\nn\\
& X\in(x_1,\inf)\implies [R,R_0,1,-1] \qor [R_0,R,-1,1],\nn\\
& X\in(-\inf,x_2)\implies [R_0,R,1,-1] \qor [R,R_0,-1,1].
\end{align}
Hence if the weights $\xi_i$ satisfy $R\in(-1,1)$, the two integral expressions \eqref{eq:shadowco11} and \eqref{eq:shadowco21} hold simultaneously and the normalization factor is
\begin{equation}
N_{123}^{-1}=\shadow(\op_1\op_2[\opshadow_3])= 2^{2\D_3-2}\frac{\pi}{|\xi_3|}(1-R)^{-1+\D_{23,1}}(1+R)^{-1+\D_{31,2}},
\end{equation}
matching with the result \eqref{eq:opeblocknormalization}.

\subsection{Check of the second term in bubble integral}\label{sec:secondtermbubble}
To check the result \eqref{eq:bubblelast}, we calculate the second term in \eqref{eq:bubble2} directly. Assuming $r_3r_4<0$, the $\d$-distribution localized on the variety \eqref{eq:secondvariety} is
\begin{equation}
\d(A_1)\d(A_2)=J_0\d(r_{3}+r_{4})\d\ppair{x_{2}-\frac{r_{13,2} x_1 x_3+r_{23,1} x_1x_4 -2 r_3 x_3x_4 }{r_3 \left(2 x_1-x_3-x_4\right)+r_{12} x_{34}}},
\end{equation}
where $J_0$ is a lengthy Jacobian factor
\begin{equation}
J_0=\frac{x_{12}^{2} x_{13} x_{23} x_{24}^{2}}{r_{13,2} x_{12} x_{23}^{2}+x_{24}^{2} \left(r_3 \left(3 x_1+x_2-4 x_3\right)-r_{12}x_{12}\right)}.
\end{equation}
Then we have
\begin{align}
B_1(\op_3,\op_4,x_{34},y_{34})
& =A\,\shadow(\opshadow_1\opshadow_2[\op_4])^{-1}\inttt{dx_1}{\R}|x_{13}|^{-i(s_3+s_4)-1}|x_{14}|^{i(s_3+s_4)-1}\nn\\
& =A\,\shadow(\opshadow_1\opshadow_2[\op_4])^{-1}4\pi \d(s_3+s_4)\frac{1}{|x_{34}|}\nn
\\
& =\shadow(\opshadow_1\opshadow_2[\op_4])^{-1}\d(\op_3,\opshadow_4)\vev{\op_3(\xy{3})\op_3(\xy{4})},
\end{align}
where the prefactor is
\begin{equation}
A=2^{-1+\D_3+\D_4}\pi^3 |r_3|^{-4+\D_3+\D_4}|r_{23,1}|^{2-\D_3-\D_4}e^{2\xi_3\frac{y_{34}}{x_{34}}}|x_{34}|^{-1-\D_3+\D_4}.
\end{equation}
Notice that in the first line we cannot evaluate the integral by analytic continuation of $s_3,\, s_4$  \eqref{eq:klt1} since the factor $A$ is singular at $x_{34}=0$. This justifies the result \eqref{eq:bubblelast} determined by shadow transform.

\subsection{Deriving the Casimir equations} \label{sec:derivingCasimirequations}

We may start from inserting a complete basis into the four-point functions in the radial quantization $x_4>x_3>x_2>x_1$,
\begin{equation}
\vev{\op_1 \op_2 \op_3 \op_4}=\bra{0} \op_4 \op_3 \sum |\op_0|\, \op_2 \op_1\vac, 
\end{equation}
where $|\op_0|$ denotes the projection operator with respect to the conformal family $\rep{\D_0,\xi_0,r}$
\begin{equation}
|\op_0|=\sum_{n,m\in \rep{\D_0,\xi_0,r}}G^{-1}{}_{n,m}\ket{n}\bra{m}.
\end{equation}
and $G_{n,m}=\bra{n}\ket{m},\, n,m\in \rep{\D_0,\xi_0,r}$ is the Gramian matrix of the inner product. In this way we can rewrite the conformal blocks \eqref{eq:block4gcft} as a summation over the matrix elements $\bra{n}\op_1 \op_2 \vac$, which are rescaled version of the three-point functions involving the descendants,
\begin{equation}
\label{eq:block5}
\blockdressed^{\schannel}_0(\xy{i})=\bra{0} \op_4(\xy{4}) \op_3(\xy{3}) |\op_0| \op_2(\xy{2}) \op_1(\xy{1})\vac.
\end{equation}
In principle we can use this projection operator to calculate the conformal blocks directly. When $\xi_0=0$ or $\D_0=0,-1,\dots$, the conformal family contains null states and we need to mod out them to get an invertible Gramian matrix \cite{Chen:2022jhx}. In this subsection we focus on the singlet case $r=1,\xi_0\neq 0$.

The Casimir differential operators are the representations of the abstract Casimir elements acting on the matrix elements $\bra{n}\op_1 \op_2 \vac$. For $X\in\isolie(2,1)$, the Ward identity is
\begin{align}
\bra{n} X \op_1 \op_2 \vac
& =
\bra{n}[X,\op_1(\xy{1})]\op_2(\xy{2})\vac+
\bra{n}\op_1(\xy{1})[X,\op_2(\xy{2})]\vac\nn\\
& =(X^{(1)}+X^{(2)})\bra{n}\op_1(\xy{1})\op_2(\xy{2})\vac,
\label{eq:Wardgcft1}
\end{align}
where the differential operators $X^{(i)}$ are from $X^{(i)}\vev{\dots \op_i(\xy{i})\dots }=\vev{\dots [X,\op_i(\xy{i})]\dots }$.

The Galilean conformal algebra $\isolie(2,1)$ admits two algebraically independent Casimir elements \eqref{eq:casimirPoincare}, and they act on the singlet $\rep{\D_0,\xi_0}$ as scalars $\lm_{i}$,
\begin{equation}
C_1=\lm_1=\xi_0^2, \hs 
C_2=\lm_2=2\xi_0 (\D_0-1).
\label{eq:casimirsinglet}
\end{equation}
In other words, the Casimirs commute with the projection operators $C_{i}|\op_0|=|\op_0|C_{i}$. By repeatedly using \eqref{eq:Wardgcft1} we get
\begin{equation}
C^{(1+2)}_i\bra{n}\op_1 \op_2 \vac
=
\lm_i \bra{n}\op_1 \op_2 \vac,
\label{eq:Casimirgcft5}
\end{equation}
where the differential operators $C^{(1+2)}_i$ are of the forms
\begin{align}
C^{(1+2)}_1
& =
-x_{12}^2\frac{\p^2}{\p y_{1} \p y_{2}}+2x_{12}\ppair{\xi_1 \pdv{y_2}-\xi_2 \pdv{y_1}} +(\xi_1+\xi_2)^2,
\\
C^{(1+2)}_2
& = 
2 x_{12} y_{12} \frac{\p^2}{\p y_{1} \p y_{2}} 
+ x_{12}^{2}\ppair{\frac{\p^2}{\p x_{1} \p y_{2}}+\frac{\p^2}{\p x_{2} \p y_{1}}}
\nn\\
&\peq 
+ 2x_{12} \ppair{\D_1 \pdv{y_2}- \D_2 \pdv{y_1} - \xi_1 \pdv{x_2} +\xi_2 \pdv{x_1}}
\nn\\
&\peq  + 2y_{12} \ppair{ - \xi_1 \pdv{x_2} +\xi_2 \pdv{x_1}}
+ 2(\D_1+\D_2-1)(\xi_1+\xi_2).
\end{align}
Inserting \eqref{eq:Casimirgcft5} into the expression of the conformal blocks \eqref{eq:block5}, we get the Casimir equations,
\begin{equation}
\ppair{C^{(1+2)}_i-\lm_{i}}\blockdressed^{\schannel}_0=0,\qquad i=1,2.
\label{eq:singletCasimir}
\end{equation}
Then plugging the stripped conformal blocks \eqref{eq:block1gcft} into these equations, we get the Casimir equations of the stripped conformal blocks
\begin{equation}
\ppair{\cC_i-\lm_{i}}\blockstripped^{\schannel}_0(\xy)=0,\qquad i=1,2.
\end{equation}
To simplify the calculations we work in the slope coordinates $(x,k)=(x,\frac{y}{x})$, and use the same symbol $g^{\schannel}(x,k)$ denoting the \schannel conformal block. In the slope coordinates we have
\begin{align}
\cC_1
& =
(1-x)\pdv[2]{k}
+(-\xi_{12}+\xi_{34}) x \pdv{k}
+\xi_{12} \xi_{34}x,
\\
\cC_2
& =
x k \pdv[2]{k}
+2x(x-1)\frac{\p^2}{\p x \p k}\nn
\\
& \peq 
+\ppair{2+(-\D_{12}+\D_{34})x+(\xi_{12}-\xi_{34})x k }\pdv{k}
+(\xi_{12}-\xi_{34}) x^2 \pdv{x}\nn
\\
& \peq 
+\ppair{\D_{12}\xi_{34}+\D_{34}\xi_{12}-\xi_{12}\xi_{34} k }x.
\end{align}

\subsection{Solving the Casimir equations}\label{sec:solvingCasimirEquations}
In this appendix we solve the Casimir equations and get the conformal block of singlet exchanged operator. For convenience we introduce the notation $R_{ij}\equiv\frac{\xi_{ij}}{\xi_{0}}$. Notice that under the shadow symmetry 
\begin{equation}
(\D_0,\xi_0) \to (2-\D_0,-\xi_0), \hs{3ex} R_{ij}\to -R_{ij}.
\end{equation}
The Casimir equations are
\begin{align}
& \ppair{\cC_1-\lm_{1}}\blockstripped^{\schannel}_0(\xy)=0,\label{Casimir1}\\
& \ppair{\cC_2-\lm_{2}}\blockstripped^{\schannel}_0(\xy)=0.\label{Casimir2}
\end{align}
The first Casimir equation \eqref{Casimir1} can be solved by setting $F(x,k)=F_{1}(x)e^{k F_{2}(x)}$, and the solutions are
\begin{equation}
F(x,k)=c_1 f_{+}(x)\exp\bpair{\frac{k}{1-x}\ppair{\frac{\xi_{12}-\xi_{34}}{2}x+ h_{+}(x)}} + c_2 f_{-}(x)\exp\bpair{\frac{k}{1-x}\ppair{\frac{\xi_{12}-\xi_{34}}{2}x+ h_{-}(x)}},\label{solutionCasimir1}
\end{equation}
where 
\begin{equation}
h_{\pm}^2(x)= \xi_{0}^2-(\xi_{0}^2+\xi_{12}\xi_{34})x +\frac{1}{4}(\xi_{12}+\xi_{34})^2 x^2.
\end{equation}
Before solving the second Casimir equation \eqref{Casimir2}, there are two subtleties to be stressed. Firstly, to manifest the shadow symmetry $\xi_0\to -\xi_0, h_{\pm}\to h_{\mp}$, and to match the result with $\xi_{12}=\xi_{34}=0$, we need to choose the branch cut of $\xi_i$'s as 
\begin{equation}
h_{\pm}(x)\equiv \pm \xi_0  H(x)= \pm \xi_0 \sqrt{1-(1+R_{12}R_{34})x +\frac{1}{4}(R_{12}+R_{34})^2 x^2}. 
\label{eq:functionH}
\end{equation}
This does not change the solutions, and can be understood from the shadow formalism.
% Secondly, when the exchanged operator is degenerate, $\xi_0=0$, and $\xi_{12}\xi_{34}\neq 0$, $h(x)$ has unusual \schannel OPE behaviour $\sim \exp k(a x+b\sqrt{x})$. We deal with the case $\xi_0=0$ in [ref \tba].
Secondly, $H(x)$ contains two branch points with respect to $x$
\begin{equation}
x_{\pm}=2\frac{1+R_{12}R_{34} \pm \sqrt{(1-R_{12}^2)(1-R_{34}^2)}}{(R_{12}+R_{34})^2},
\label{eq:rootsOfHx}
\end{equation}
and the condition $x_{\pm}\notin (0,1)$ rules out the following regions
\begin{equation}
R_{12},\,R_{34}>1, \qand R_{12},R_{34}<-1.
\end{equation}

Let us continue solving the Casimir equations. With the solution \eqref{solutionCasimir1}, the second Casimir equation \eqref{Casimir2} is reduced to a first-order differential equation of $f_{\pm}(x)$
\begin{equation}
\ppair{f'(x)/f(x)}_{\pm}=-\frac{ A_1(x)h_{\pm}(x) +A_2(x)}{8 (1-x)x h_{\pm}^2(x)},
\end{equation}
where
\begin{align}
A_{1}(x)
& =
2(\D_{12}+\D_{34})(\xi_{12}+\xi_{34}) x^2
- 4\ppair{2(\D_{0}-1)\xi_{0}+\D_{34}\xi_{12}+\D_{12}\xi_{34}} x
+ 8(\D_{0}-1)\xi_{0},
\nn
\\
A_{2}(x)
& =
(\D_{12}-\D_{34})(\xi_{12}+\xi_{34})^2 x^3
- 4(1+\D_{12}-\D_{34})(\xi_{0}^{2}+\xi_{12}\xi_{34})x^2
\nn
\\
& \peq + 4\ppair{(3+\D_{12}-\D_{34})\xi_{0}^{2}+\xi_{12}\xi_{34}} x 
-8\xi_{0}^{2}.
\end{align}
Hence the solutions of the two Casimir equations are
\begin{equation}
F(x,k)=c_1 \blockstripped_{+}(x,k) + c_2 \blockstripped_{-}(x,k),
\end{equation}
where
\begin{equation}
\blockstripped_{\pm}(x,k)=\exp\bpair{\frac{k}{1-x}\ppair{\half (\xi_{12}-\xi_{34})x \pm \xi_{0}H(x)}-\intt{dx}\frac{  A_1(x)h_{\pm}(x) +A_2(x)}{ 8(1-x)x h_{\pm}^2(x)}}.
\end{equation}

By checking the \schannel OPE limit $x, k\to 0$, and redefining the normalization to ensure the exchanged primary operator contributes one: $\blockstripped_{\D_0,\xi_0}^{\schannel}\sim x^{\D_0} e^{-k \xi_0 }$, we find that the second solution $\blockstripped_{-}(x,k)$ can be identified to be the physical block
\begin{align}
\blockstripped_{\D_0,\xi_0}^{\schannel}(x,k)
& =
\frac{N(\D_0,\xi_0)}{H(x)}
\exp\bpair{\frac{k}{1-x}\ppair{\half (\xi_{12}-\xi_{34})x - \xi_{0}H(x)}}
\nn
\\
& \peq
\cdot x^{\D_0}
\bpair{\xi_{0}^{2}-\half(\xi_{0}^{2}+\xi_{12}\xi_{34}) x + \xi_{0}^{2}H(x)}^{1-\D_0} 
\nn
\\
& \peq
\cdot \bpair{ \xi_{0}^{2}-\xi_{12}\xi_{34} + \half (2\xi_{0}^{2}- \xi_{12}^{2}- \xi_{34}^{2})x + (\xi_{12}-\xi_{34}) \xi_{0}H(x)}^{\half(\D_{12}-\D_{34})}
\nn
\\
& \peq 
\cdot \bpair{ \xi_{0}^{2}+\xi_{12}\xi_{34} -\half (\xi_{12}+\xi_{34})^2 x + (\xi_{12}+\xi_{34}) \xi_{0}H(x)}^{\half(\D_{12}+\D_{34})}
\label{eq:strippedblock1}
\end{align}
where the normalization factor is
\begin{equation*}
N(\D_0,\xi_0)=2^{\D_0-1} \xi_{0}^{2\D_{0}-3} \bpair{(\xi_0+\xi_{12})(\xi_0-\xi_{34})}^{-\half(\D_{12}-\D_{34})} \bpair{(\xi_0+\xi_{12})(\xi_0+\xi_{34})}^{-\half(\D_{12}+\D_{34})},
\end{equation*}
and the other solution $\blockstripped_{+}(x,k)$ is proportional to the shadow block $\blockstripped_{2-\D_0,-\xi_0}^{\schannel}(x,k)$, thus the two solutions respect the shadow symmetry.

\subsection{Calculation of the projector of conformal partial waves}\label{sec:completeness1}
In this subsection we show that for four identical external operators the projection operator of conformal partial waves \eqref{eq:projector1} is proportional to
\begin{equation}
\cP^{1'2'3'4'}_{1234}\sim \th(1-x_{12,34})\d(x_{12,34}-x'_{12,34})\d(k_{12,34}-k'_{12,34}).
\label{eq:projector11}
\end{equation}
The overall coefficient is irrelevant to our discussion and has been omitted. 
For the $\f_1\f_2\f_2\f_1$-type external operators, there seems no closed-form of this projector because the associated shadow variety reduces to higher order equations of $\set{x_0,x'_0,y_0,y'_0}$, and the remaining integration is hard to calculate.

Inserting the definition of conformal partial wave \eqref{eq:shadowCPW1} into the projection operator \eqref{eq:projector1} we have,
\begin{align}
&\peq \cP^{1'1'1'1'}_{1111}
\nn
\\
&=
\inttt{\frac{dr_{0}ds_{0}}{4\pi^{2} \cN(\D_{0},\xi_{0})}}{\R\times \Rpositive}
\intt{dx_{0}dy_{0}dx'_{0}dy'_{0}}
\vev{\opshadow'_1\opshadow'_1\opshadow'_{0}}\vev{\op'_0\opshadow'_1\opshadow'_1}
\vev{\op_1\op_1\op_0}\vev{\opshadow_0\op_1\op_1}
\nn
\\
&=
\inttt{\frac{dr_{0}ds_{0}}{4\pi^{2} \cN(\D_{0},\xi_{0})}}{\R\times \Rpositive}
\intt{dx_{0}dy_{0}dx'_{0}dy'_{0}}
\vpair{F_{0}}^{i\, s^0}
F_1 
e^{i(A_0 y_{0}r_{0}+A'_0 y'_{0}r_{0}+A_{1}r_{0}+A_2)}
\nn
\\
&=\intt{dx_{0} dx'_{0}}
\d(A_0)\d(A'_{0})\d(A_{1})\d(F_{0})F_1 e^{iA_2},
\label{eq:projectorcalculation}
\end{align}
where $\op'_{i}=\op_{i}(x'_{i},y'_{i})$. In the second line the integral region of $s_{0}$ has been changed from $(0,\inf)$ to $\R$ using \eqref{eq:inversionformula}, and we need to select the correct $\d$-distribution related to the conformal block in the partial wave. In the last line the substitution $r_{0}y_{0}\to y_{0},\,r_{0}y_{0}\to y_{0} $ is used to separate the first three terms in the exponential part, and the factor $r_{0}^2$ from $\cN$ gets canceled so that the integration of $y_{0},\, y'_{0},\, r_{0}$ gives three $\d$-distributions. The $s_{0}$-dependence is collected into $F_{0}$, and the integration gives $\d(F_{0})$. The divergent volume factor is kept since we haven't done the gauge fixing procedure. One can check that after fixing $(x'_{i},y'_{i})$ to the standard conformal frame and renormalizing $\op'_{1}(x'_{4},y'_{4})$ as \eqref{eq:innerproductfourpoint}, the integral is finite.

In total the shadow variety $\cV$ is defined by four equations,
\begin{alignat}{2}
&A_{0}:   \hs    && \frac{1}{x_{01}}+\frac{1}{x_{02}}-\frac{1}{x_{03}}-\frac{1}{x_{04}}=0,
\nn\\
&A'_{0}: &&             \frac{1}{x'_{01}}+\frac{1}{x'_{02}}-\frac{1}{x'_{03}}-\frac{1}{x'_{04}}=0,
\nn
\\
&A_{1}:&&  
\frac{y_{1}}{x_{01}}+\frac{y_{2}}{x_{02}}-\frac{y_{3}}{x_{03}}-\frac{y_{4}}{x_{04}}
-\frac{y_{12}}{x_{12}}+\frac{y_{34}}{x_{34}}
\nn
\\
&&&
=\frac{y'_{1}}{x'_{01}}+\frac{y'_{2}}{x'_{02}}-\frac{y'_{3}}{x'_{03}}-\frac{y'_{4}}{x'_{04}}
-\frac{y'_{12}}{x'_{12}}+\frac{y'_{34}}{x'_{34}},
\nn
\\
&F_{0}:&&\frac{x_{12}x_{30}x_{40}x'_{34}x'_{10}x'_{20}}{x_{34}x_{10}x_{20}x'_{12}x'_{30}x'_{40}}=\pm 1,
\nn
\end{alignat}
in which there are two irreducible components of $F_{0}$.

The rest integrals in \eqref{eq:projectorcalculation} can be done by solving $x_{0},\, x'_{0}$ from $A_{0}=A'_{0}=0$, giving rise to the overall factor in \eqref{eq:projector11}, then one irreducible component of $\d(F_{0})\d(A_{1})$ is proportional to $\d(x_{12,34}-x'_{12,34})\d(k_{12,34}-k'_{12,34})$.
The Heaviside function is originated from the reality condition of the shadow variety. The solution of $A_{0}=0$ is 
\begin{equation}
x_{0,\pm}=\frac{x_{1}x_{2}-x_{3}x_{4}\pm\sqrt{x_{13}x_{23}x_{14}x_{24}}}{x_{13}-x_{24}},
\end{equation}
and the projection operator \eqref{eq:projectorcalculation} is nonvanishing only if the roots are real:
\begin{equation}
x_{13}x_{23}x_{14}x_{24}>0,\implies x_{12,34}<1.
\end{equation}

Another component of $\d(F_{0})\d(A_{1})$ is related to the first one by
\begin{equation}
x_{12,34}\to \frac{x_{12,34}}{x_{12,34}-1}, \hs y_{12,34}\to -\frac{y_{12,34}}{(1-x_{12,34})^2}.
\end{equation}
This is exactly the permutation symmetry of the four-point functions with identical external operators, and in result the projected four-point functions respect this symmetry automatically.

\newpage
\printbibliography
\end{document}